\documentclass[12pt,preprint]{aastex}

\newcommand\hms{h~~m~~s}
\newcommand\dms{$~~~^{\circ}~~~^{\prime}~~~^{{\prime}{\prime}}$}

\received{}
\accepted{}
\journalid{}{}
\articleid{}{}

\slugcomment{Astronomical Journal, submitted}

\shortauthors{Landolt \& Uomoto}
\shorttitle{$UBVRI$ photometry of spectrophotometric standards}

\begin{document}

\title{OPTICAL MULTI-COLOR PHOTOMETRY OF \\ SPECTROPHOTOMETRIC STANDARD STARS}

\author{Arlo U.~Landolt\altaffilmark{1,2}}
\affil{Department of Physics \& Astronomy, Louisiana State University, Baton Rouge, LA 70803-4001}
\email{landolt@phys.lsu.edu}

\and

\author{Alan K.~Uomoto\altaffilmark{1}}
\affil{Carnegie Observatories, 813 Santa Barbara St., Pasadena, CA 91101-1292}
\email{au@ociw.edu}

\altaffiltext{1}{Visiting Astronomer, Kitt Peak National Observatory.}
\altaffiltext{2}{Visiting Astronomer, Cerro Tololo Inter-American Observatory, National Optical 
Astronomical Observatories, which are operated by the Association of Universities for Research in 
Astronomy, under contract with the National Science Foundation.}

\begin{abstract}
\label{sec:abstract}

Photoelectric data on the Johnson-Kron-Cousins $UBVRI$ broadband photometric system are provided for a set 
of stars which have been used as spectrophotometric standard stars at the Hubble Space Telescope.

\end{abstract}


\section{Introduction}
\label{sec:introduction}

A document originally prepared by \citet{Bohlin1990} and published as \citet{Turnshek1990}, provided 
two lists of stars to be used as spectrophotometric standard stars by the Hubble Space Telescope (HST) 
instruments.  One table listed some stars which either had known ultraviolet fluxes or which could have 
their ultraviolet fluxes measured with the International Ultraviolet Explorer (IUE) satellite.  A second 
table identified stars which might prove useful standard stars but which were too faint to be observed with 
the IUE.  These stars could be used to intercompare calibrations for the different HST instruments.

The papers by \citet{Bohlin1990} and \citet{Turnshek1990} discussed the considerations that led to 
selecting these spectrophotometric standard star candidates.  Their goal was to construct a list of 
stars lengthy enough to encompass all HST instrument calibration requirements but short enough to 
minimize data collection efforts.  An effort also was made to identify objects accessible to both HST 
and ground-based instrumentation.  They chose at least two stars for each calibration requirement to 
avoid possible variability in any individual star.  They avoided strong-lined stars because objects with 
such spectra complicate absolute calibration effects.  Finally, stars covering a large magnitude range 
were included to allow instrumentation linearity checks.

All of the stars in this program were observed at the behest of HST staff, with the majority of the stars 
being taken from lists in \citet{Bohlin1990} and \citet{Turnshek1990}.  It may be noted that an early 
version of the photometry presented in this paper is the basis for the photometry of stars in common with 
those in the HST Calspec database.

\section{Observations}
\label{sec:observations}

The Kitt Peak National Observatory (KPNO) 1.3-m telescope was scheduled for this program for 101 nights 
in the interval 1985 September to 1991 June.  Of those scheduled nights, 49, or 48.5 percent, provided 
usable photometric data.

The broad-band $UBVRI$ photometric observations all were obtained with the same RCA 31034A-02 (KPNO 
serial no. H 18862) type photomultiplier used in a pulse counting mode.  The photomultiplier always 
was kept in cold box no. 51 and was operated at -1600 volts.  The 1.3-m telescope was operated in its 
chopping mode; ten seconds were spent on a star, and then ten seconds on the sky, over a twenty second 
interval of time.  The data were recorded on magnetic tape, and were reduced on the IBM 3090 computer 
at the Louisiana State University System Network Computer Center.

The KPNO ``J" $UBVRI$ filter set was used throughout the data acquisition process at KPNO, with one 
exception.  The 1985 September and December observing runs made use of an ultraviolet $U$ filter 
combination of Corning 9863, plus a solid CuSO$_{4}$ crystal.  The same $BVRI$ filters were used 
throughout the program.  Their specifications, plus the specification for the $U$ filter used for all 
except the late 1985 September and December runs, as laid down by \citet{Bessell1979}, are given in 
Table \ref{tab:table1}.

On average, 23 $UBVRI$ standard stars, as defined by \citet{Landolt1983}, were observed each night 
together with the program stars.  Standard stars were observed in groups of four or five periodically 
throughout the night.  Each such group, physically close together on the sky, contained stars in as wide 
a color range as possible.  An attempt was made to ensure that the standard star observations 
encompassed as wide a range in air mass as did the program stars.  Almost all program star measures were 
taken at less than 1.5 air masses.  The exception was the star AGK+81$^{\circ}$266 whose northern 
declination meant that data for it were obtained between 1.56 and 1.75 air masses.

A complete data set for a star consisted of a series of measures $IRVBUUBVRI$.  Throughout the process, 
the sky was sampled once per second via the telescope's chopping mode.  A 17.7 second of arc diaphragm 
was used (because that was the most reasonable size diaphragm available given the instrument setup).  
Counting intervals, i.e., the time spent on a star, ranged from no less than ten seconds for the 
brightest stars to sixty seconds for the faintest stars.  The longest integrations were constrained by 
the lack of an automatic guiding mode; one had to depend upon the telescope drive to keep the star 
centered for the duration of the observation.  Fortunately, the KPNO 1.3-m was a very stable telescope!  
Data reduction procedures followed the precepts outlined by \citet{SchulteCrawford1961}.

Extinction coefficients were extracted from three or four standard stars possessing a range in color 
index which were followed over to an air mass of 2.1, or so.  Each night's data were reduced using the 
primary extinction coefficients derived from that night, whenever possible.  Average secondary 
extinction coefficients, for a given run, were used.  The average extinction coefficient values found 
over the seventy month observational interval of this project are given in Table \ref{tab:table2}.  It 
is interesting to compare these $UBV$ extinction coefficient values with those from earlier years' data 
obtained at KPNO \citep{Landolt1967, Landolt1973}, and this is done in Table 3.  One notes that, within 
the errors of the data, extinction coefficients essentially have remained unchanged, although, of 
course, there exist on occasion wide variations from night to night, and even during a night 
\citep{Landolt2007}.  Therefore, mean extinction coefficients only should be used with great caution.

A more detailed description of extinction coefficient behavior and the data reduction procedures 
employed by the author may be found in \citet{Landolt2007}.

The final computer printout for each night's reductions contained the magnitude and color indices for 
each of the standard stars.  Since the time of observation was recorded for each measurement, it was 
possible to plot the residuals in the $V$ magnitude and in the different color indices for each 
standard star against Universal Time for a given night.  These plots permitted small corrections to be 
made to all program star measures.  The corrections usually were less than a few hundredths of a 
magnitude.  Such corrections took into account small changes in both atmospheric and instrumental 
conditions which occurred during the course of a night's observations.

A problem was discovered near the end of the observing session in 1986 November in the sense that 
frost had formed on the coldbox's Fabry lens at some point during the course of the run.  The 
subsequent data analysis showed no discernible effect on the derived values of the program stars' 
color indices.  However, small trends did appear in the $V$ magnitudes.  To be on the safe side, all 
of the data from that observing run were discarded.

\section{Discussion}
\label{sec:discussion}

A total of 32 stars, distributed over the sky, made up this observational program.  The data were 
reduced night by night with the results having been tied into the $UBVRI$ photometric system defined by 
\citet{Landolt1983} standard stars.  A thorough check was made to ensure that the $U$ data obtained 
during the 1985 September and December observing sessions were on the same $U$ filter system with which 
the remaining and majority of the data were acquired.

A check on the accuracy of the magnitude and color index transformations was made via a comparison of 
the magnitudes and color indices of the stars from \citet{Landolt1983} which were used as standards 
herein, with the magnitudes and color indices of these same standard stars obtained during this project.  
The comparisons, the delta quantities, were in the sense data from this program $minus$ corresponding 
magnitudes and color indices from \citet{Landolt1992}, since this latter paper was a successor to 
\citet{Landolt1983}.

\begin{figure}
\plotone{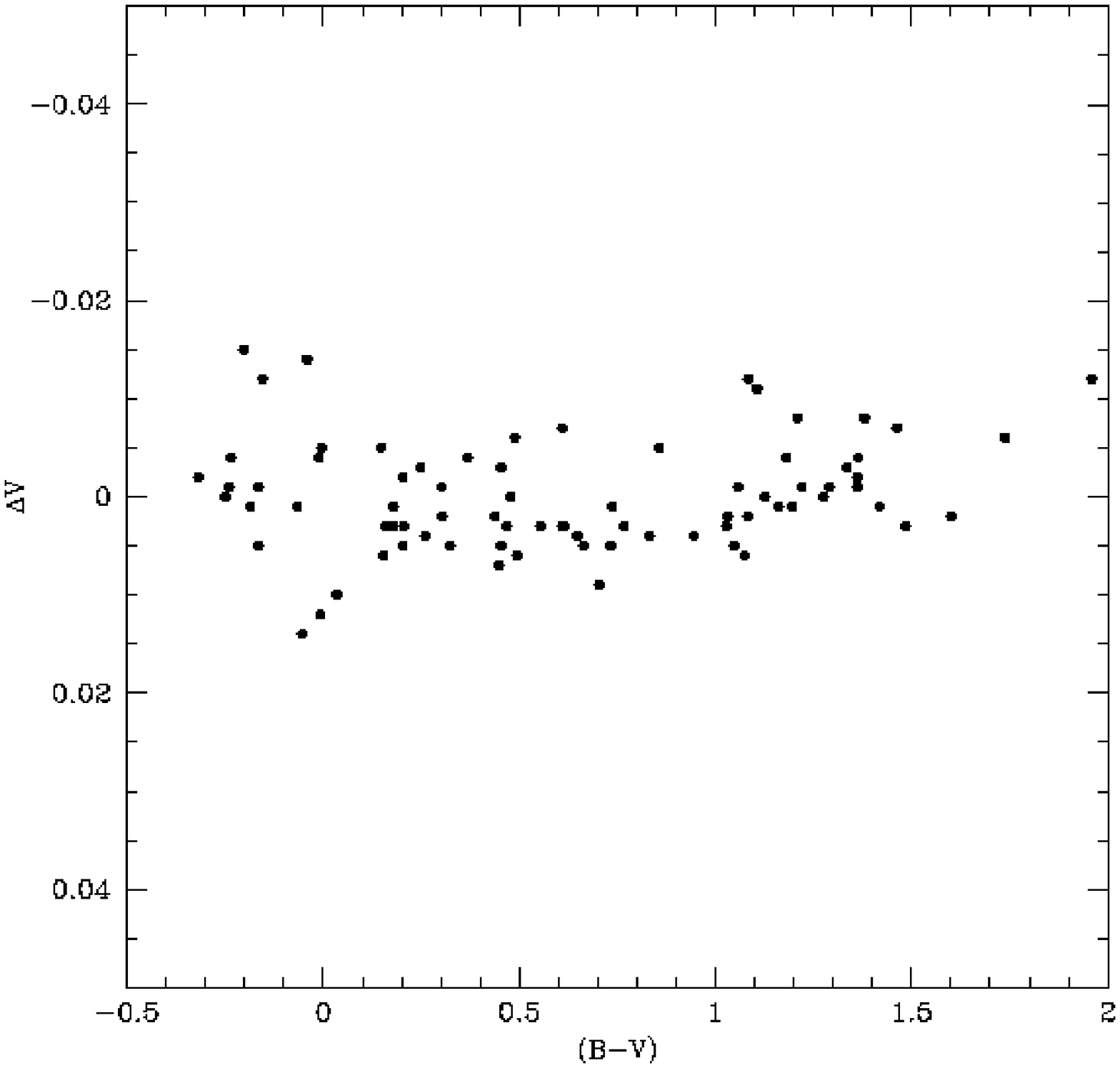}
\caption{A comparison of the $V$ magnitudes tied into \citet{Landolt1983} standard stars as a function of the \citet{Landolt1992} 
equatorial standard's $(B-V)$ color indices.}
\label{fig:figure01}
\end{figure}

\begin{figure}
\plotone{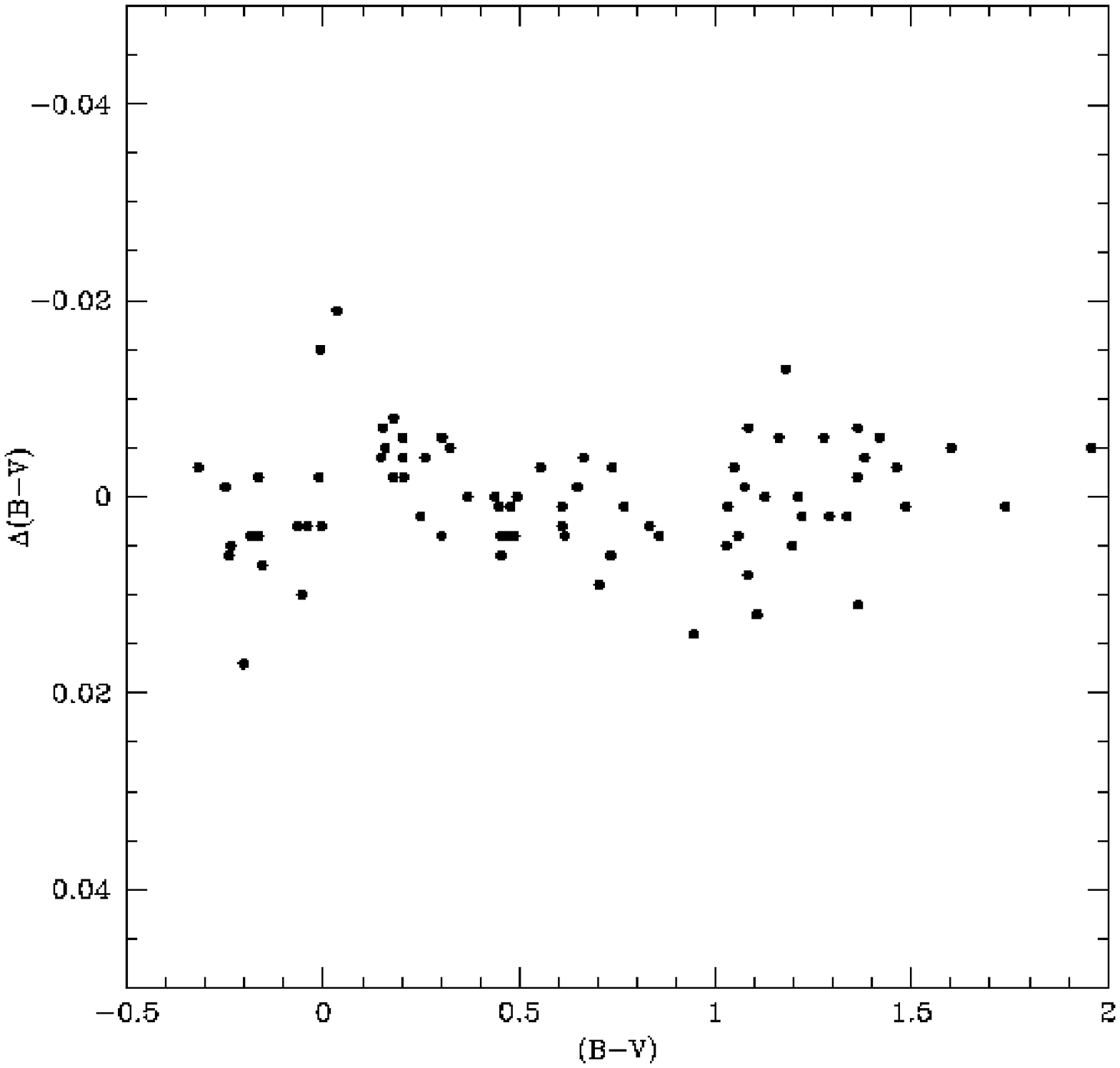}
\caption{A comparison of the $(B-V)$ color indices tied into \citet{Landolt1983} standard stars as a function of the 
\citet{Landolt1992} equatorial standard's $(B-V)$ color indices.}
\label{fig:figure02}
\end{figure}

\begin{figure}
\plotone{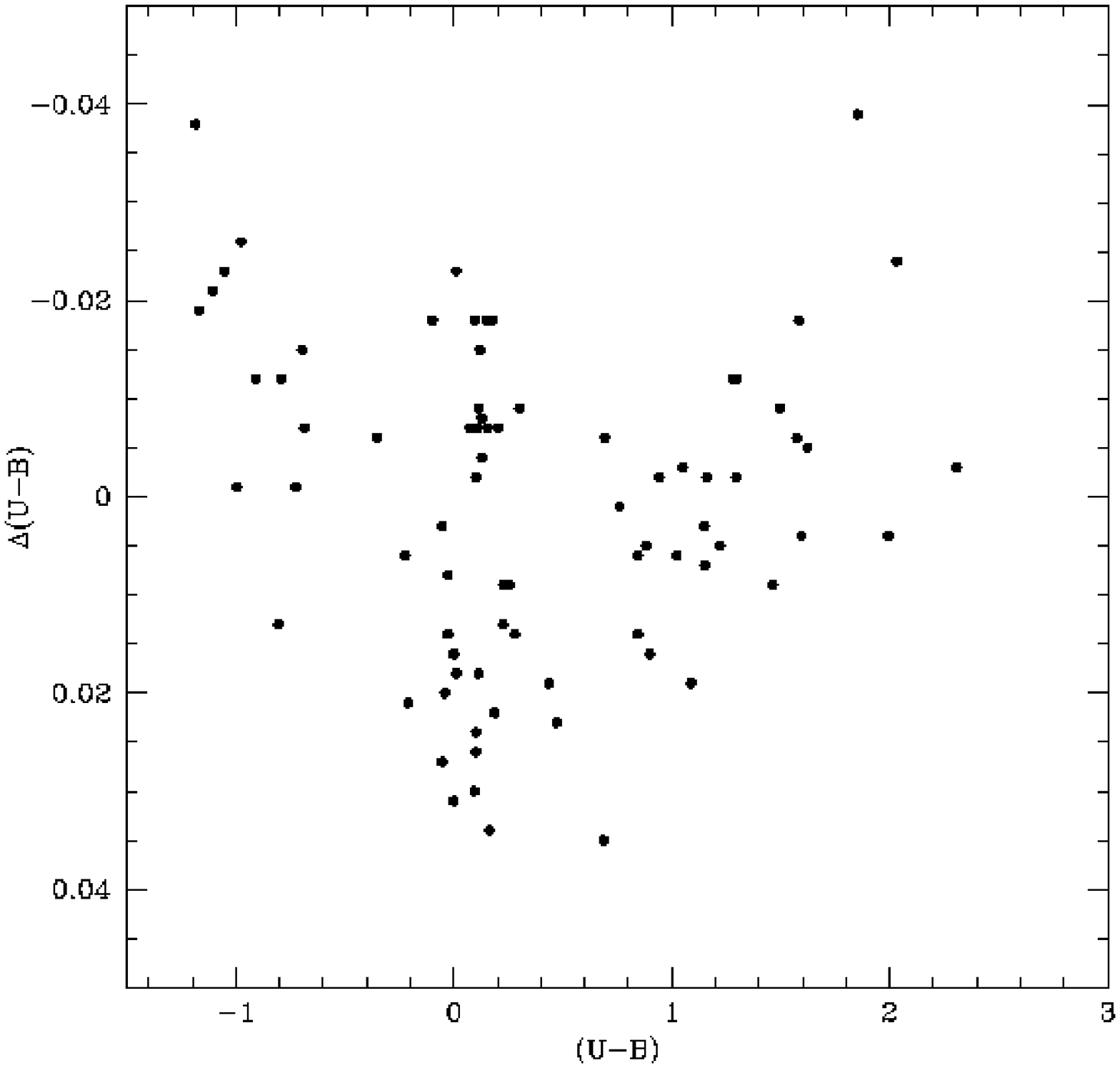}
\caption{A comparison of the $(U-B)$ color indices tied into \citet{Landolt1983} standard stars as a function of the 
\citet{Landolt1992} equatorial standard's $(U-B)$ color indices.}
\label{fig:figure03}

\clearpage
\end{figure}

\begin{figure}
\plotone{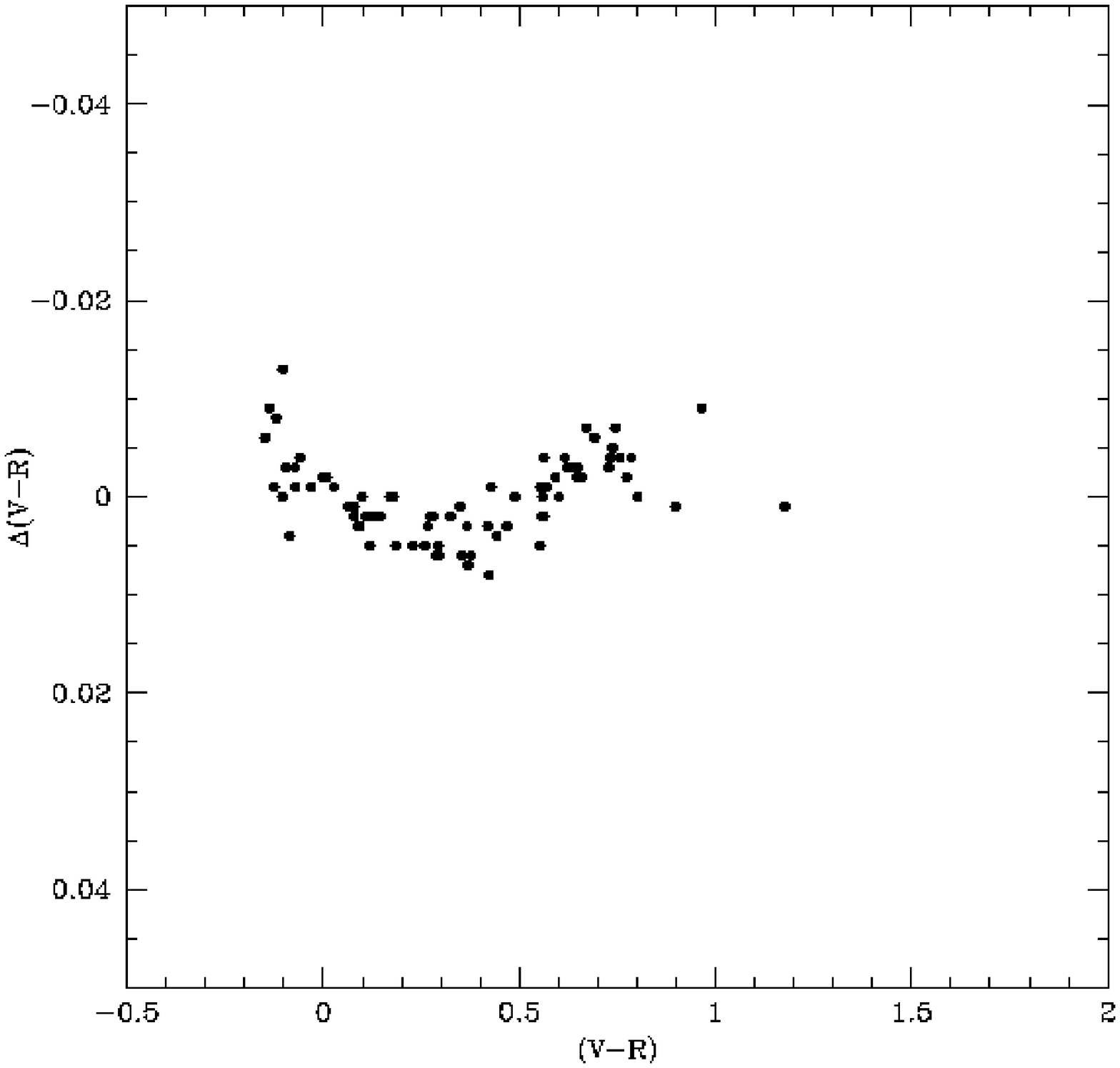}
\caption{A comparison of the $(V-R)$ color indices tied into \citet{Landolt1983} standard stars as a function of the 
\citet{Landolt1992} equatorial standard's $(V-R)$ color indices.}
\label{fig:figure04}

\clearpage
\end{figure}

\begin{figure}
\plotone{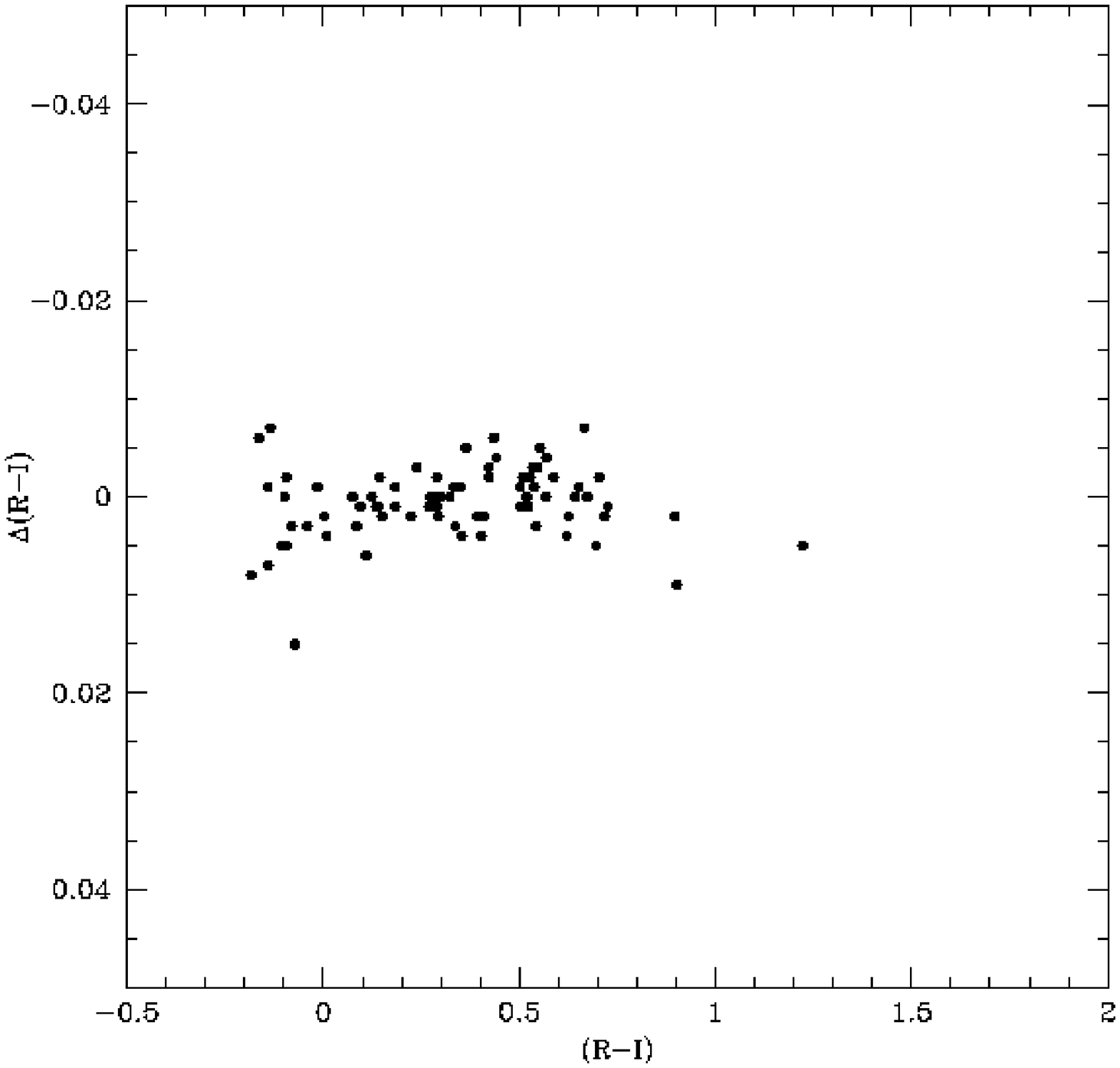}
\caption{A comparison of the $(R-I)$ color indices tied into \citet{Landolt1983} standard stars as a function of the 
\citet{Landolt1992} equatorial standard's $(R-I)$ color indices.}
\label{fig:figure05}
\end{figure}

\begin{figure}
\plotone{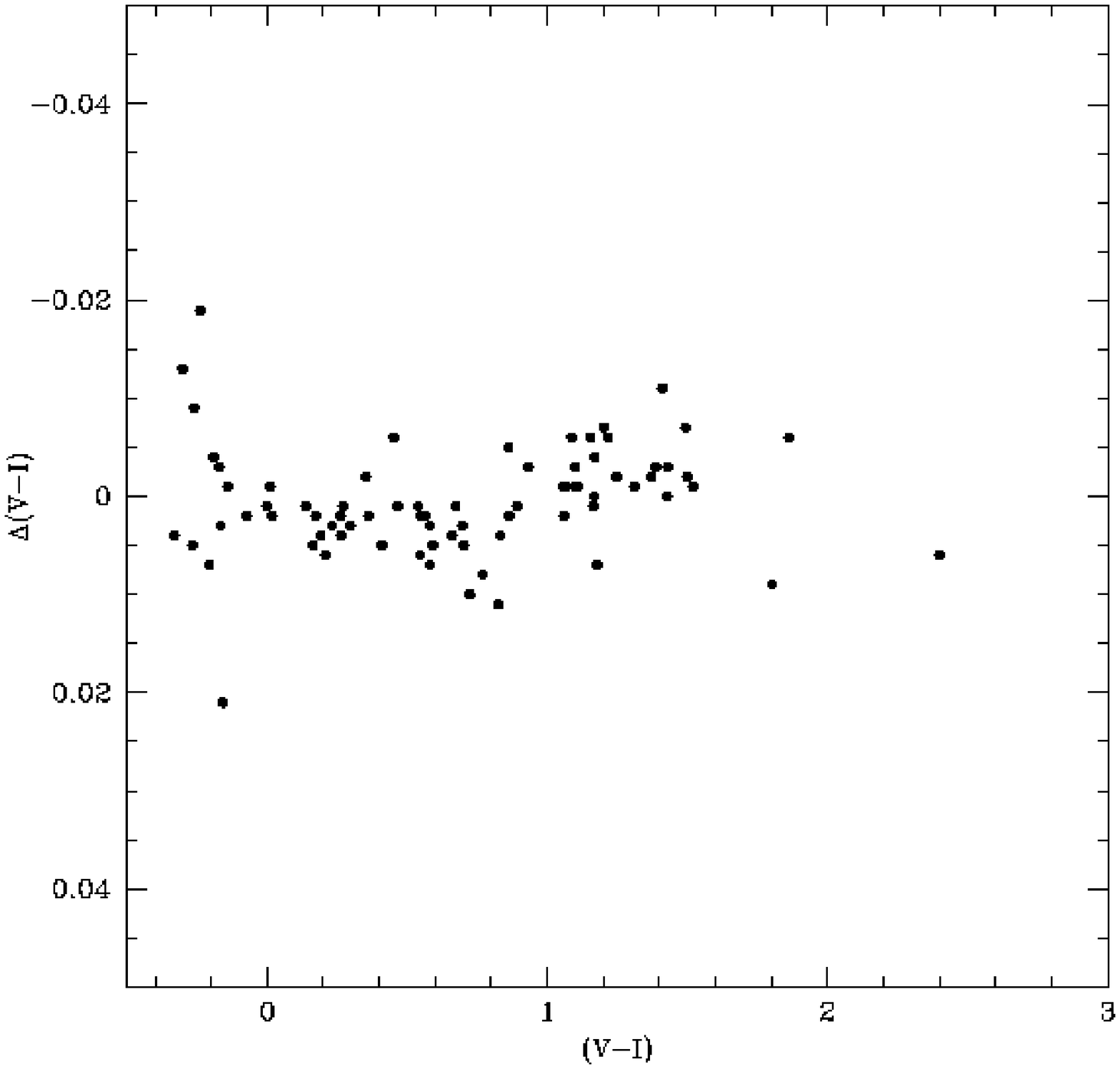}
\caption{A comparison of the $(V-I)$ color indices tied into \citet{Landolt1983} standard stars as a function of the 
\citet{Landolt1992} equatorial standard's $(V-I)$ color indices.}
\label{fig:figure06}
\end{figure}

Figures \ref{fig:figure01}-\ref{fig:figure06} illustrate the plots of the delta quantities on the 
ordinates versus the color indices on the abscissas.  Nonlinearities are apparent in the figures.  
Inspection of each figure allowed the nonlinear ``breakpoints" to be chosen.  They are indicated below 
in association with the appropriate nonlinear transformation relation, which relations were derived by 
least squares from the data appearing in Figures \ref{fig:figure01}-\ref{fig:figure06}.

The nonlinear transformation relations, then, had the form, where c = catalogue and obs = observed:

\begin{eqnarray}
(B-V)_{c}&=&+0.00268 + 1.02847(B-V)_{obs}~~~~~~~~~~~~~~~~~~~~~~~~(B-V)<+0.1, \nonumber \\
& & \pm0.00322 \pm0.01879 \nonumber \\
& & \nonumber \\
(B-V)_{c}&=&+0.00709 + 0.98474(B-V)_{obs}~~~~~~~~~~~~~~~~~~~+0.1<(B-V)<+1.0, \nonumber \\
& & \pm0.00163 \pm0.00314 \nonumber \\
& & \nonumber \\
(B-V)_{c}&=&-0.00835 + 1.00688(B-V)_{obs}~~~~~~~~~~~~~~~~~~~~~~~~(B-V)>+1.0, \nonumber \\
& & \pm0.00679 \pm0.00518 \nonumber \\
& & \nonumber \\ 
V_{c}&=&-0.00036 - 0.01444(B-V)_{c} + V_{obs}~~~~~~~~~~~~~~~~~~~~(B-V)<+0.1, \nonumber \\
& & \pm0.00379 \pm0.02213 \nonumber \\
& & \nonumber \\
V_{c}&=&-0.00112 - 0.00271(B-V)_{c} + V_{obs}~~~~~~~~~~~~~~~+0.1<(B-V)<+1.0, \nonumber \\
& & \pm0.00163 \pm0.00312 \nonumber \\
& & \nonumber \\
V_{c}&=&-0.00692 + 0.00713(B-V)_{c} + V_{obs}~~~~~~~~~~~~~~~~~~~~(B-V)>+1.0, \nonumber \\
& & \pm0.00558 \pm0.00426 \nonumber \\
& & \nonumber \\
(U-B)_{c}&=&-0.01701 + 0.96496(U-B)_{obs}~~~~~~~~~~~~~~~~~~~~~~~~(U-B)<-0.2, \nonumber \\
& & \pm0.00643 \pm0.00746 \nonumber \\
& & \nonumber \\
(U-B)_{c}&=&-0.00565 + 0.99602(U-B)_{obs}~~~~~~~~~~~~~~~~~~~-0.2<(U-B)<+0.5, \nonumber \\
& & \pm0.00567 \pm0.02805 \nonumber \\
& & \nonumber \\
(U-B)_{c}&=&-0.02240 + 1.01788(U-B)_{obs}~~~~~~~~~~~~~~~~~~~~~~~~(U-B)>+0.5, \nonumber \\
& & \pm0.00771 \pm0.00565 \nonumber \\
& & \nonumber \\
(V-R)_{c}&=&+0.00133 + 0.96767(V-R)_{obs}~~~~~~~~~~~~~~~~~~~~~~~~(V-R)<+0.1, \nonumber \\
& & \pm0.00073 \pm0.00818 \nonumber \\
& & \nonumber \\
(V-R)_{c}&=&-0.00267 + 0.99641(V-R)_{obs}~~~~~~~~~~~~~~~~~~~+0.1<(V-R)<+0.5, \nonumber \\
& & \pm0.00268 \pm0.00451 \nonumber \\
& & \nonumber \\
(V-R)_{c}&=&-0.00129 + 1.00502(V-R)_{obs}~~~~~~~~~~~~~~~~~~~~~~~~(V-R)>+0.5, \nonumber \\
& & \pm0.00307 \pm0.00432 \nonumber \\
& & \nonumber \\
(R-I)_{c}&=&-0.00155 + 0.99765(R-I)_{obs}~~~~~~~~~~~~~~~~~~~~~~~~(R-I)<+0.1, \nonumber \\
& & \pm0.00125 \pm0.01193 \nonumber \\
& & \nonumber \\
(R-I)_{c}&=&-0.00258 + 1.00789(R-I)_{obs}~~~~~~~~~~~~~~~~~~~+0.1<(R-I)<+0.5, \nonumber \\
& & \pm0.00155 \pm0.00506 \nonumber \\
& & \nonumber \\
(R-I)_{c}&=&+0.00753 + 0.98853(R-I)_{obs}~~~~~~~~~~~~~~~~~~~~~~~~(R-I)>+0.5, \nonumber \\
& & \pm0.00228 \pm0.00347 \nonumber \\
& & \nonumber \\
(V-I)_{c}&=&-0.00116 + 0.98201(V-I)_{obs}~~~~~~~~~~~~~~~~~~~~~~~~(V-I)<+0.1, \nonumber \\
& & \pm0.00350 \pm0.01750 \nonumber \\
& & \nonumber \\
(V-I)_{c}&=&-0.00228 + 0.99807(V-I)_{obs}~~~~~~~~~~~~~~~~~~~+0.1<(V-I)<+1.0, \nonumber \\
& & \pm0.00166 \pm0.00288 \nonumber \\
& & \nonumber \\
(V-I)_{c}&=&+0.00683 + 0.99628(V-I)_{obs}~~~~~~~~~~~~~~~~~~~~~~~~(V-I)>+1.0. \nonumber \\
& & \pm0.00397 \pm0.00291 \nonumber \\
\nonumber
\end{eqnarray}

\begin{figure}
\plotone{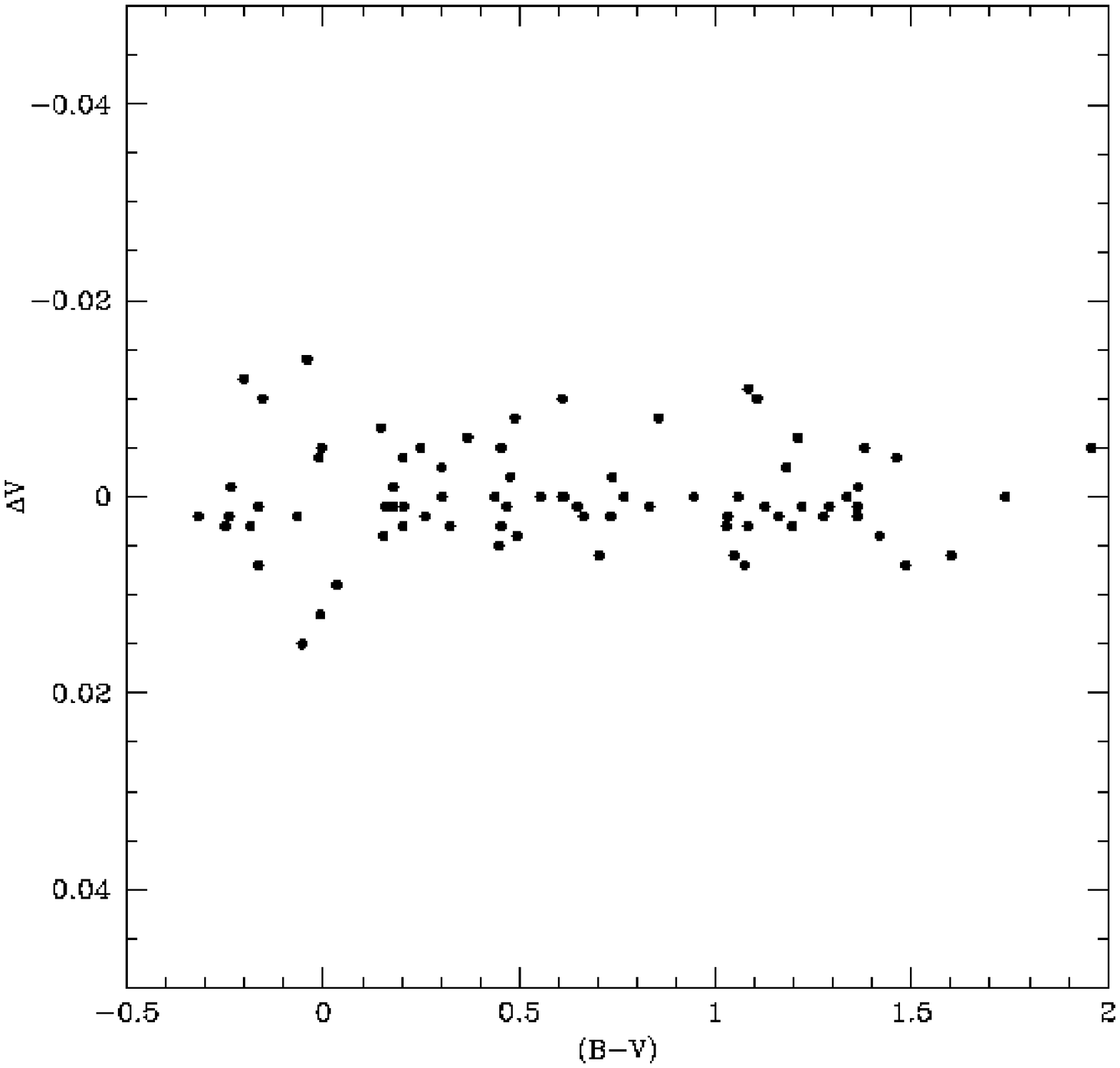}
\caption{A comparison of the $V$ magnitudes tied into \citet{Landolt1992} standard stars as a function of the 
\citet{Landolt1992} equatorial standard's $(B-V)$ color indices.}
\label{fig:figure07}
\end{figure}

\begin{figure}
\plotone{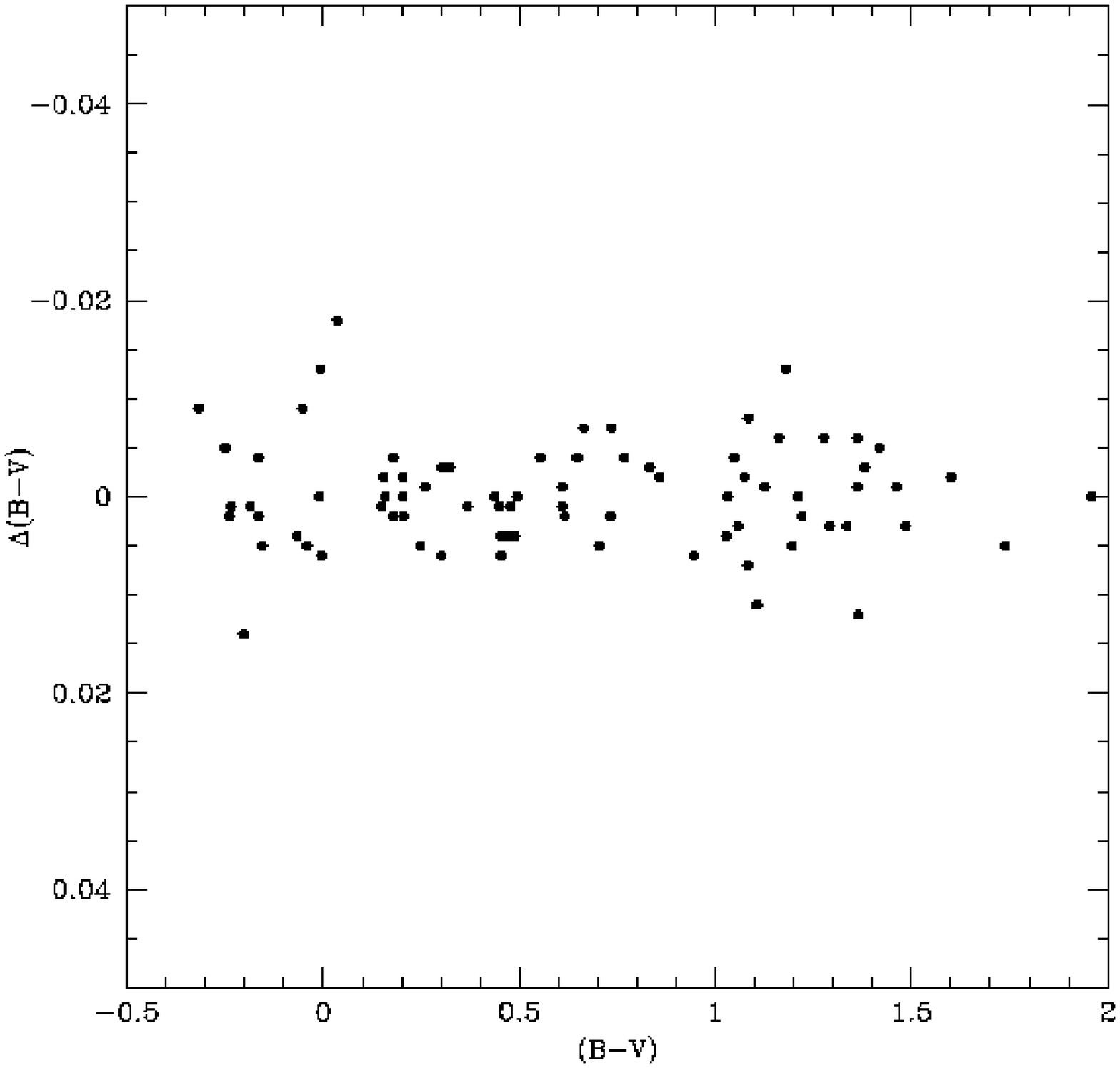}
\caption{A comparison of the $(B-V)$ color indices tied into \citet{Landolt1992} standard stars as a function of the 
\citet{Landolt1992} equatorial standard's $(B-V)$ color indices.}
\label{fig:figure08}
\end{figure}

\begin{figure}
\plotone{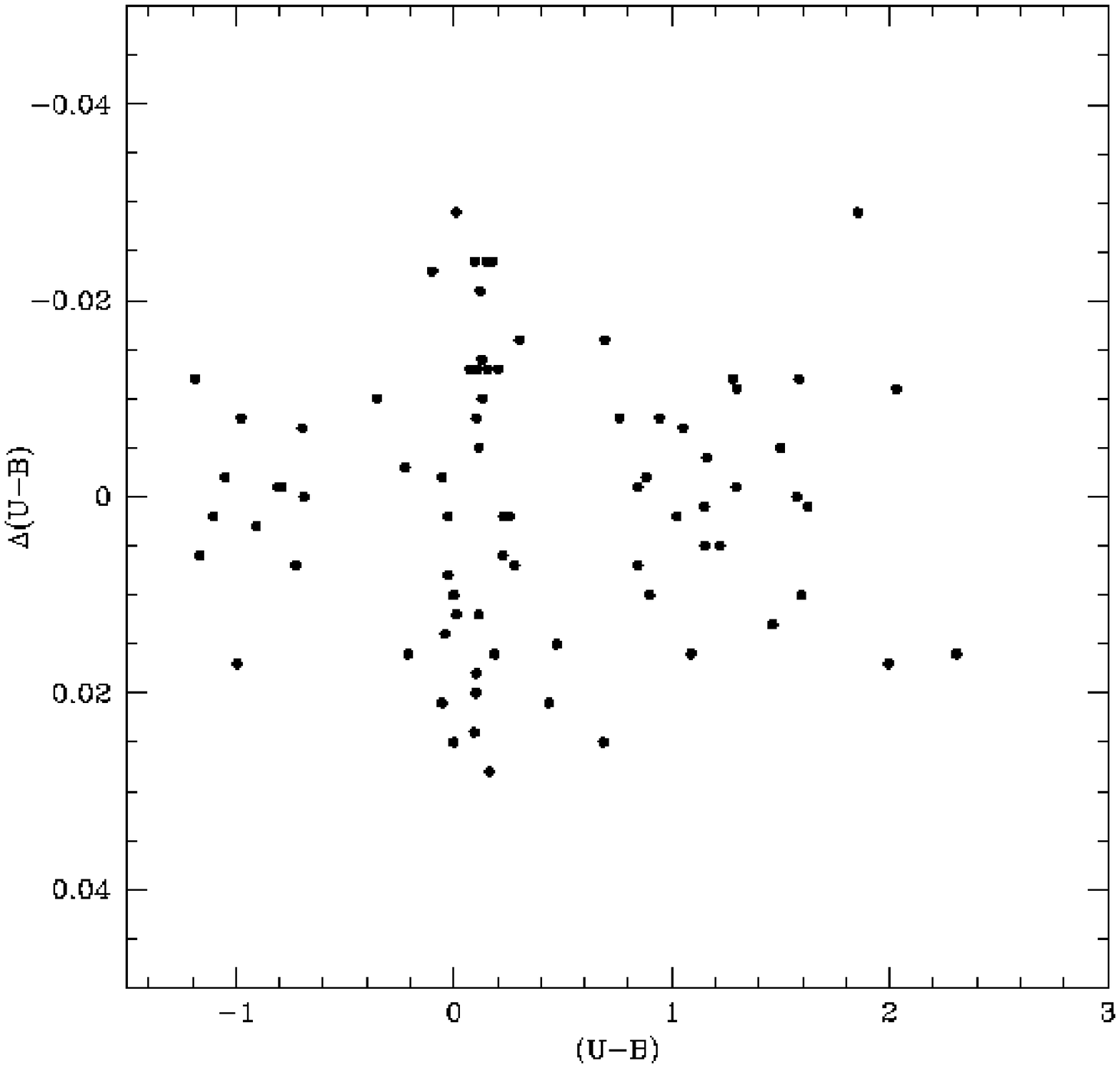}
\caption{A comparison of the $(U-B)$ color indices tied into \citet{Landolt1992} standard stars as a function of the 
\citet{Landolt1992} equatorial standard's $(U-B)$ color indices.}
\label{fig:figure09}
\end{figure}

\begin{figure}
\plotone{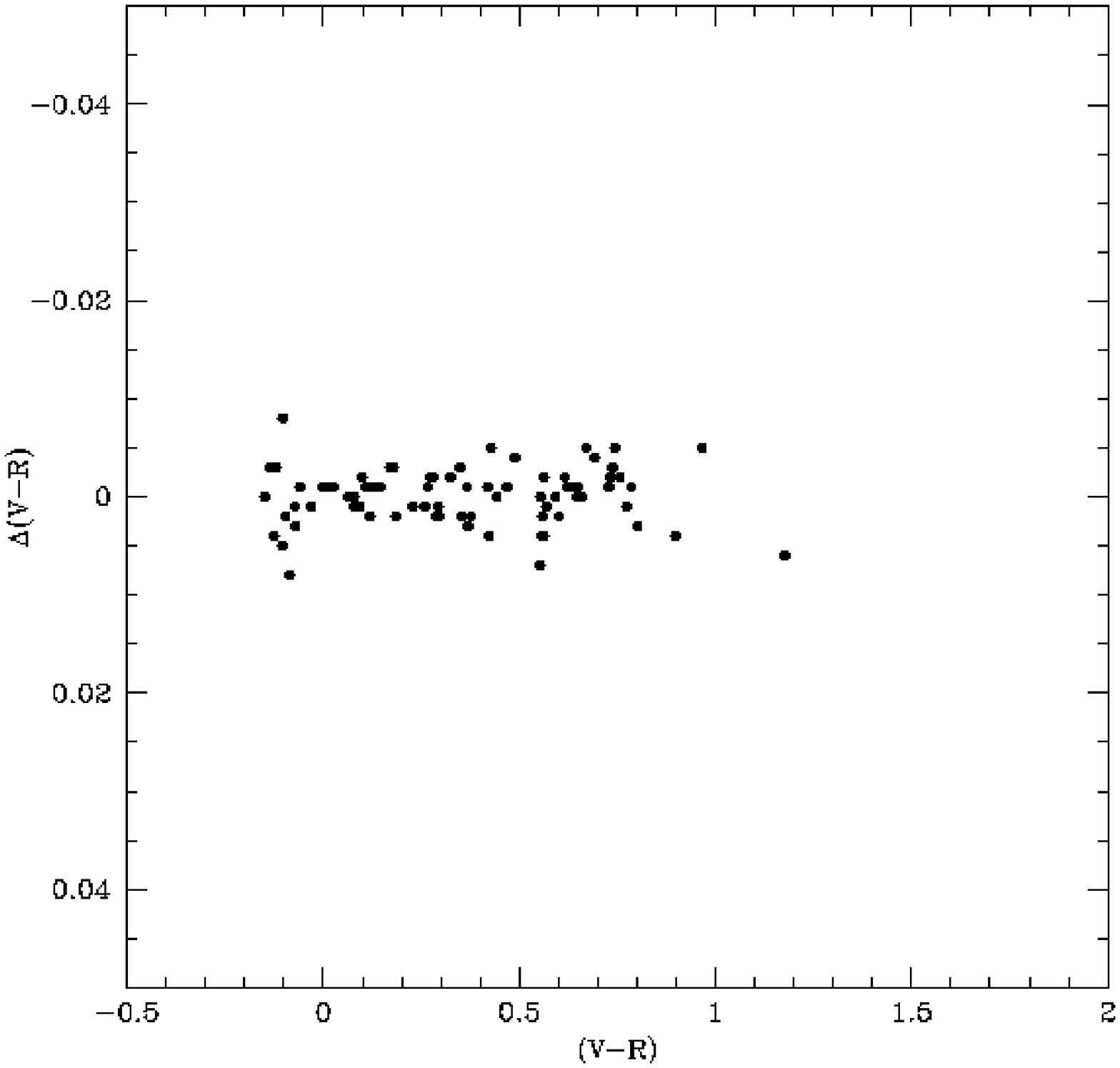}
\caption{A comparison of the $(V-R)$ color indices tied into \citet{Landolt1992} standard stars as a function of the 
\citet{Landolt1992} equatorial standard's $(V-R)$ color indices.}
\label{fig:figure10}
\end{figure}

\begin{figure}
\plotone{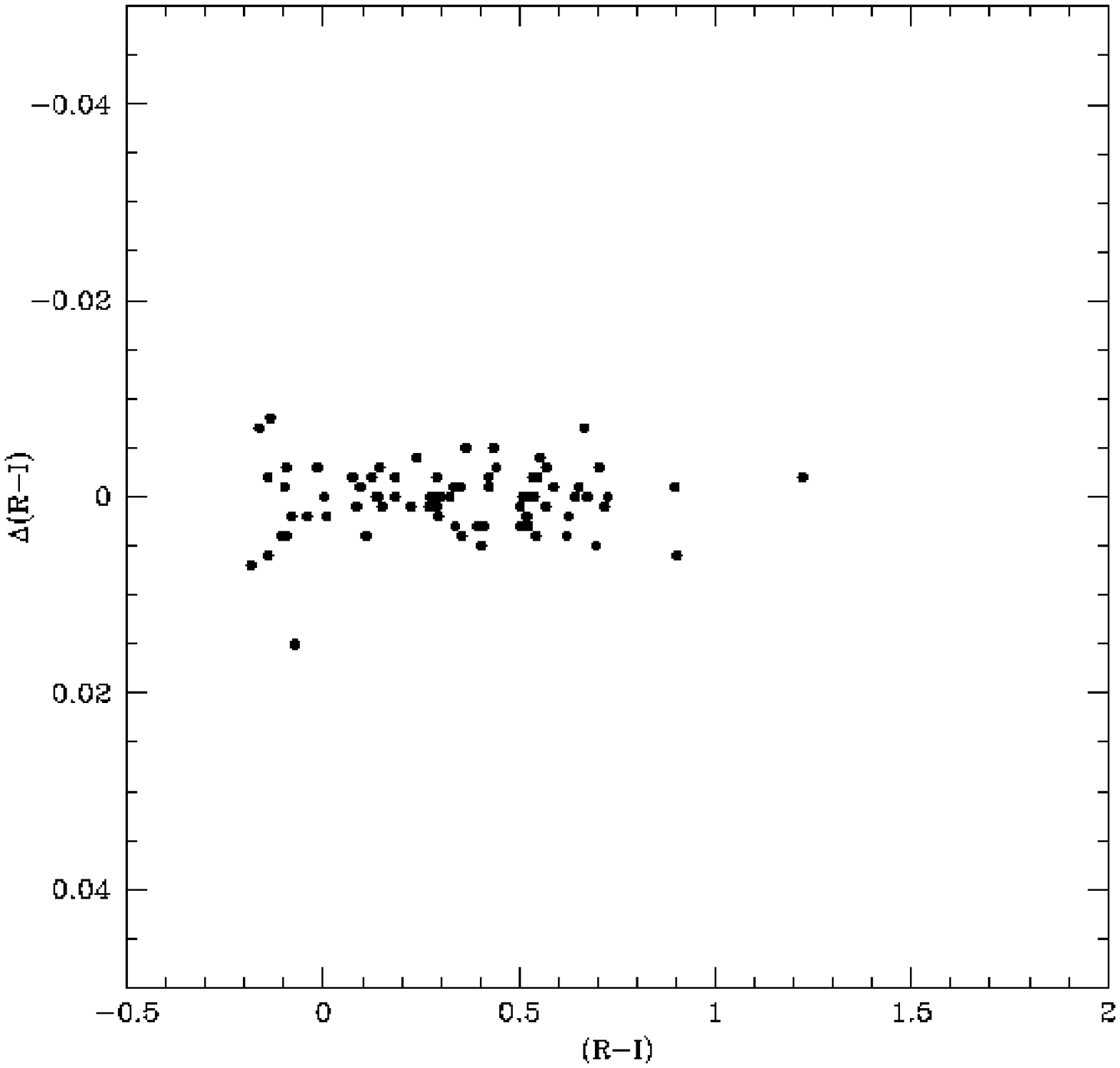}
\caption{A comparison of the $(R-I)$ color indices tied into \citet{Landolt1992} standard stars as a function of the 
\citet{Landolt1992} equatorial standard's $(R-I)$ color indices.}
\label{fig:figure11}
\end{figure}

\begin{figure}
\plotone{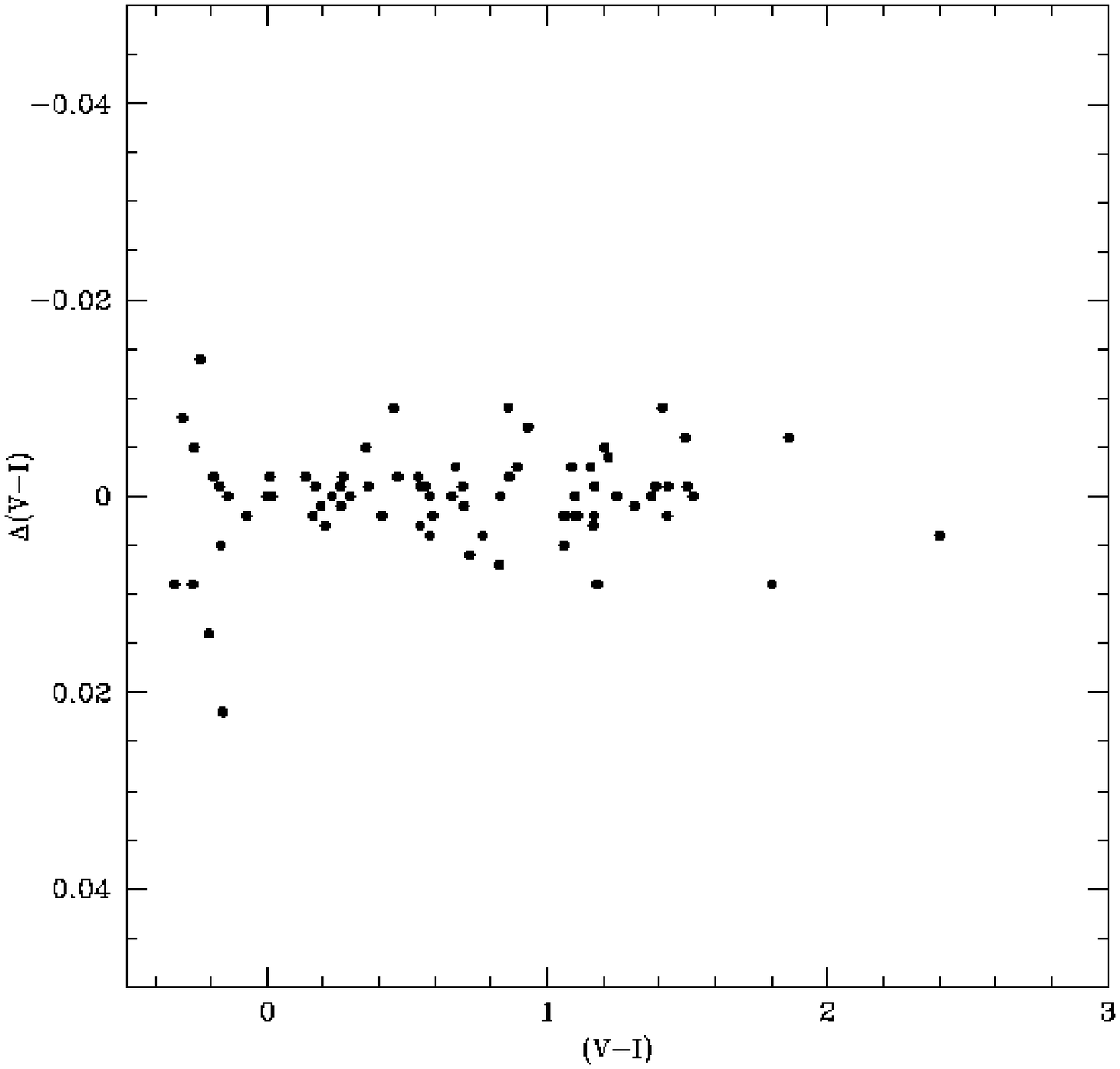}
\caption{A comparison of the $(V-I)$ color indices tied into \citet{Landolt1992} standard stars as a function of the 
\citet{Landolt1992} equatorial standard's $(V-I)$ color indices.}
\label{fig:figure12}
\end{figure}

Once these relations were applied to the recovered magnitudes and color indices of the standard stars 
used in this project, the data were on the broadband $UBVRI$ photometric system defined by the 
standard stars in \citet{Landolt1992}.  Next the now corrected for nonlinear transformation standard 
star magnitudes and color indices once again were compared to the published values in the sense 
corrected values $minus$ published magnitudes and color indices.  The fact that the nonlinear effects 
have been corrected successfully is illustrated in Figures \ref{fig:figure07}-\ref{fig:figure12}. 
Hence, the data in this paper have been transformed to the photometric system defined in 
\citet{Landolt1992}.

Two of the stars herein (BPM 16274 and HD 49798) are too far south to be observed from 
KPNO.  Hence these stars were observed at the Cerro Tololo Inter-American Observatory (CTIO) as part of 
a standard star observational program there \citep{Landolt1992}.  The CTIO $UBVRI$ data were tied into 
the same standard stars \citep{Landolt1983} as were the northern data.  The data reductions were handled 
in the same fashion as were the KPNO reductions.  Several stars on the KPNO program could be observed 
from both hemispheres, and this was done to further check that data from the two observatories were tied 
together as best as one could do.

\begin{figure}
\plotone{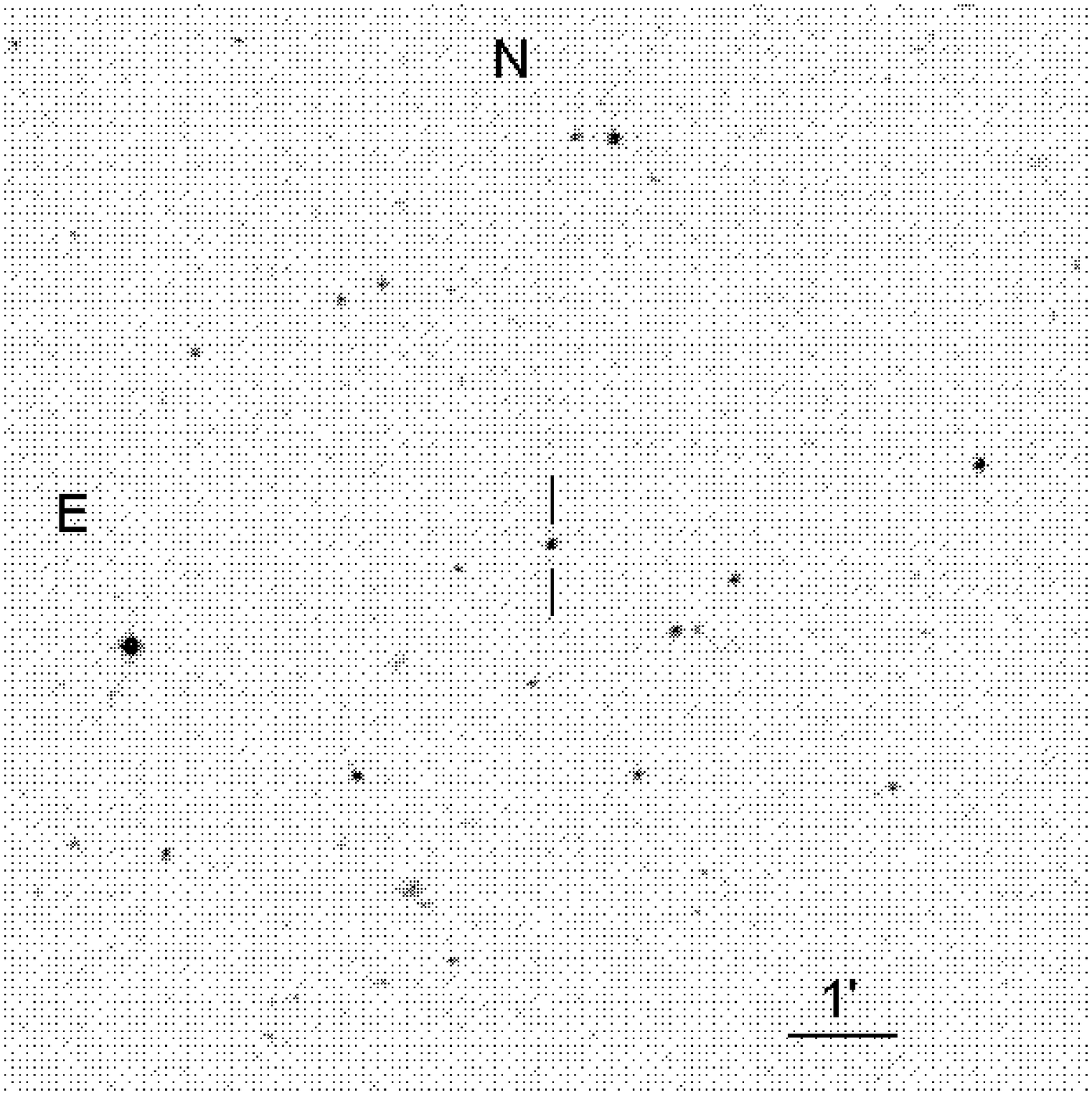}
\caption{The field, 10 arc minutes on a side, of the star G$\,$158-100.}
\label{fig:figure13}
\end{figure}

\begin{figure}
\plotone{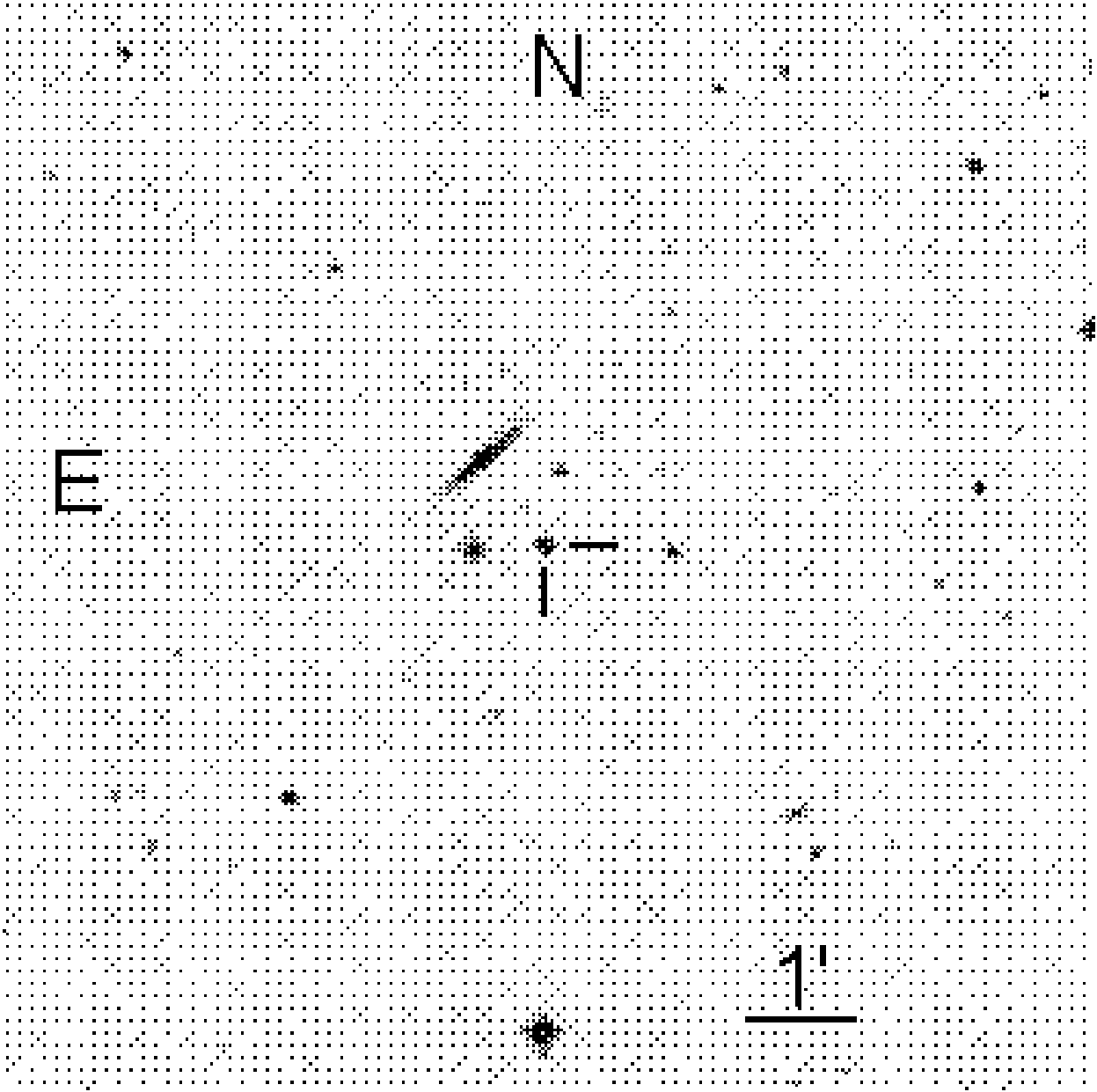}
\caption{The field, 10 arc minutes on a side, of the star BPM$\,$16274.}
\label{fig:figure14}
\end{figure}

\begin{figure}
\plotone{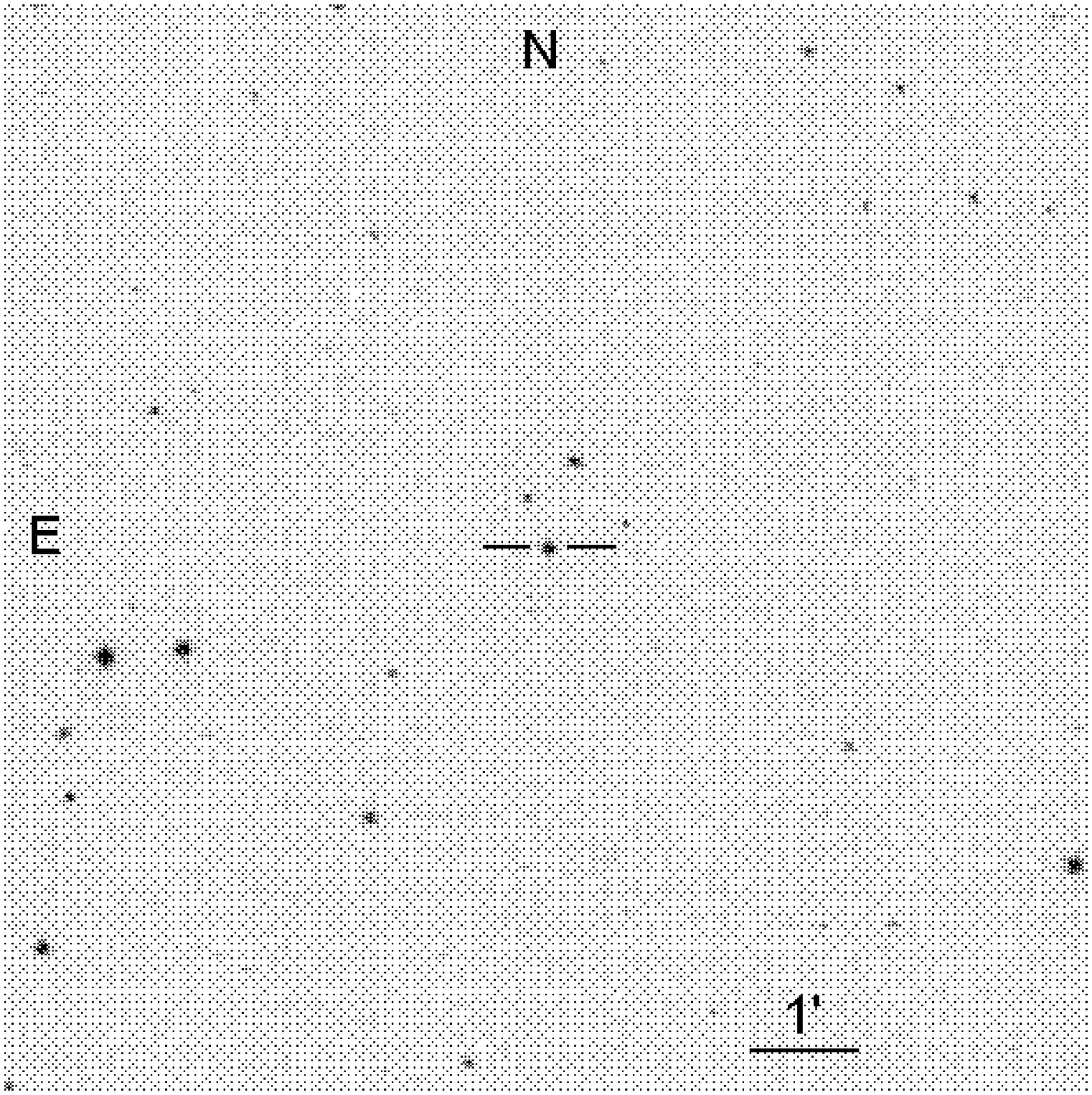}
\caption{The field, 10 arc minutes on a side, of the star HZ$\,$4.}
\label{fig:figure15}
\end{figure}

\begin{figure}
\plotone{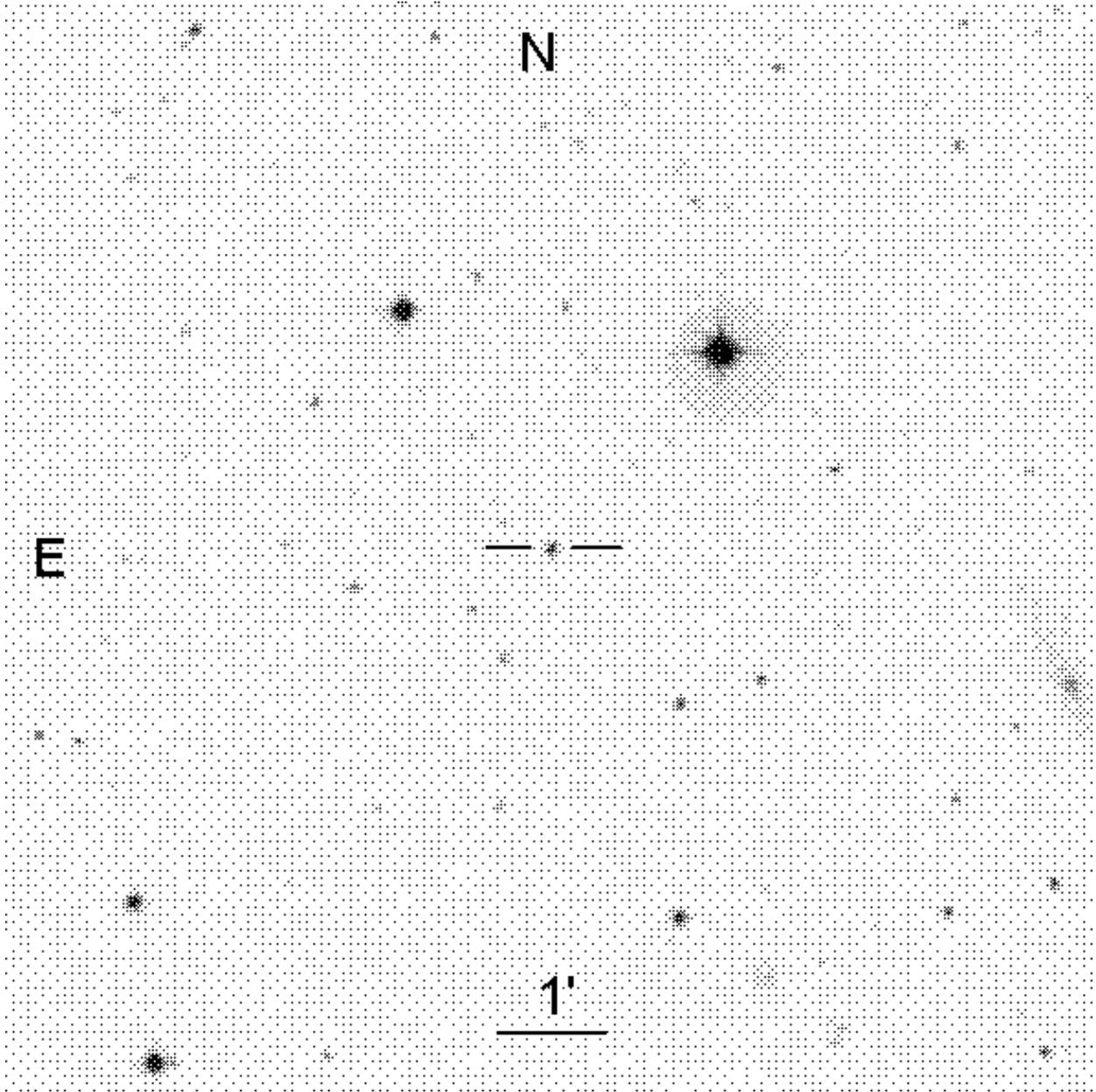}
\caption{The field, 10 arc minutes on a side, of the star LB$\,$227.}
\label{fig:figure16}
\end{figure}

\begin{figure}
\plotone{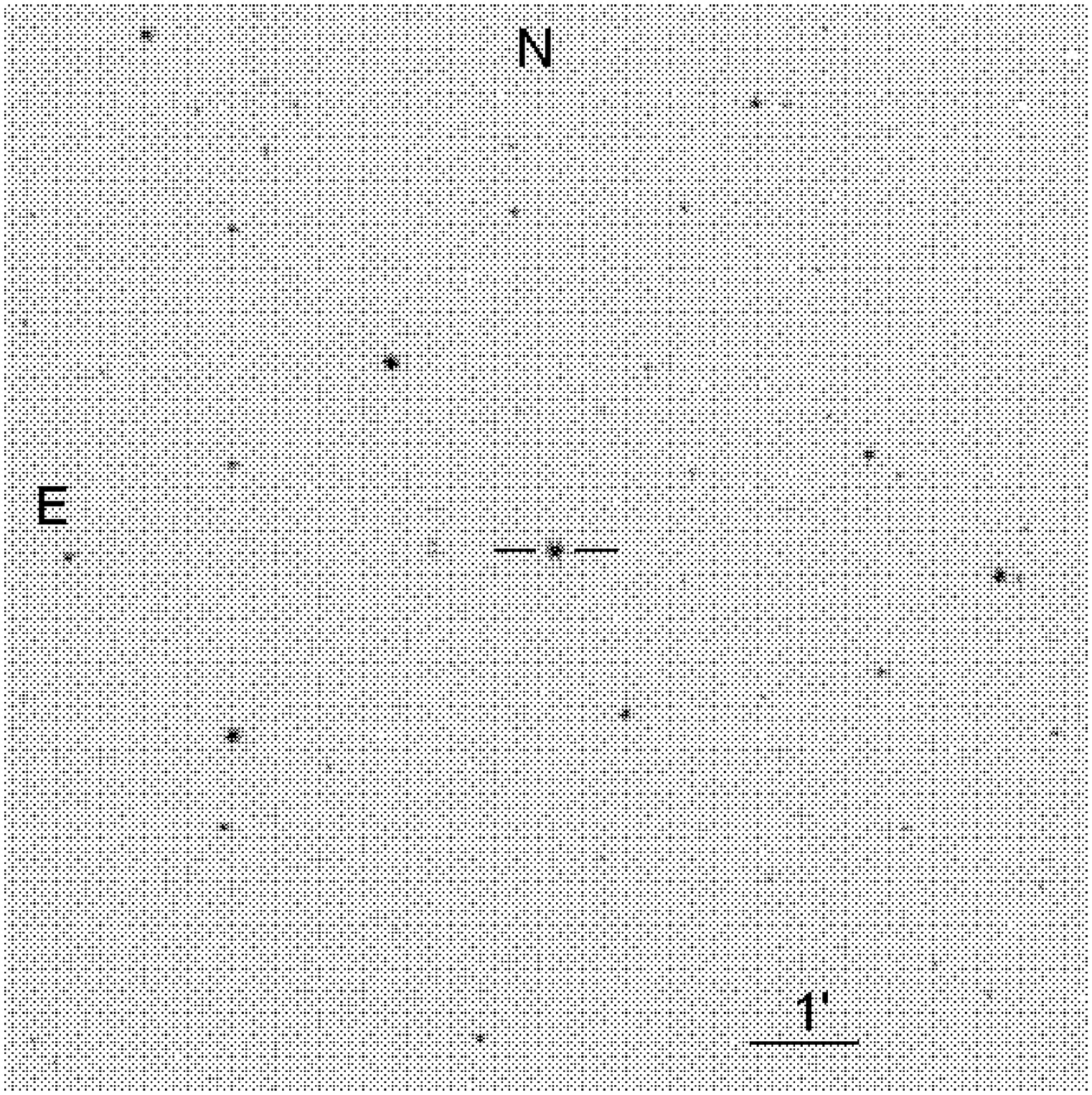}
\caption{The field, 10 arc minutes on a side, of the star HZ$\,$2.}
\label{fig:figure17}
\end{figure}

\begin{figure}
\plotone{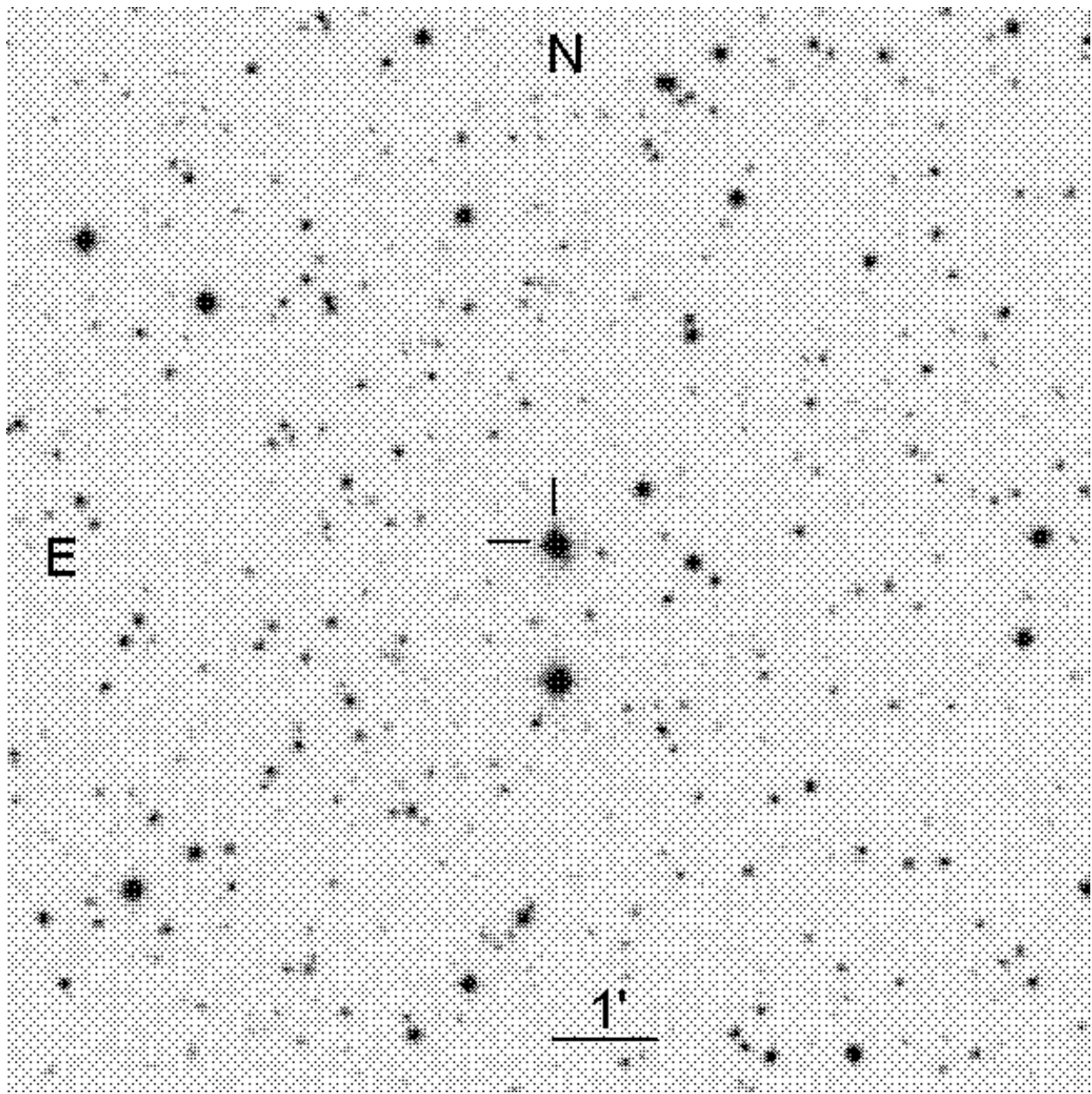}
\caption{The field, 10 arc minutes on a side, of the star G$\,$191-B2B.}
\label{fig:figure18}
\end{figure}

\begin{figure}
\plotone{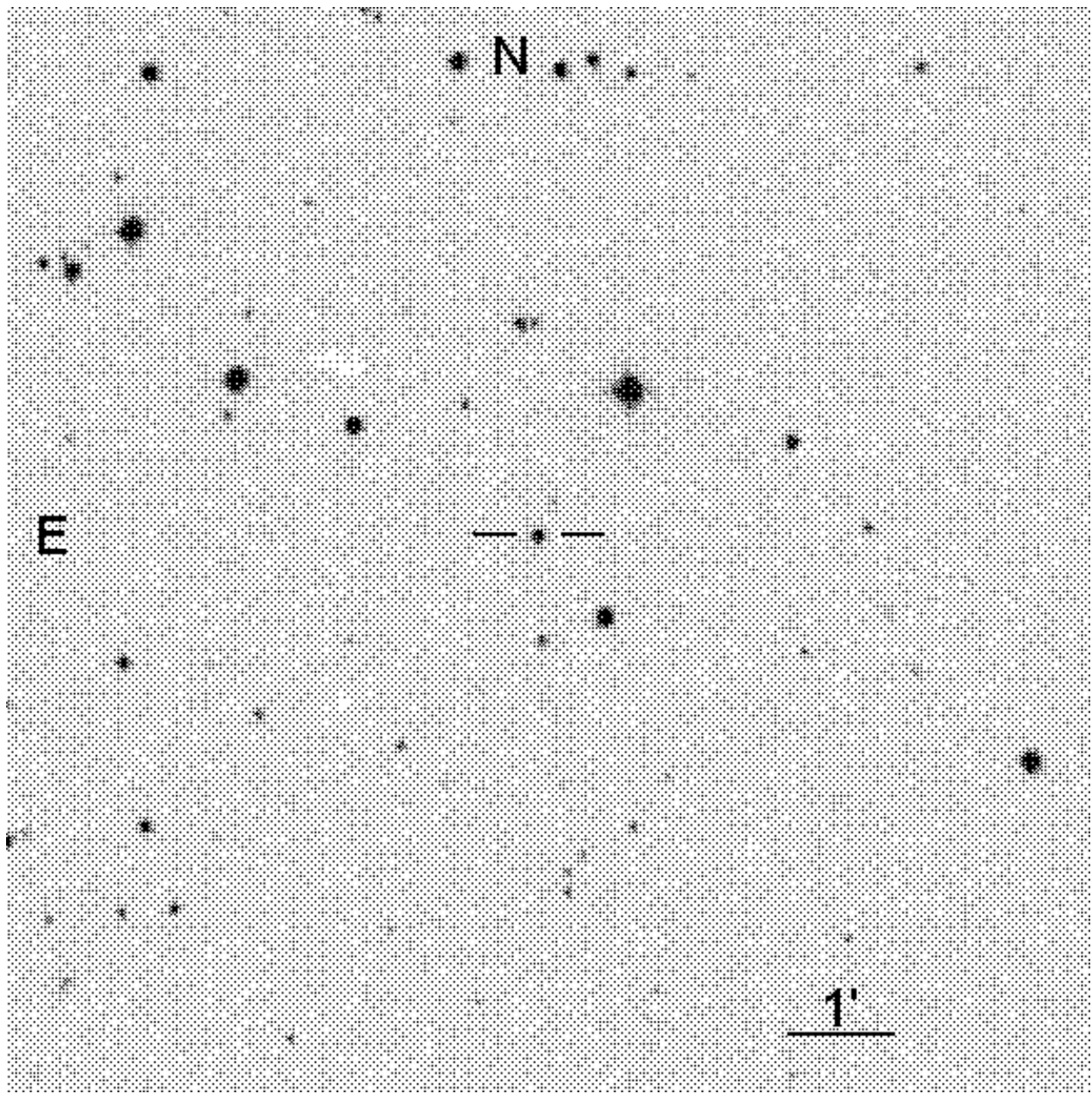}
\caption{The field, 10 arc minutes on a side, of the star G$\,$193-74.}
\label{fig:figure19}
\end{figure}

\begin{figure}
\plotone{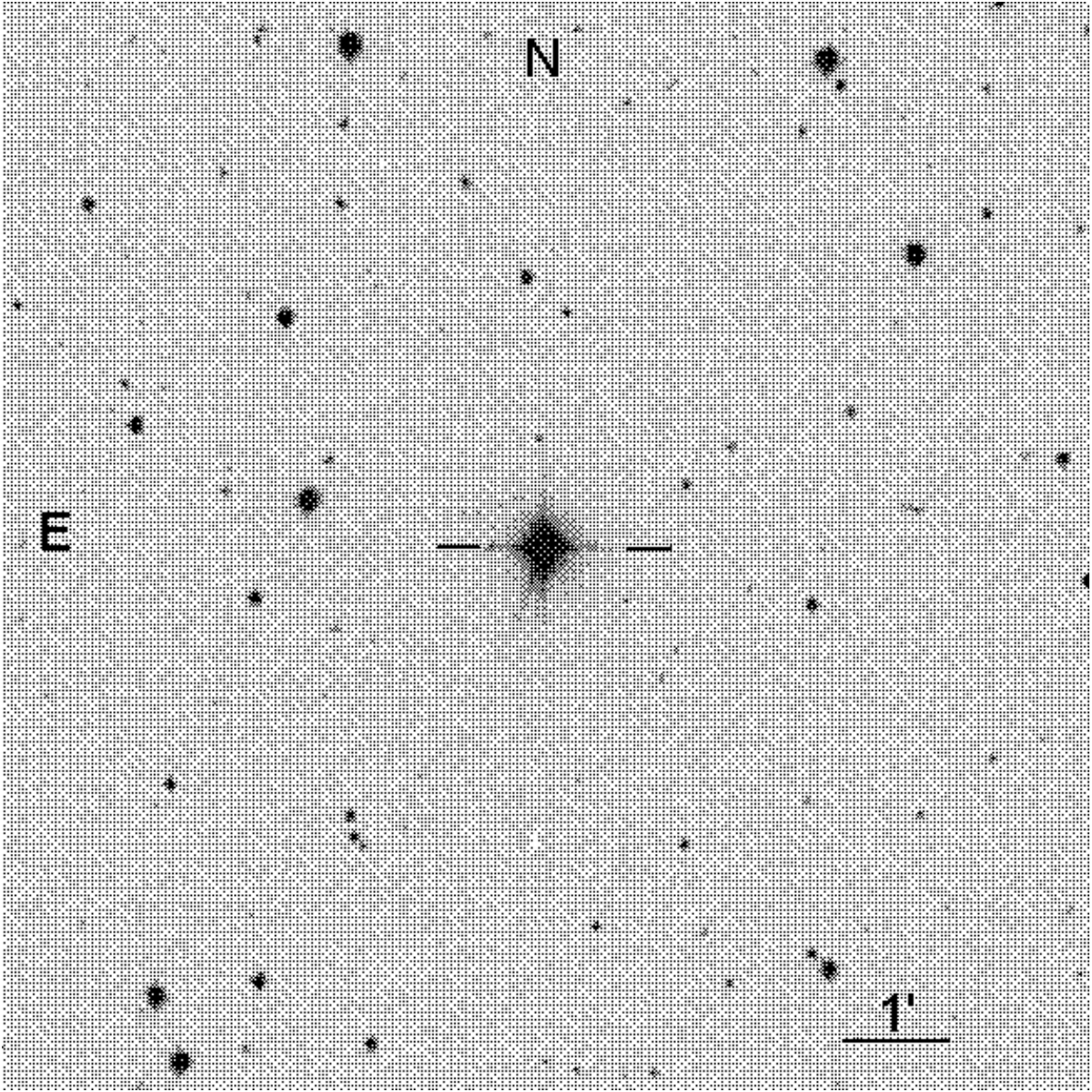}
\caption{The field, 10 arc minutes on a side, of the star BD+75$^{\circ}$325.}
\label{fig:figure20}
\end{figure}

\begin{figure}
\plotone{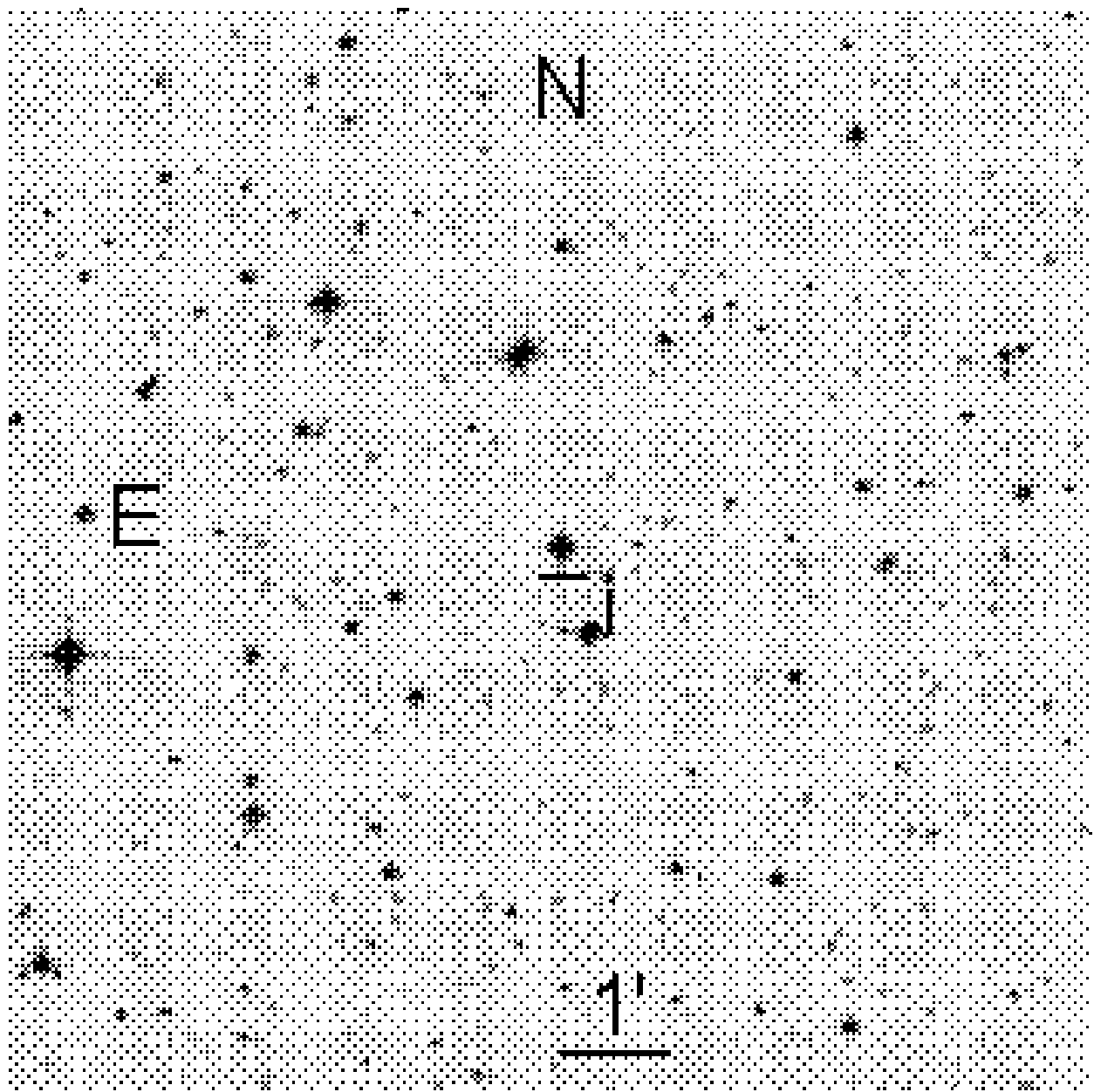}
\caption{The field, 10 arc minutes on a side, of the star LDS$\,$235B.}
\label{fig:figure21}
\end{figure}

\begin{figure}
\plotone{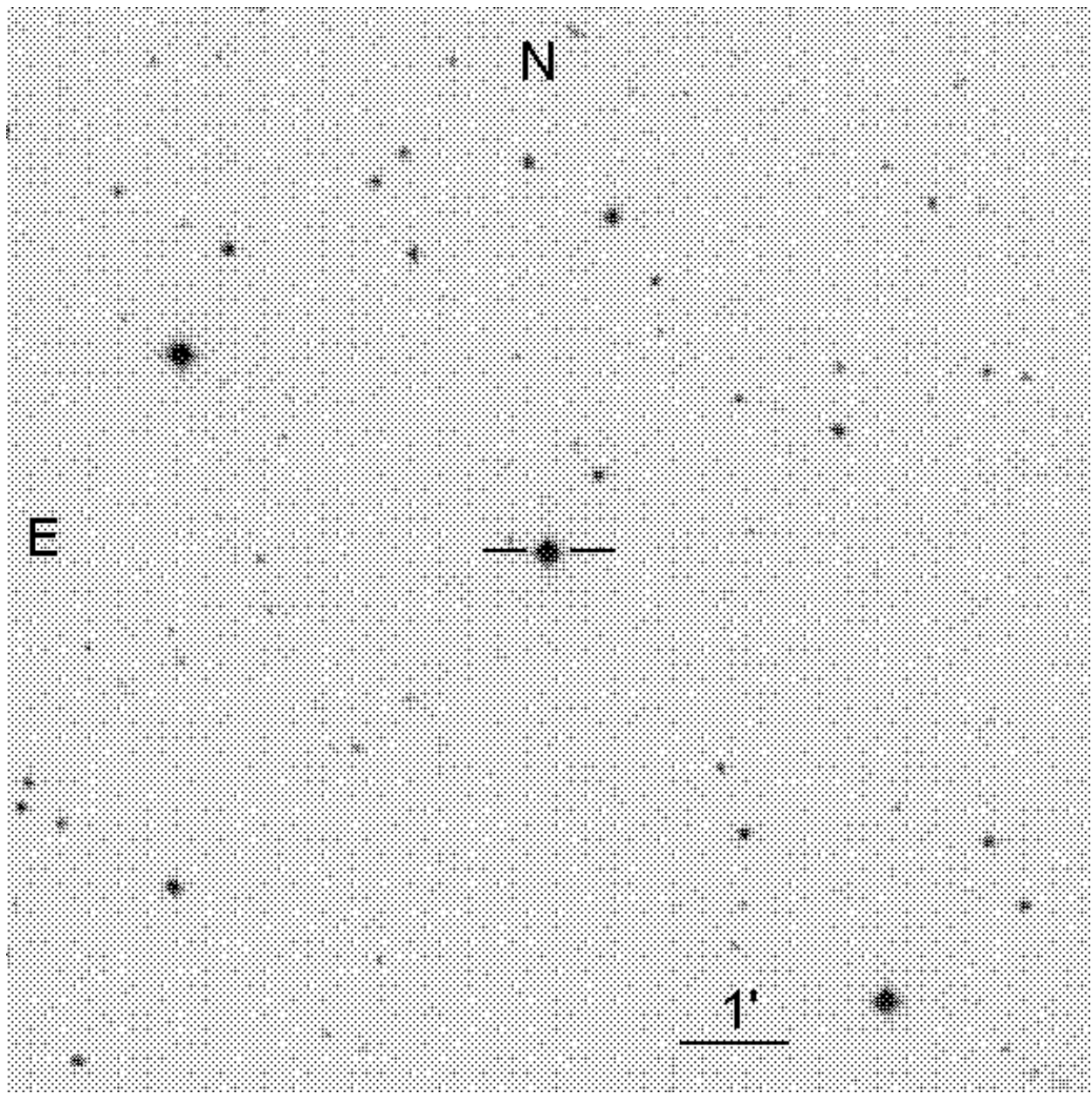}
\caption{The field, 10 arc minutes on a side, of the star AGK+81$^{\circ}$266.}
\label{fig:figure22}
\end{figure}

\begin{figure}
\plotone{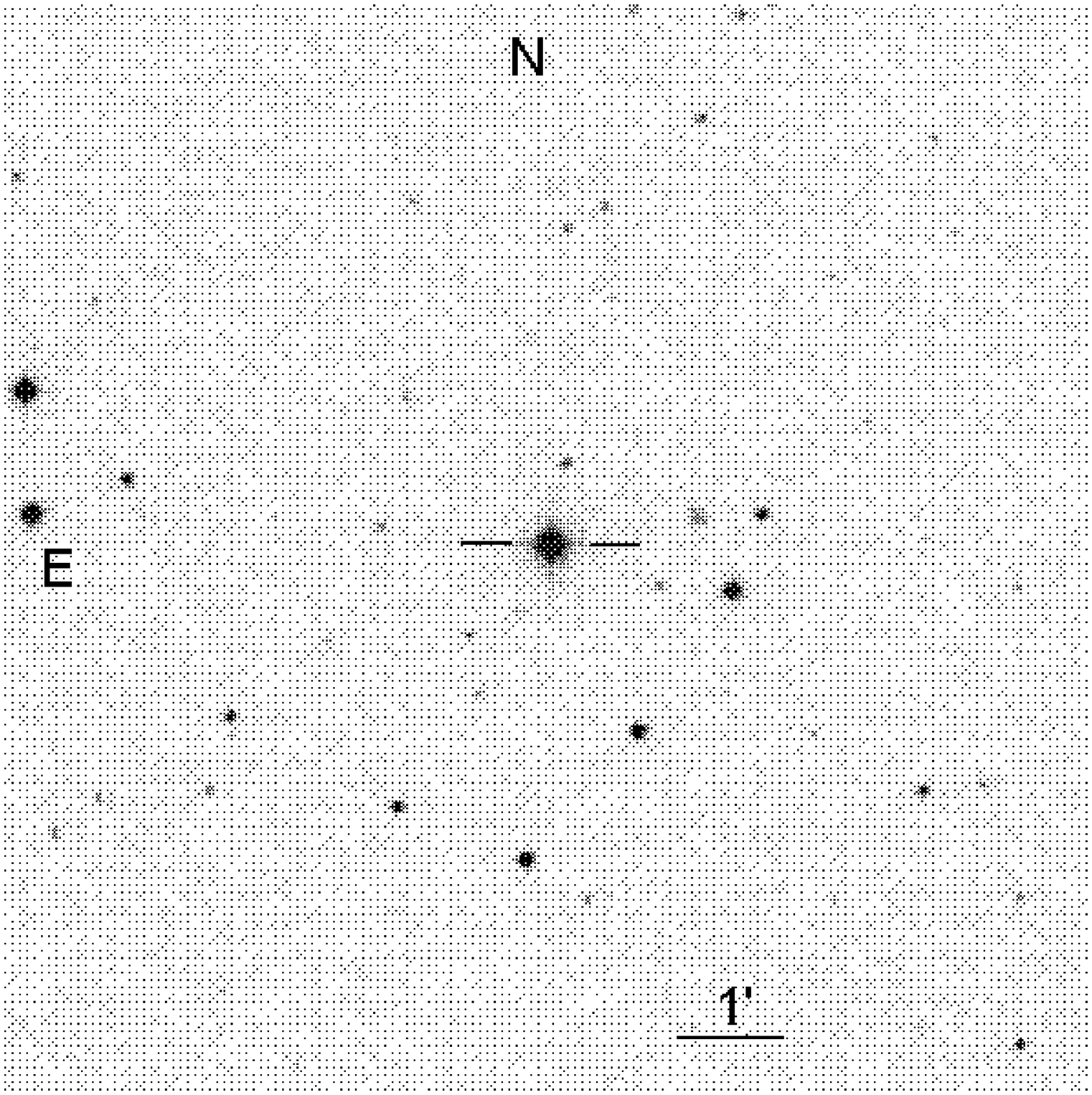}
\caption{The field, 10 arc minutes on a side, of the star Feige$\,$34.}
\label{fig:figure23}
\end{figure}

\begin{figure}
\plotone{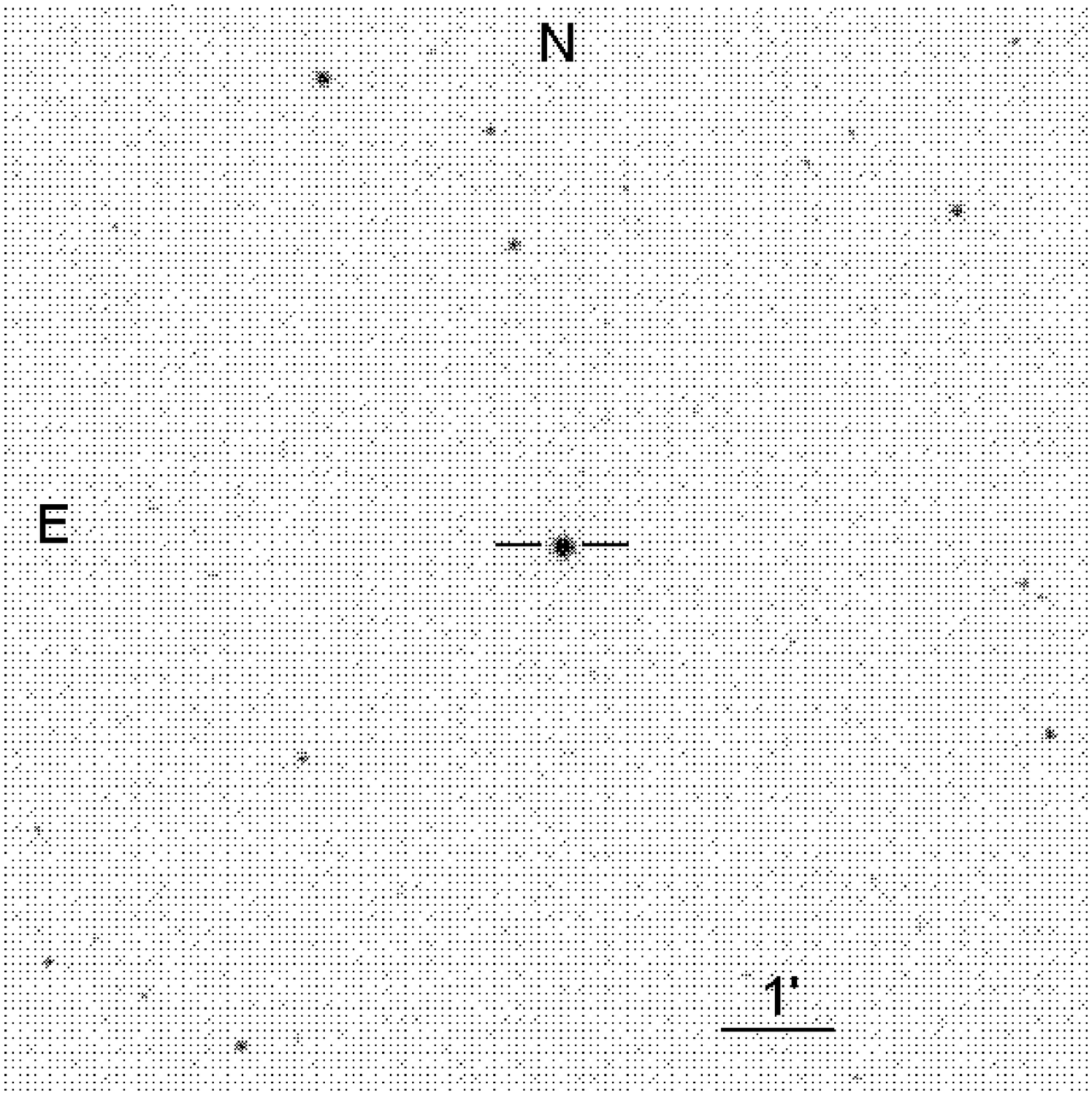}
\caption{The field, 10 arc minutes on a side, of the star GD$\,$140.}
\label{fig:figure24}
\end{figure}

\begin{figure}
\plotone{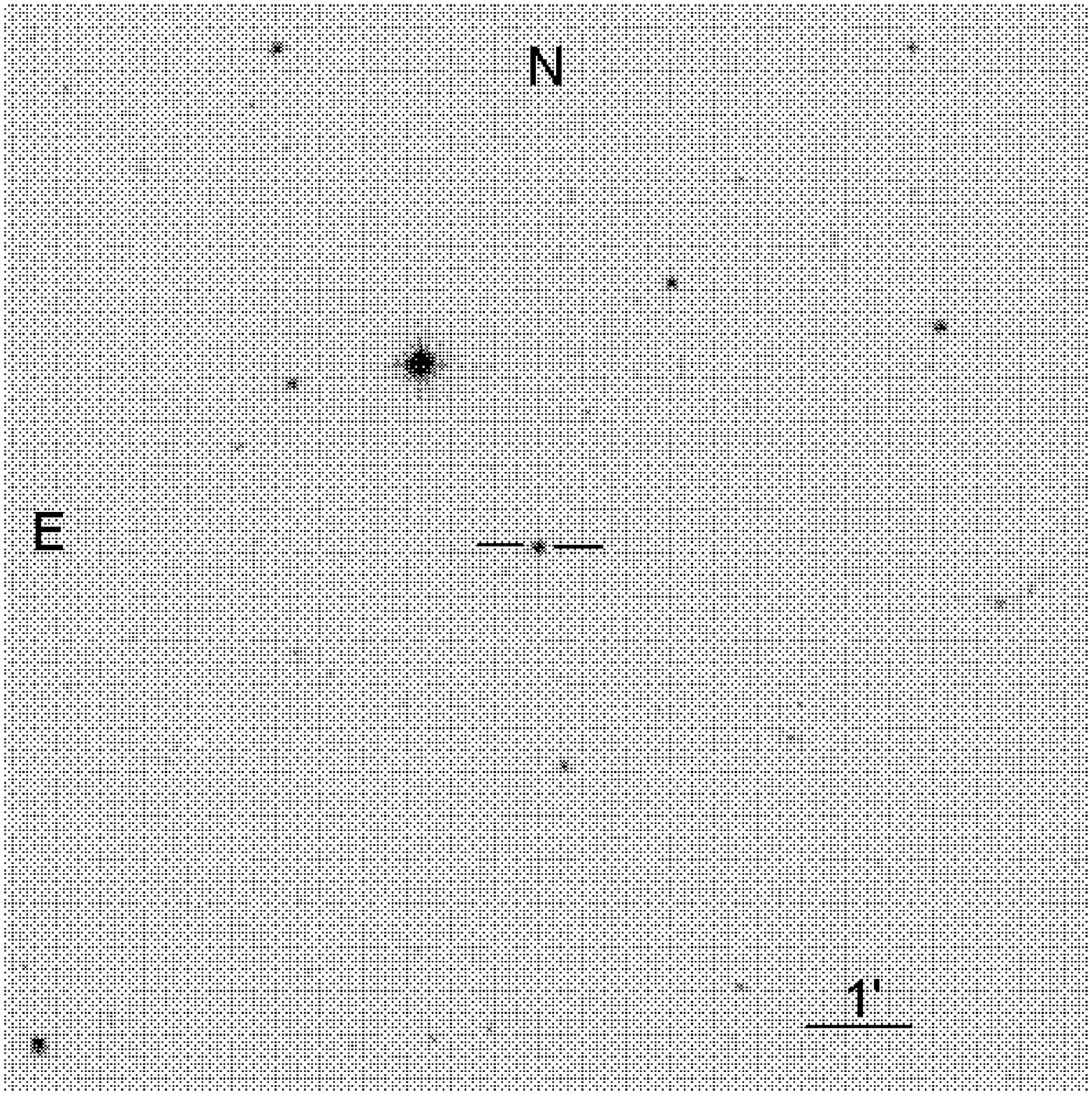}
\caption{The field, 10 arc minutes on a side, of the star HZ$\,$21.}
\label{fig:figure25}
\end{figure}

\begin{figure}
\plotone{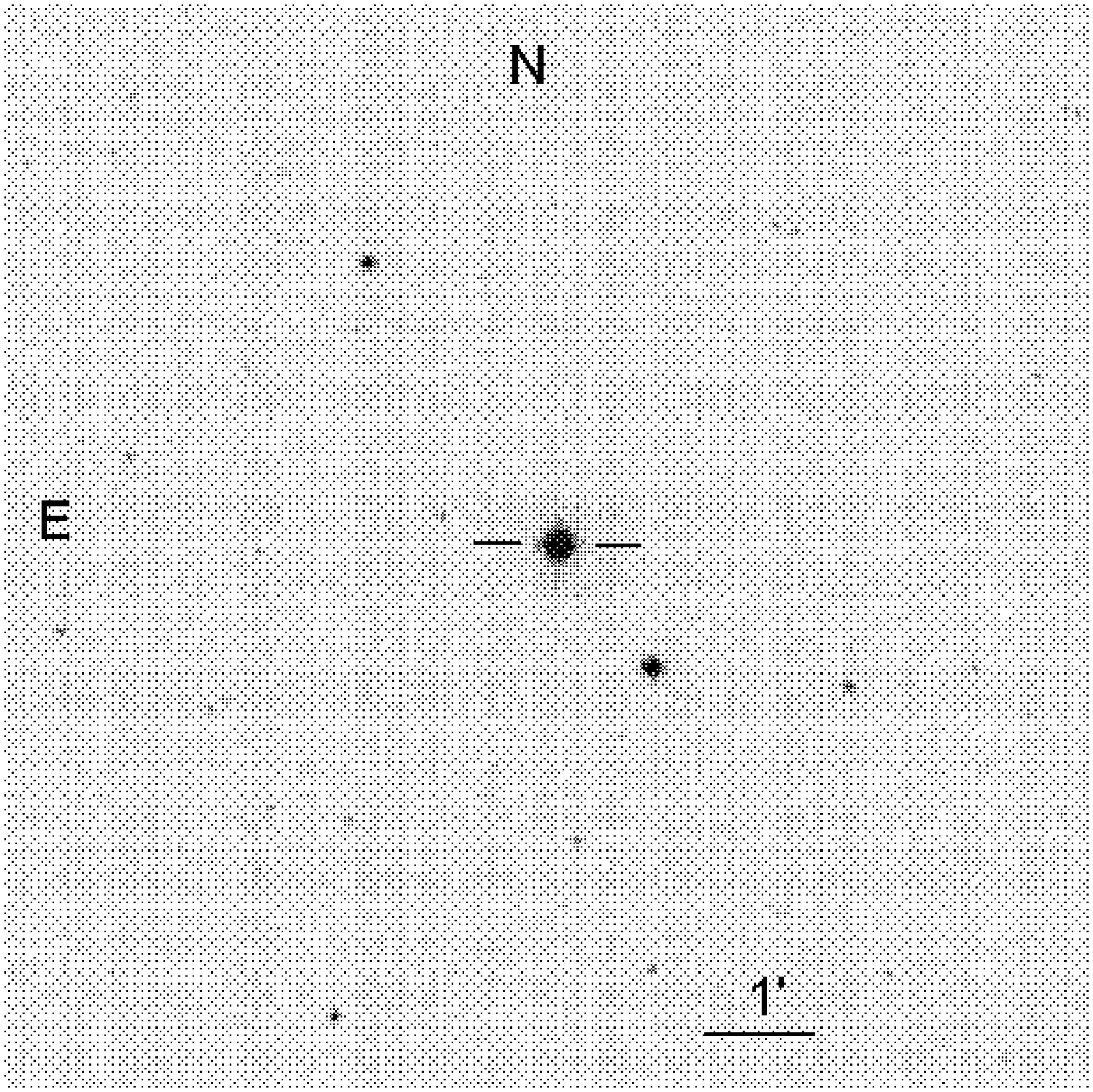}
\caption{The field, 10 arc minutes on a side, of the star Feige$\,$66.}
\label{fig:figure26}
\end{figure}

\begin{figure}
\plotone{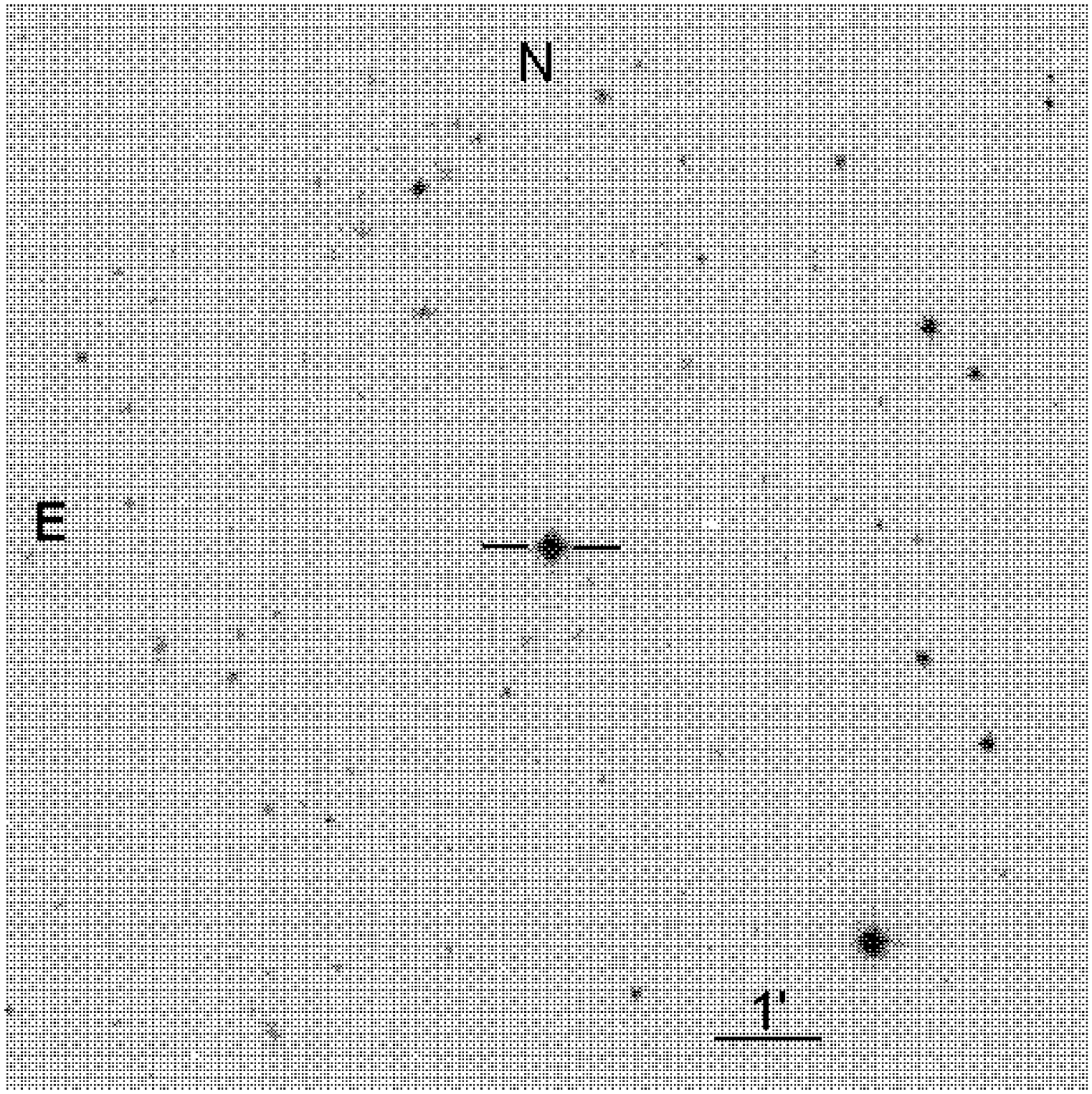}
\caption{The field, 10 arc minutes on a side, of the star Feige$\,$67.}
\label{fig:figure27}
\end{figure}

\begin{figure}
\plotone{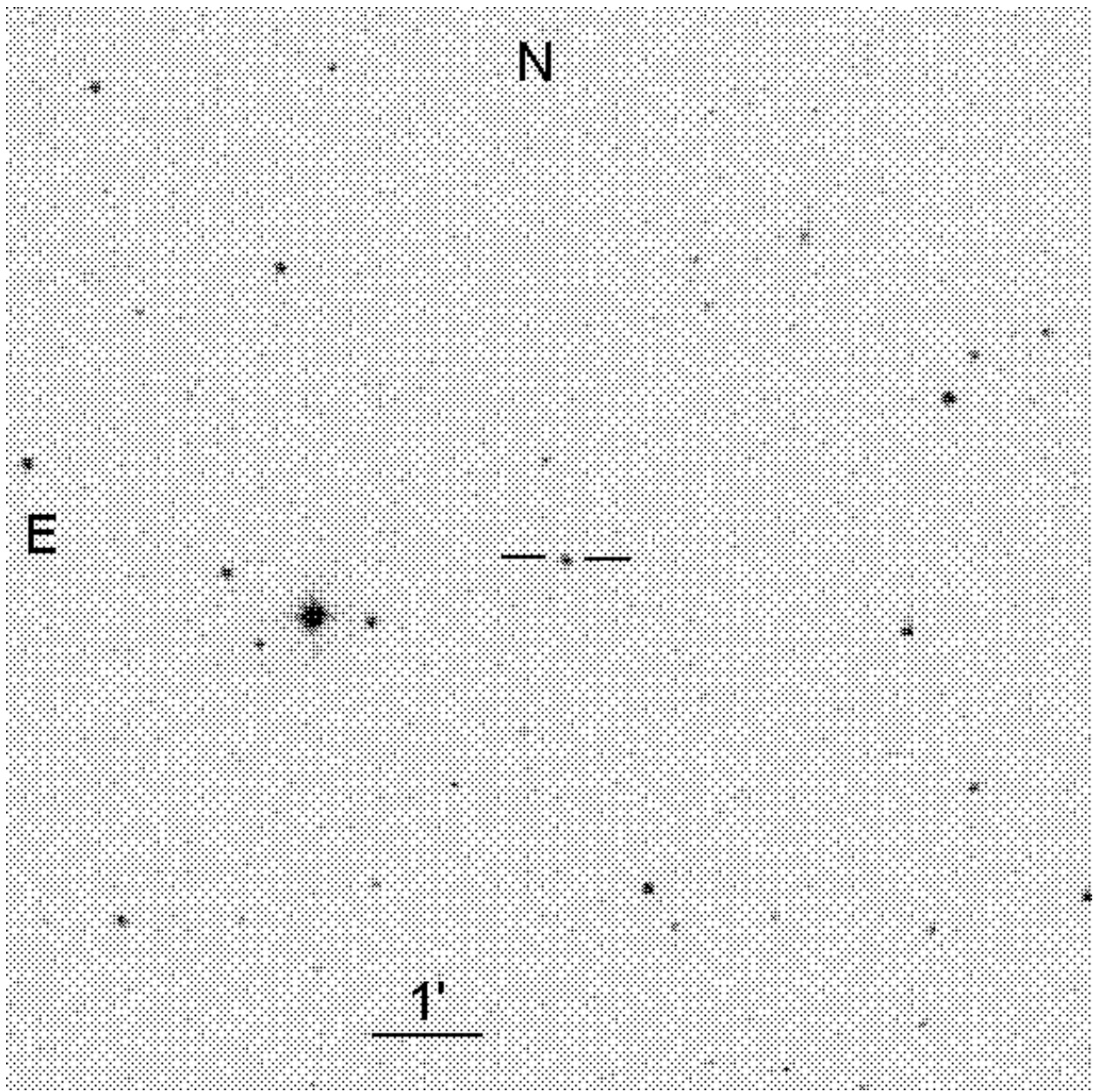}
\caption{The field, 10 arc minutes on a side, of the star G$\,$60-54.}
\label{fig:figure28}
\end{figure}

\begin{figure}
\plotone{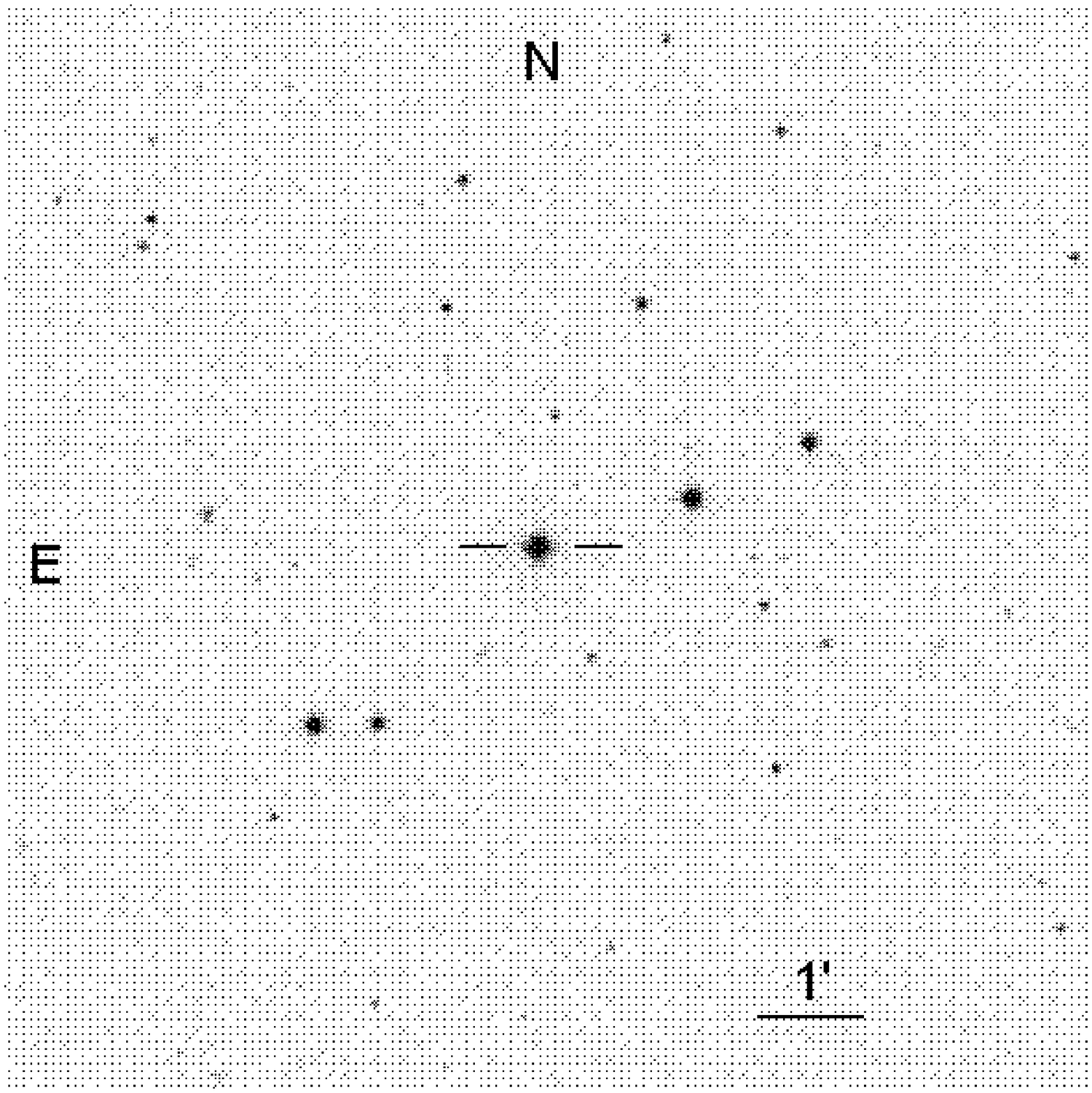}
\caption{The field, 10 arc minutes on a side, of the star HZ$\,$44.}
\label{fig:figure29}
\end{figure}

\begin{figure}
\plotone{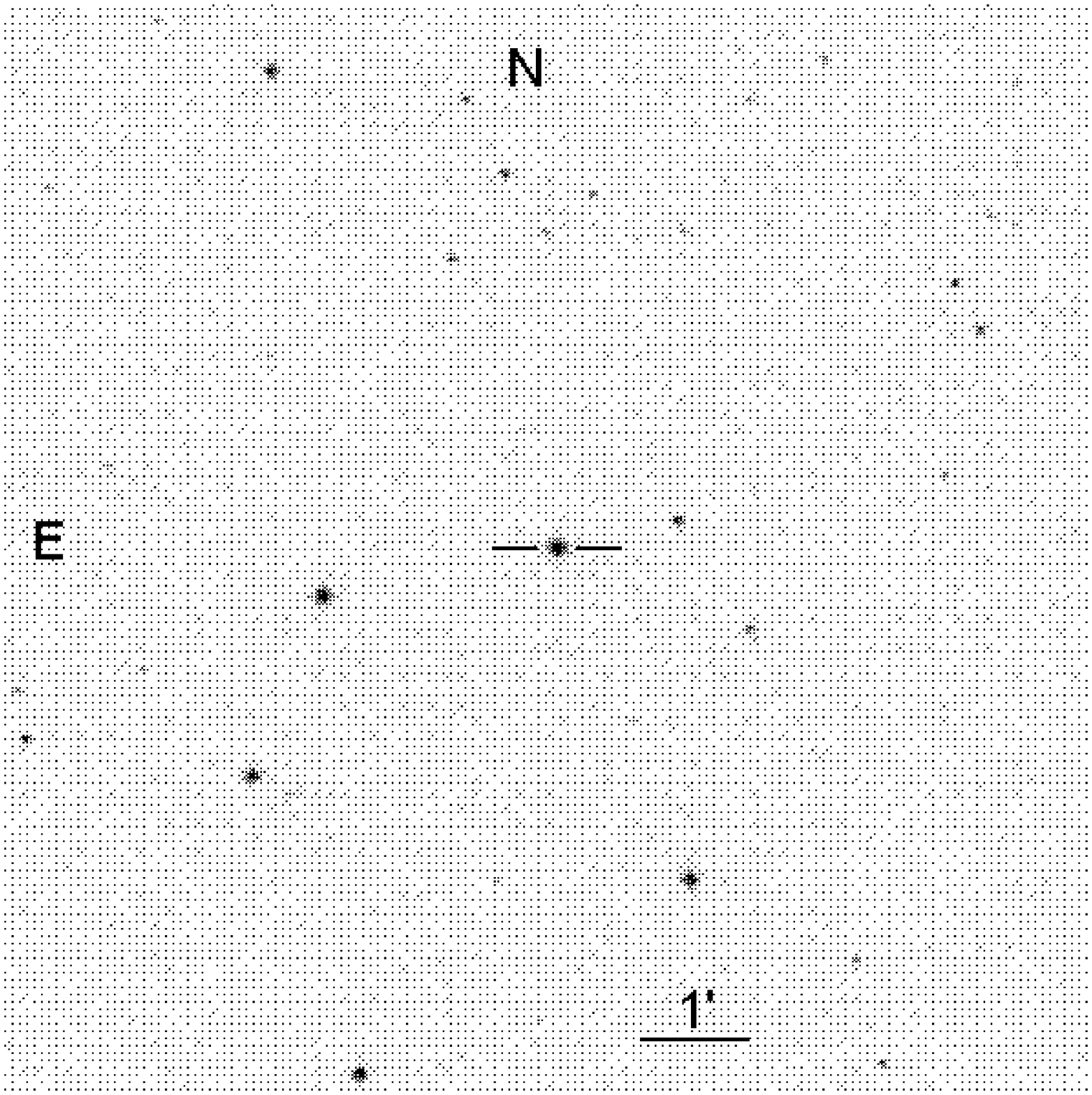}
\caption{The field, 10 arc minutes on a side, of the star GRW+70$^{\circ}$5824.}
\label{fig:figure30}
\end{figure}

\begin{figure}
\plotone{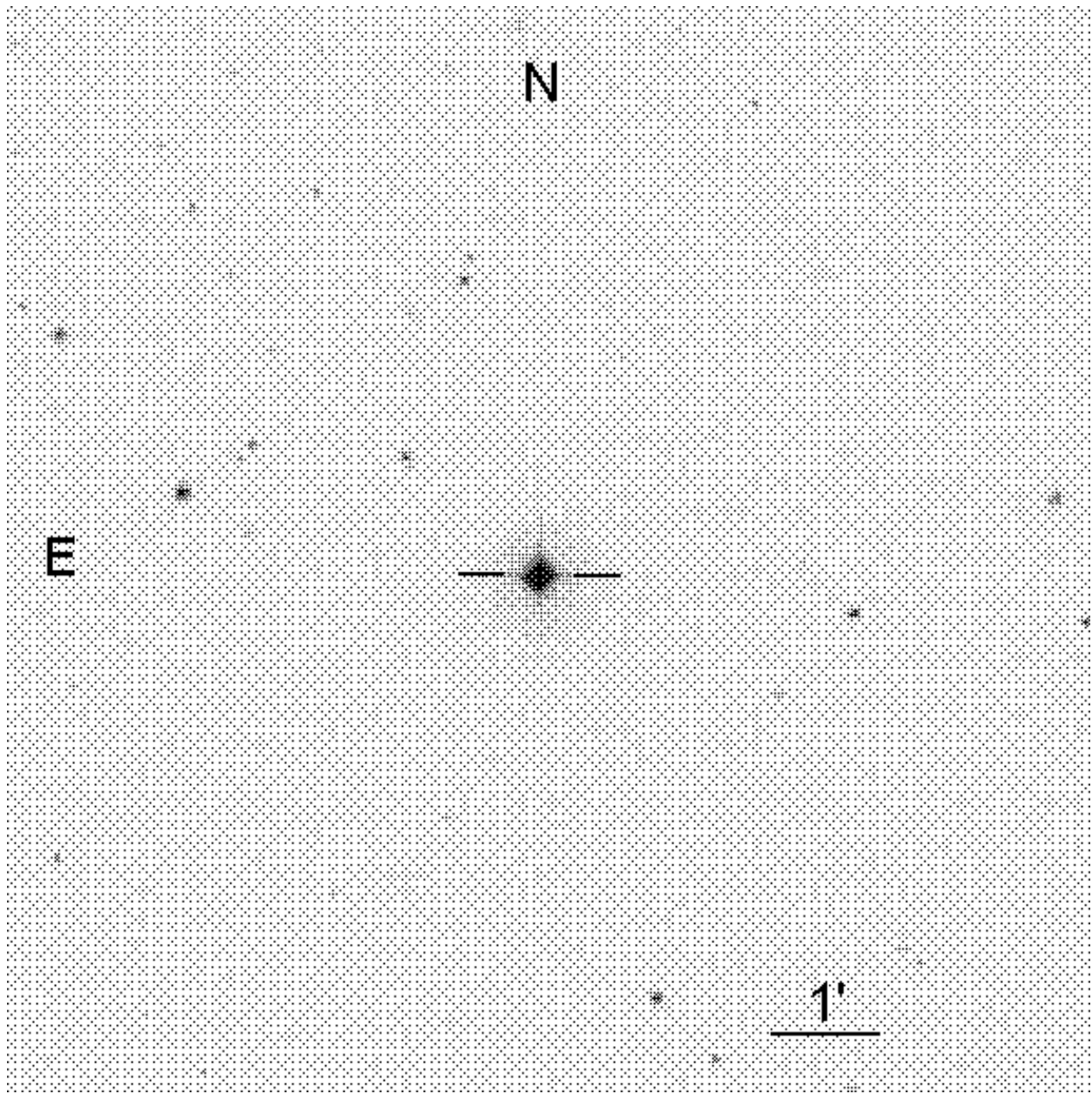}
\caption{The field, 10 arc minutes on a side, of the star BD+26$^{\circ}$2606.}
\label{fig:figure31}
\end{figure}

\begin{figure}
\plotone{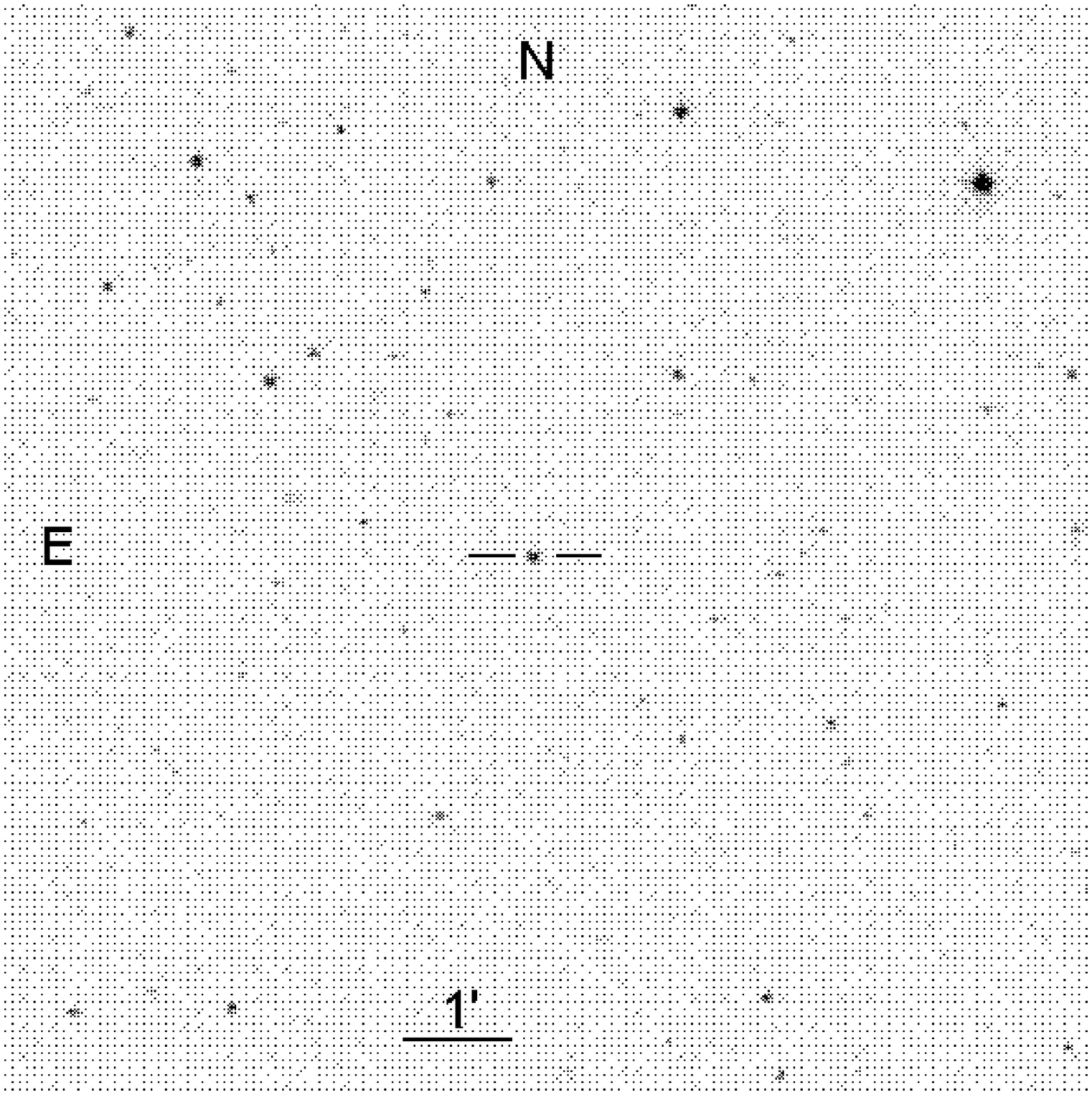}
\caption{The field, 10 arc minutes on a side, of the star GD$\,$190.}
\label{fig:figure32}
\end{figure}

\begin{figure}
\plotone{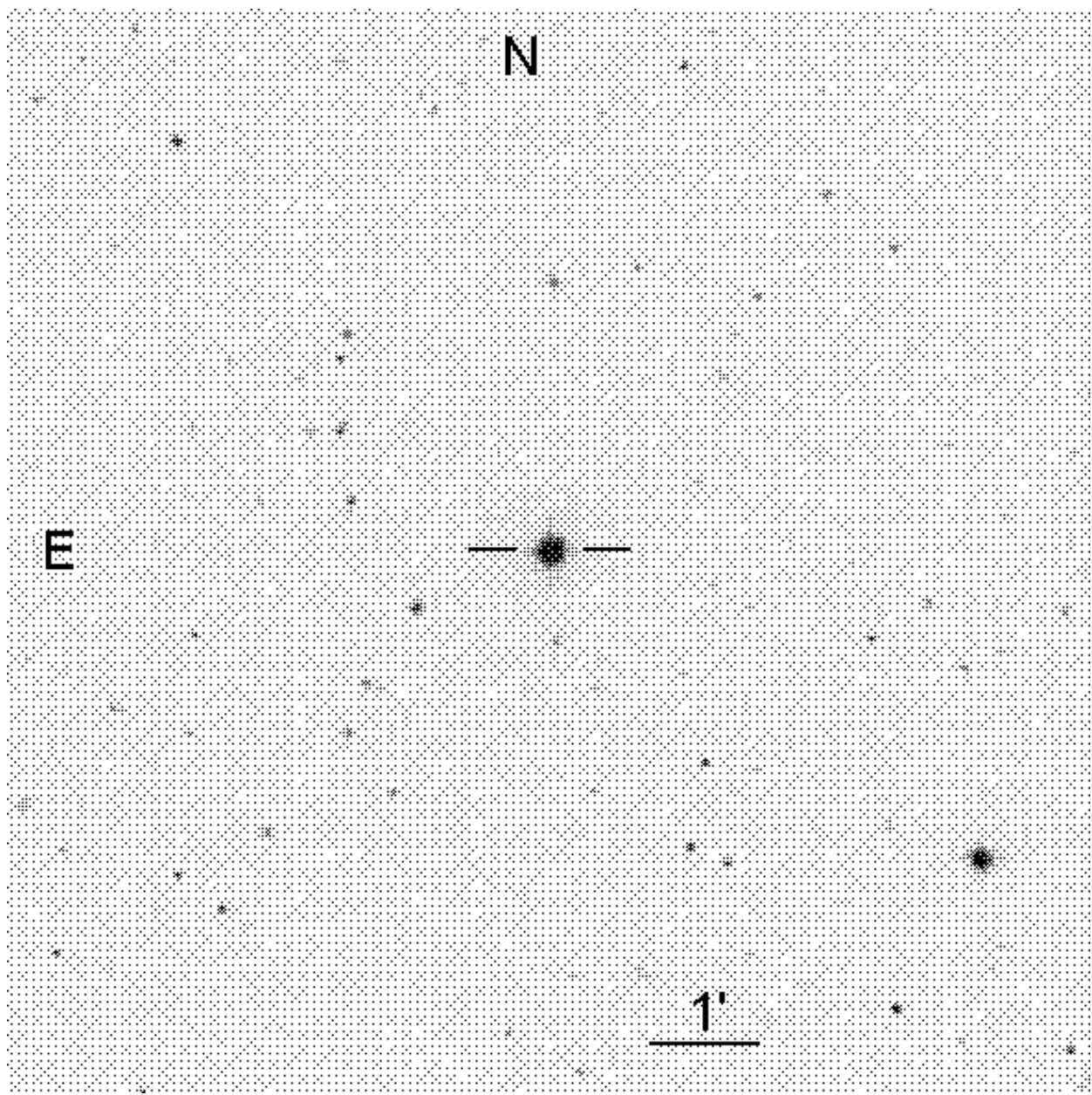}
\caption{The field, 10 arc minutes on a side, of the star BD+33$^{\circ}$2642.}
\label{fig:figure33}
\end{figure}

\begin{figure}
\plotone{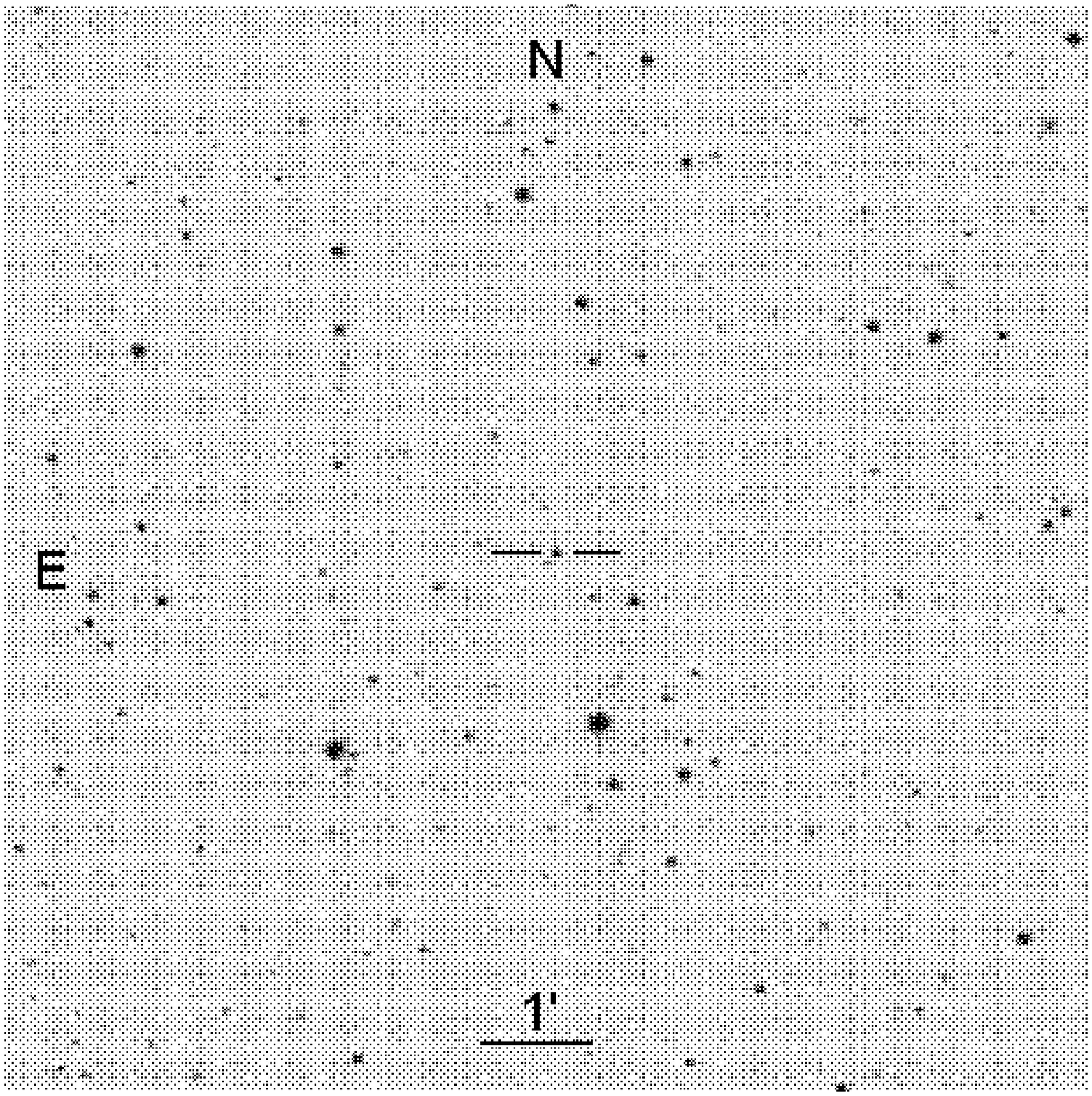}
\caption{The field, 10 arc minutes on a side, of the star G$\,$138-31.}
\label{fig:figure34}
\end{figure}

\begin{figure}
\plotone{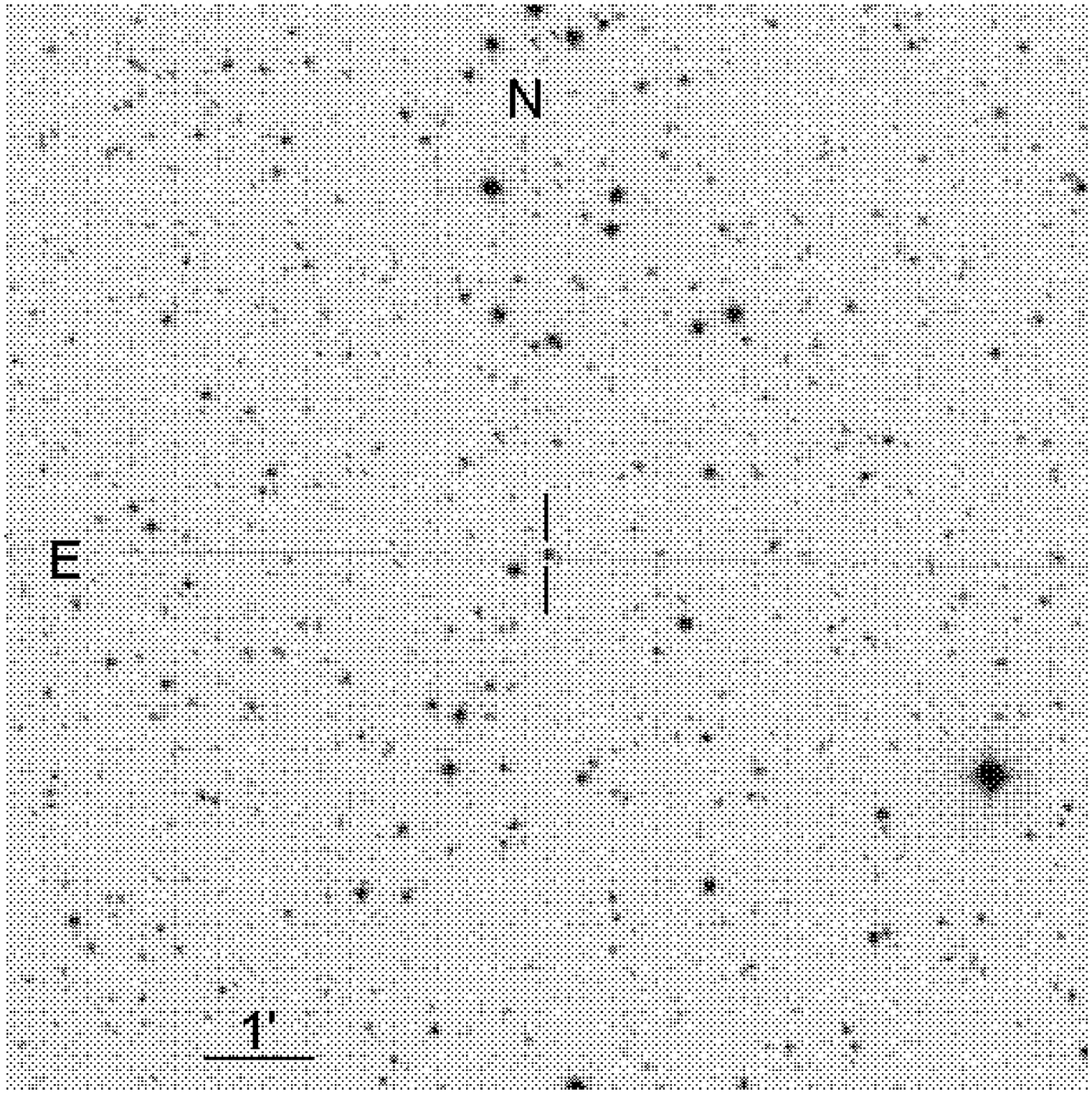}
\caption{The field, 10 arc minutes on a side, of the star G$\,$24-9.}
\label{fig:figure35}
\end{figure}

\begin{figure}
\plotone{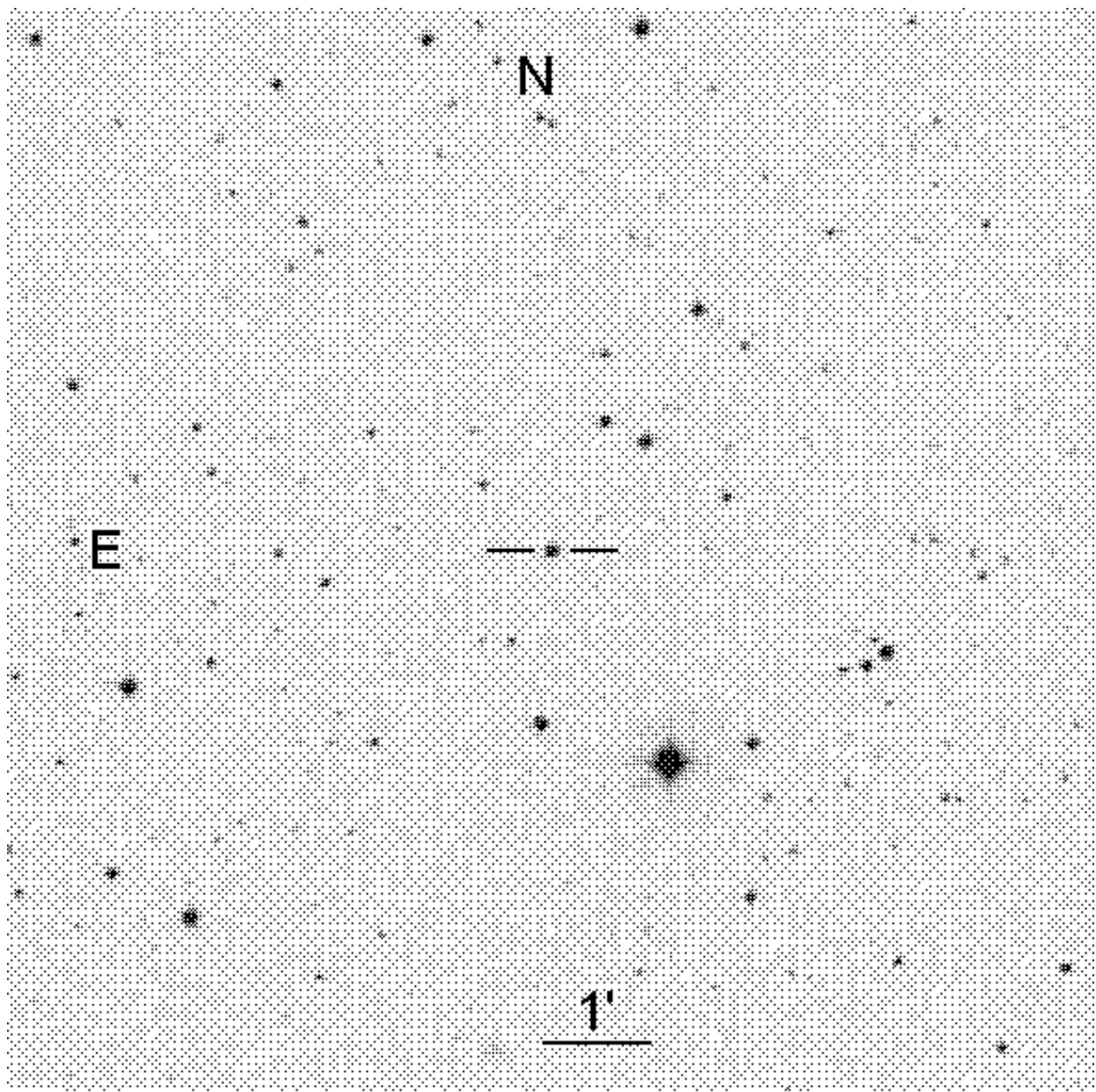}
\caption{The field, 10 arc minutes on a side, of the star LDS$\,$749B.}
\label{fig:figure36}
\end{figure}

\begin{figure}
\plotone{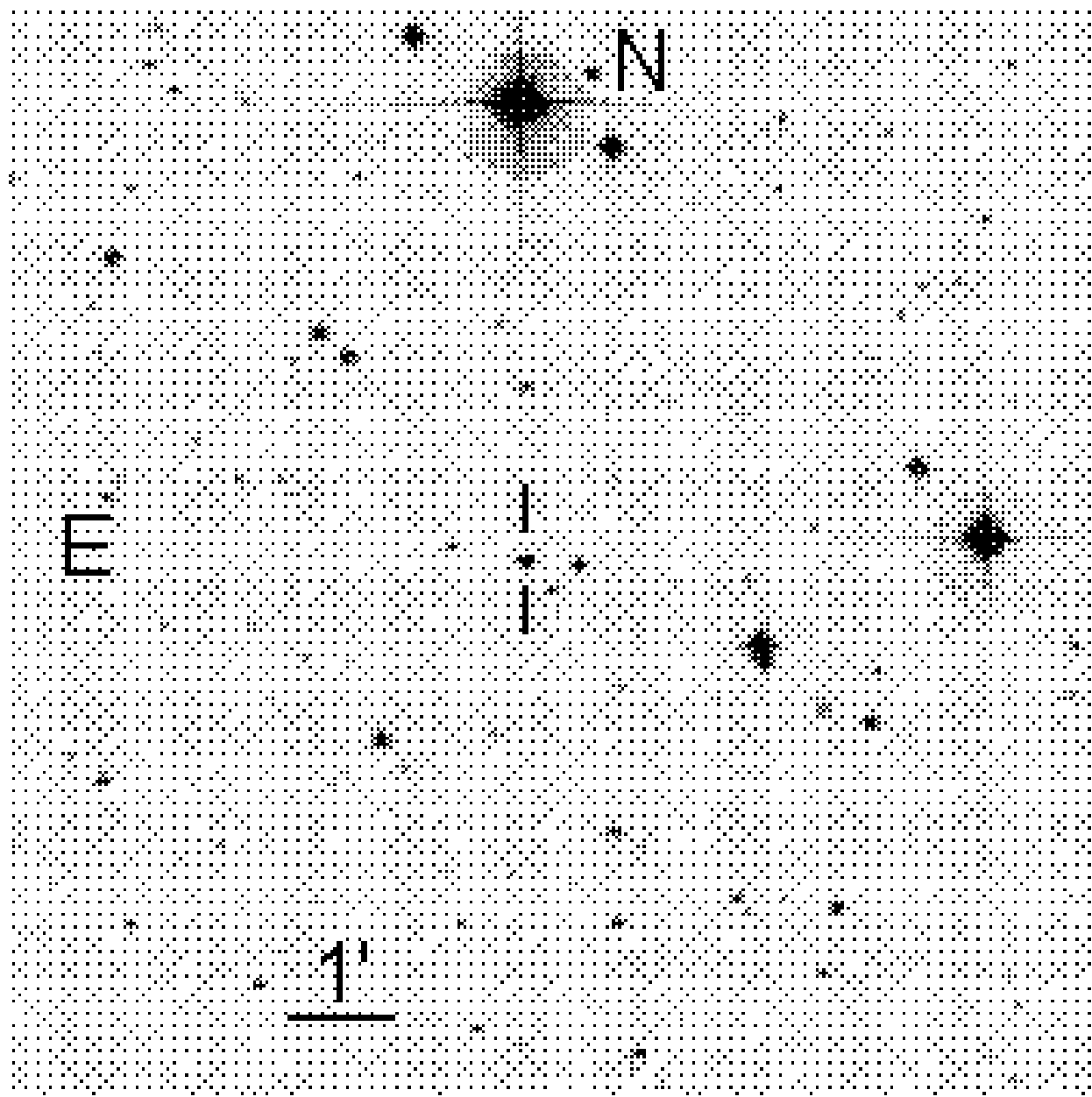}
\caption{The field, 10 arc minutes on a side, of the star L$\,$930-80}
\label{fig:figure37}
\end{figure}

\begin{figure}
\plotone{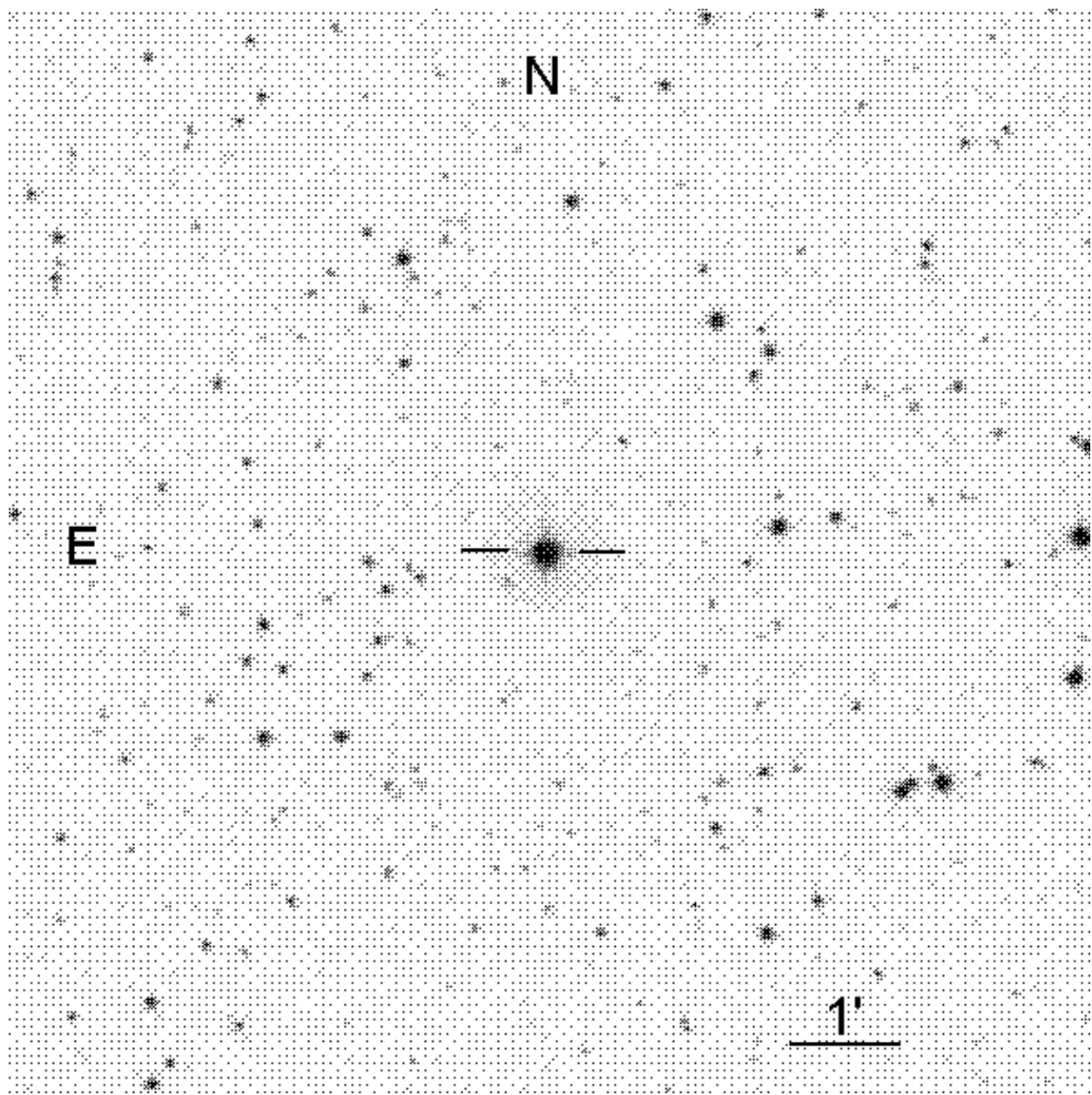}
\caption{The field, 10 arc minutes on a side, of the star BD+28$^{\circ}$4211.}
\label{fig:figure38}
\end{figure}

\begin{figure}
\plotone{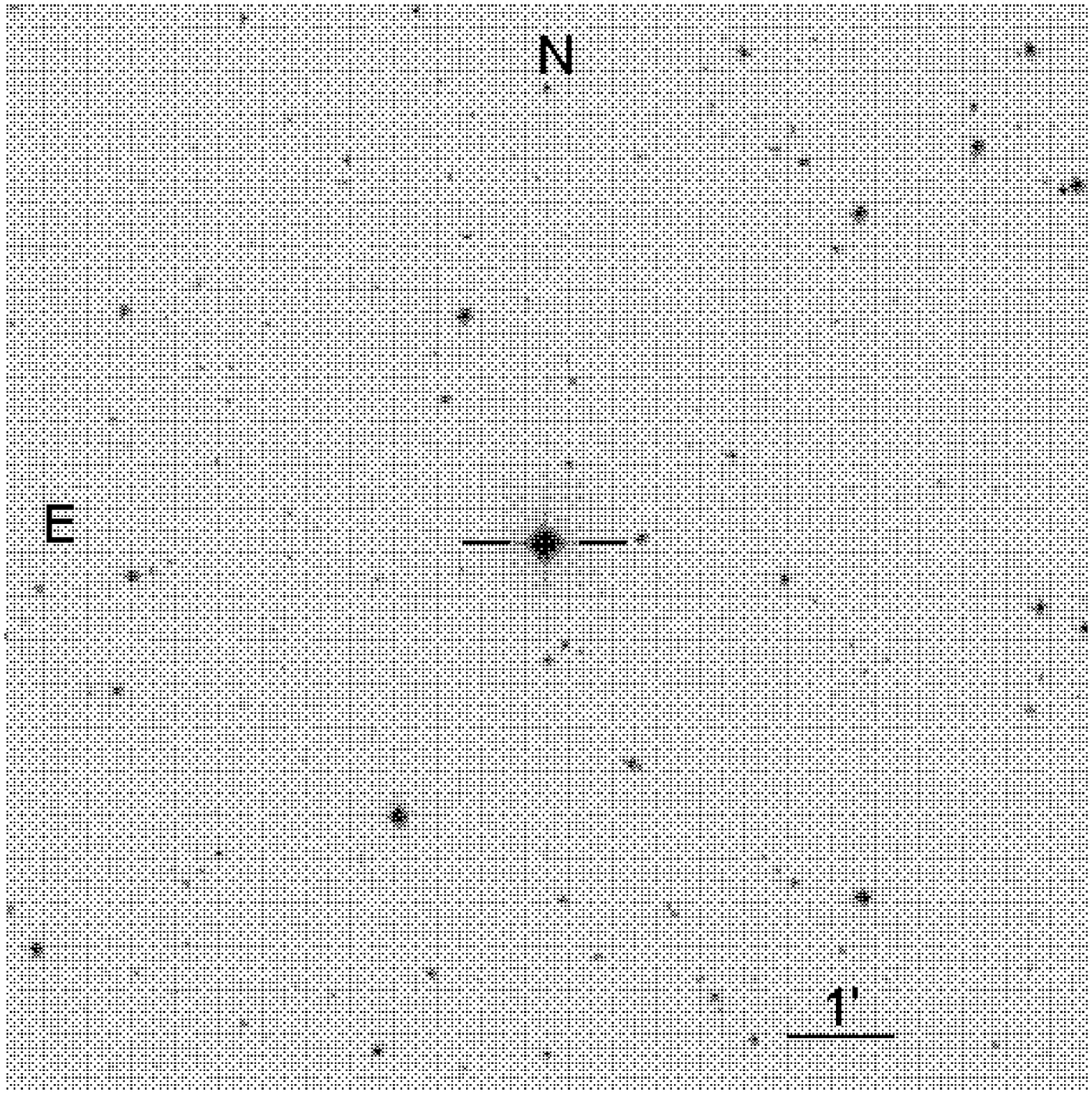}
\caption{The field, 10 arc minutes on a side, of the star BD+17$^{\circ}$4708.}
\label{fig:figure39}
\end{figure}

\begin{figure}
\plotone{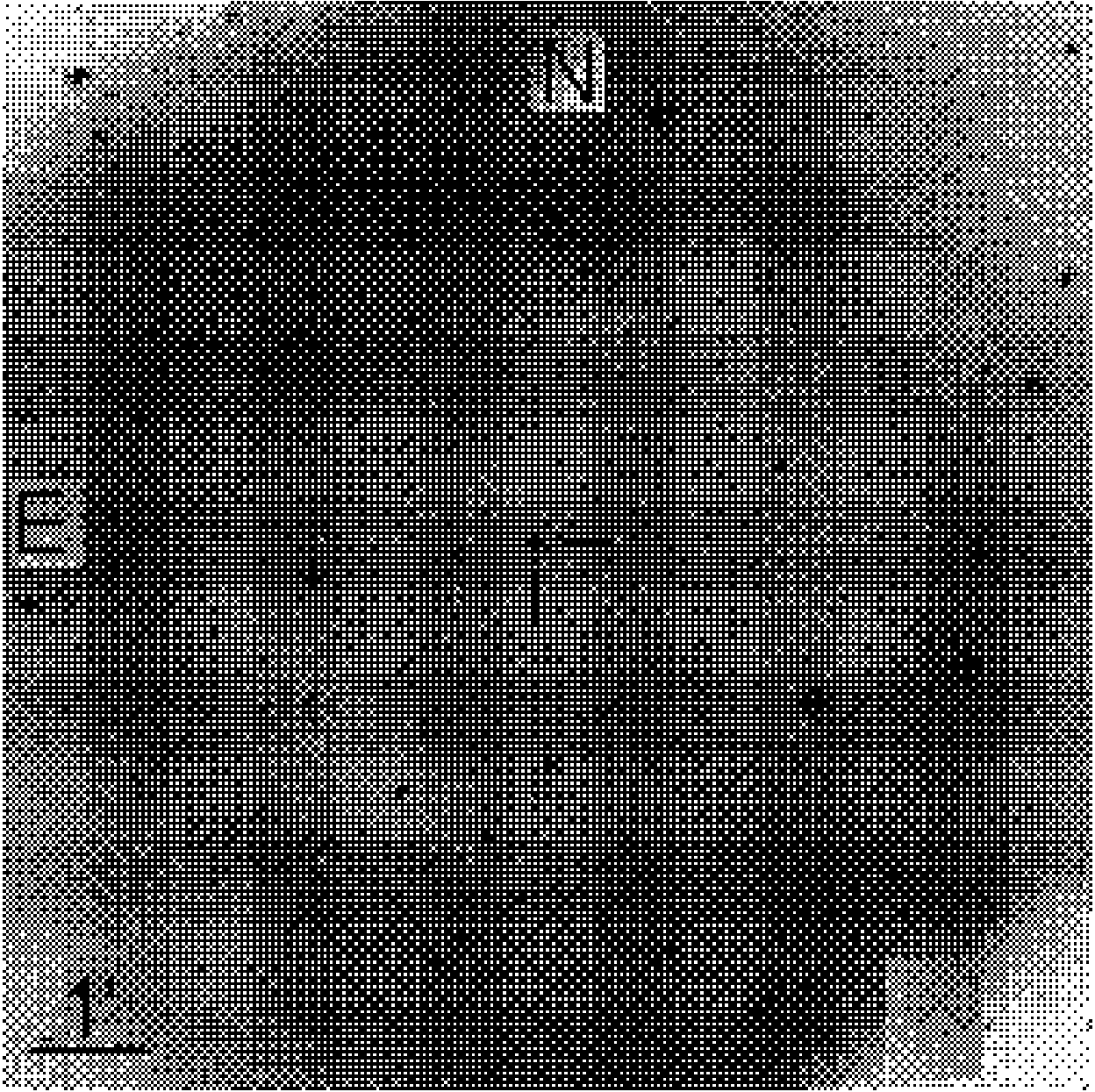}
\caption{The field, 10 arc minutes on a side, of the planetary nebula NGC$\,$7293.}
\label{fig:figure40}
\end{figure}

\begin{figure}
\plotone{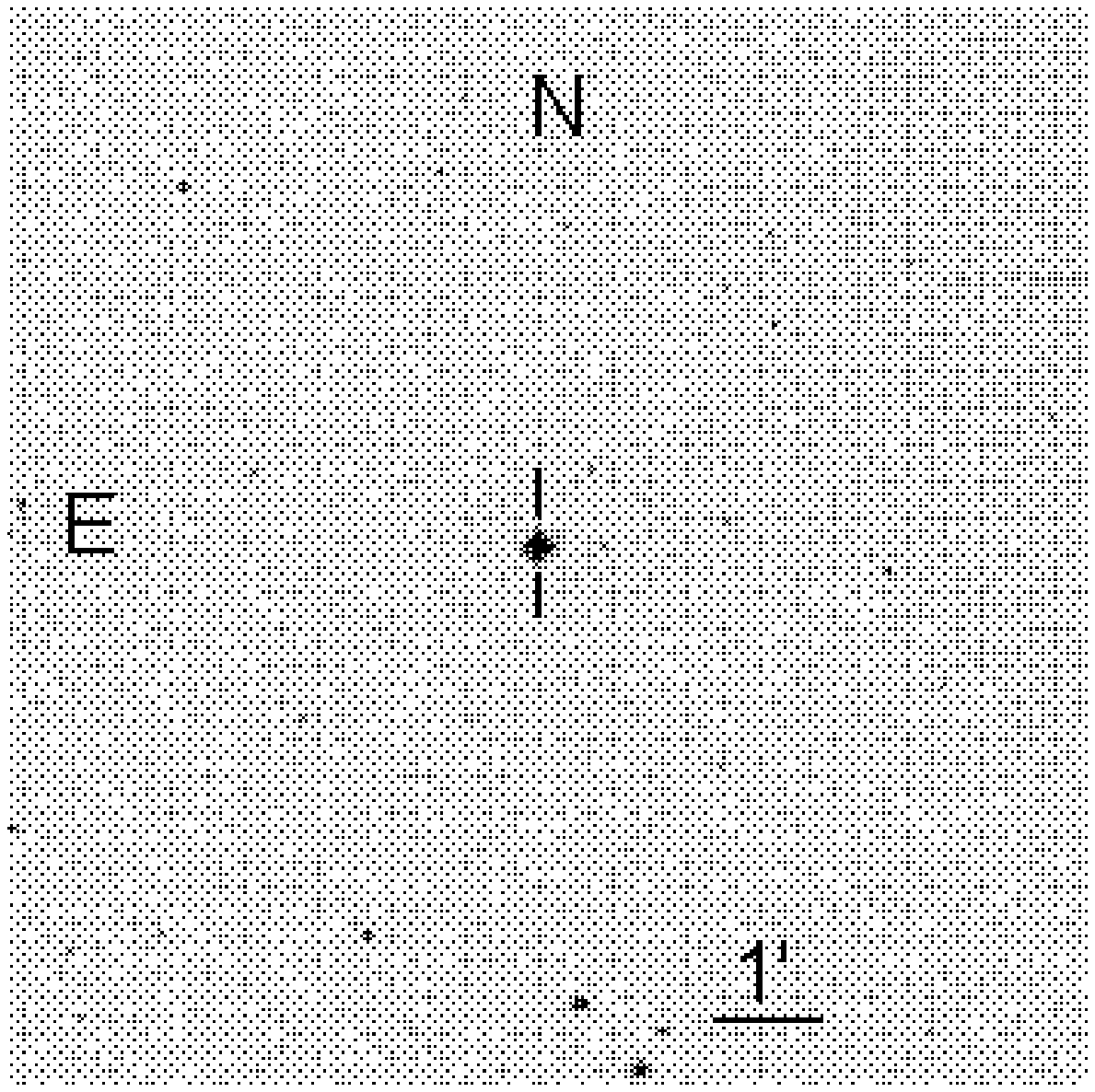}
\caption{The field, 10 arc minutes on a side, of the star Feige$\,$110.}
\label{fig:figure41}
\end{figure}

\begin{figure}
\plotone{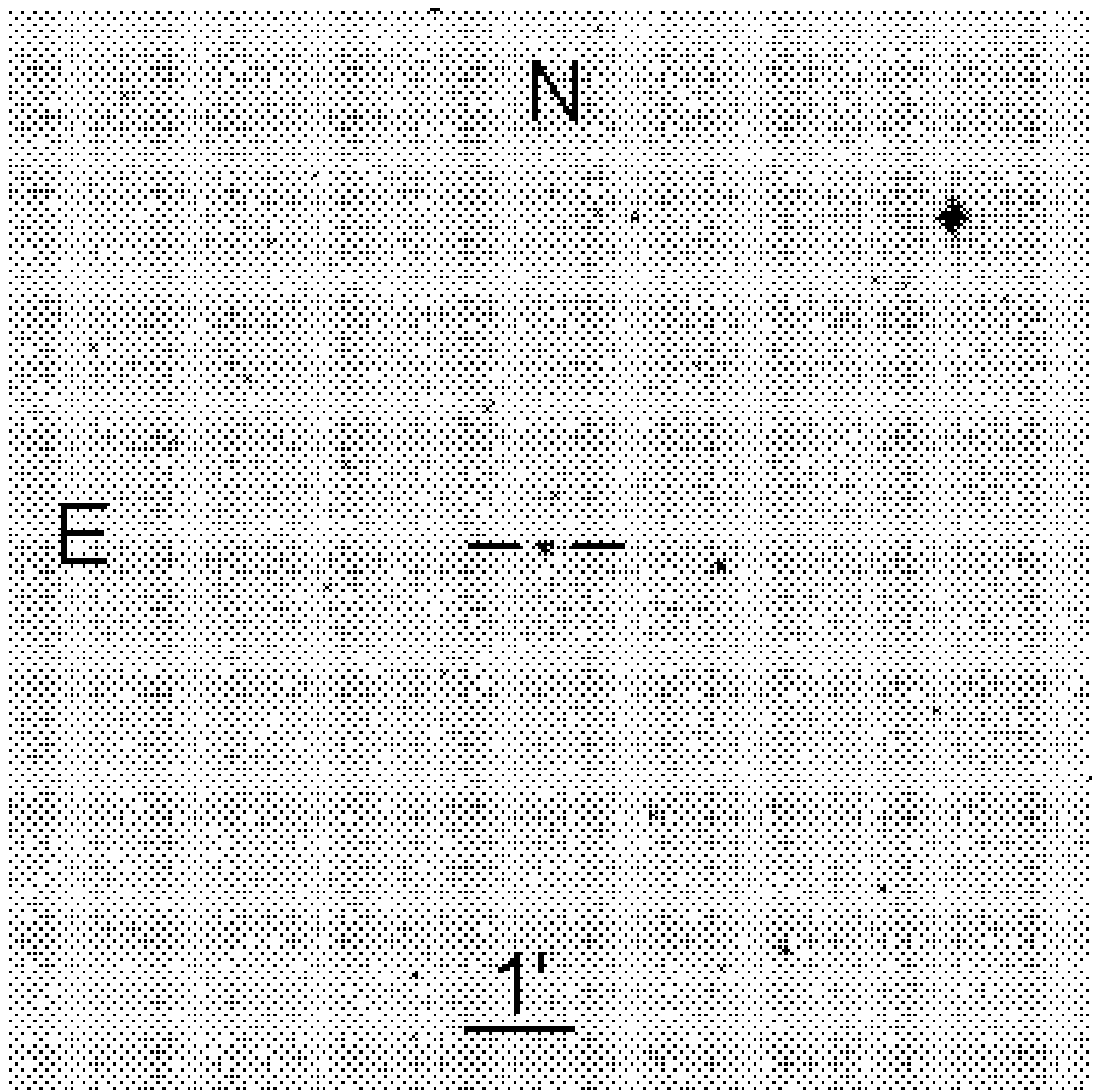}
\caption{The field, 10 arc minutes on a side, of the star LTT$\,$9491.}
\label{fig:figure42}
\end{figure}

\begin{figure}
\plotone{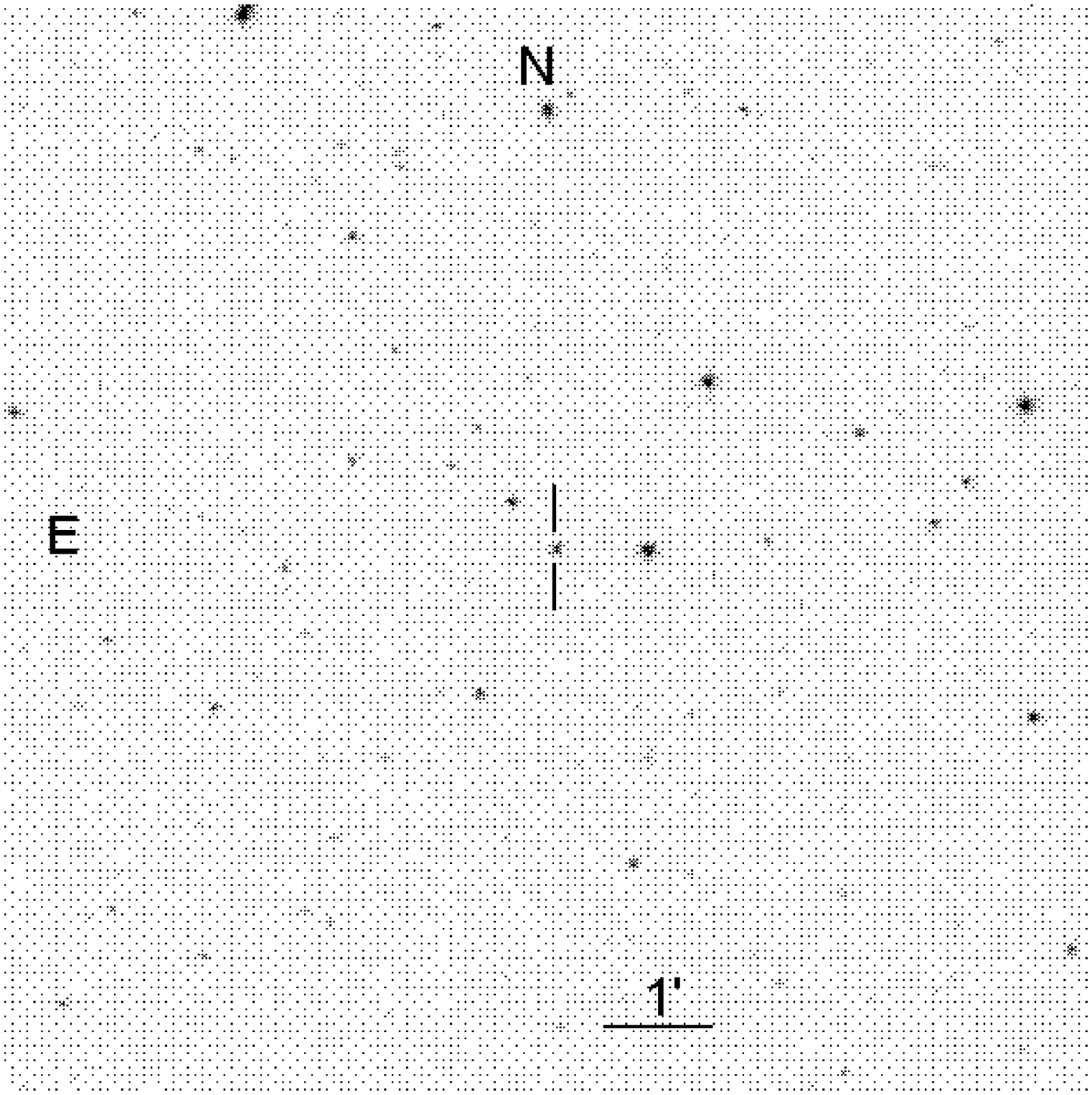}
\caption{The field, 10 arc minutes on a side, of the star GD$\,$248.}
\label{fig:figure43}
\end{figure}

The final magnitude and color indices for the stars in this program are tabulated in Table 
\ref{tab:table4}.  Each star was observed an average of 35 times on 17 nights.  Most of the stars' 
identifications were provided by the Space Telescope Science Institute (STScI) staff; a few were taken 
from the literature.  Finding charts are provided herein via Figures 
\ref{fig:figure13}-\ref{fig:figure43}.

The coordinates for the stars in Table \ref{tab:table4} were computed by STScI staff for the equinox 
J2000.  Proper motion terms were included where necessary.

Columns 4 - 9 in Table \ref{tab:table4} give the final magnitude and color indices on the $UBVRI$ 
photometric system as defined by \citet{Landolt1992}.  Column 10 indicates the number of times, $n$, 
that each star was observed.  Column 11 gives the number of nights, $m$, that each star was observed.  
The numbers in columns 4 - 9 are mean magnitudes and color indices.  Hence, the errors tabulated in 
columns 12 - 17 are mean errors of the mean magnitude and color indices \citep[see][p. 
450]{Landolt1983}.

\section{Comments on Individual Stars}
\label{sec:comments}

G 24-9: One of the program stars, G24-9, was found to be variable in light.  It was reported to be quite 
faint \citep{Landolt1985}, at $V=18.3$, on 1985 October 7.11 UT.  This observation has not been included 
in the averaged magnitude and color indices in Table \ref{tab:table4}.  Such a large drop in brightness 
of some 2.6 magnitudes would point toward the occurrence of an eclipse.  The observation led to G24-9's 
designation as V1412 Aql \citep{Kholopov1989}.  Confirmation appeared in the literature 
\citep{CarilliConner1988, ZuckermanBecklin1988}.  In addition, on other occasions, the $V$ magnitude of 
G24-9 seems to show more scatter than was evidenced for other stars of similar brightness in this 
program.  The approximate 0.25 magnitude variation otherwise observed may indicate that one component is 
variable in light.  On the other hand, there is a faint nearby star, not visible on the acquisition 
monitor, and several arc seconds distant, whose relative location slowly is changing due to G24-9's 
large proper motion.  The relatively small variations may be due to that faint star's sometime presence 
within the photometer's diaphragm, and at other times its exclusion.  \citet{FilippenkoGreenstein1984} 
classified G24-9 as a DQ7 white dwarf.  G24-9's nearness to its optical companion means that under any 
circumstances it is not a good standard star anyway, especially since it has not been possible to 
properly calibrate G24-9.

BD+75$^{\circ}$325: The star BD+75$^{\circ}$325 was considered by \citet{Bartolini1982} to be a possible 
variable star of small amplitude, perhaps 0.03 magnitude.  It has been assigned the suspected variable 
star name NSV 17739 \citep{Kazarovets1998}.  Data taken on one night indicated the presence of a period 
of 0.0465116 days.  However small variations on other nights did not fit that period.  The data in this 
paper were taken on 16 nights over a period of 63 months between 1985 December 14 and 1991 March 25.  No 
more than two or three data points were taken on any one night.  The mean error of a single observation 
is about 0.04 magnitude, much too large for a star so bright.  So, the current data agree with the short 
term amplitude found by \citet{Bartolini1982}.  What is more interesting, though, is that Bartolini et 
al. quote a $V$ magnitude of 8.9 (their Table 1), whereas the current data indicate $V=9.548$, in 
agreement with the Hipparcos value of $V=9.55$ (HIP 40047).  On the other hand, if one reads off an 
average $\Delta m$ of $-0.06$ from their Figure 4, and applies that quantity to a $V$ of 9.60, taken from 
SIMBAD, for their primary comparison star, BD+74$^{\circ}$356, one finds $V=9.54$, on average, for their 
measurements of BD+75$^{\circ}$325.  Therefore, overall, the star has had no long term light variation 
of note.  The fact that both the \citet{Bartolini1982} data and the current results show a variation of 
three or four percent is a firm indication that the star is variable in light.  And, the quoted $V=8.9$ 
either is not a $V$ magnitude, or is a typo.

Feige 34: Feige 34 is listed by \citet{Thejll1995} as a binary based on a measured infrared flux excess.

HZ 44: The star HZ 44 has been assigned the suspected variable star designation NSV 19768 
\citep{Kazarovets1998} apparently on the basis on one discrepant measurement \citep{Kilkenny1977}.  The 
data herein in Table \ref{tab:table4} indicate that HZ 44 is constant in light at the level of the 
accuracy quoted.  \citet{UllaThejll1998} list HZ 44 as a suspected binary based on a measured infrared 
flux excess, but note the possibility of a filter wheel problem.

BD+17$^{\circ}$4708: This star long has been used as a primary spectrophotometric standard star 
\citep{OkeGunn1983}.  \citet{Lu1987} showed, via speckle observations, that BD+17$^{\circ}$4708 = 
G$\,$126-62 is an astrometric binary with a period of 29.6 years.

BD+33$^{\circ}$2642: A long used spectrophotometric standard star surrounded by a faint planetary 
nebula \citep{Napiwotzki1993}.  The star also exhibits radial velocity variations 
\citep{Napiwotzki2001, DeMarco2004}.

BD+26$^{\circ}$2606: This star long has been used as a primary spectrophotometric standard star 
\citep{OkeGunn1983}.  \citet{CarneyLatham1987} showed that BD+26$^{\circ}$2606 = G 166-45 was a double 
lined spectroscopic binary.  It also is a high proper motion star \citep[][and see Volume 8 of the 
Hipparcos Catalogue]{Perryman1997}.  The Hipparcos Catalogue shows a range in brightness of 0.1 
magnitude.  The present data indicate that the error of a single observation is 0.0045 x 6 = 0.027 
magnitude, a bit larger than one might expect for so bright a star.

BD+28$^{\circ}$4211: \citet{MasseyGronwall1990} reported that this star, long used as a spectrophotometric 
standard, had a companion at position angle 240$^{\circ}$ and with a separation of 2.8$^{\prime\prime}$.  
\citet{UllaThejll1998} list BD+28$^{\circ}$4211 as a suspected binary based on a measured infrared flux 
excess.

Feige 110: The star Feige 110 has been assigned the suspected variable star number NSV 14503 
\citep{Kazarovets1998} apparently on the basis of the $V$ magnitude discrepancy found by 
\citet{Graham1969} between his and those belonging to \citet{EggenGreenstein1965} (11.81 versus 11.50, 
respectively).  The $V$ magnitude of 11.832 reported herein in Table \ref{tab:table4} agrees with 
\citet{Graham1969}.  Furthermore, the small error indicated in Table \ref{tab:table4} emphasizes that 
Feige 110 is constant in light.  \citet{UllaThejll1998} list this star as a suspected binary based on a 
measured infrared flux excess, but note the possibility of a filter wheel problem.

HD 49798: Finally, a few comments on the star HD 49798 = UCAC2 12836082 [$\alpha=06^{h}48^{m}04.7^{s}$; 
$\delta=-44^{\circ}18^{\prime}58.4^{{\prime}{\prime}}$; 2000.0; $\mu_{\alpha}=-4.9\,$mas/yr, 
$\mu_{\delta}=+7.6\,$mas/yr, all from UCAC2].  It had been included in the list of spectrophotometric 
standard stars for which $UBVRI$ photometry was desirable for HST needs, but was too bright to be 
included in the main observational program.  Hence HD 49798 was observed on several nights at a CTIO 
0.4-m telescope, and at the 0.61-m Lowell telescope located at CTIO.  A total of six measures were made 
on five different nights, resulting in $V=8.287\pm0.0024$, $(B-V)=-0.270\pm0.0024$, 
$(U-B)=-1.259\pm0.0029$, $(V-R)=-0.104\pm0.0012$, $(R-I)=-0.149\pm0.0020$, and $(V-I)=-0.256\pm0.0012$.  
The errors again are mean errors of the mean.

\acknowledgements{It is a pleasure to thank the staffs of KPNO and CTIO for their hospitality and 
assistance.  Helpful comments on drafts of this paper were made by Drs. John A.~Graham and Philip 
Massey.  AUL is most indebted to Dr. David A.~Turnshek and his then colleagues at STScI, and to 
Dr. Ralph C.~Bohlin of the STScI for the finding charts and for their support and consultation 
throughout the project.  Thanks go to Dr. T.~Kinman who verified certain instrumental 
characteristics during the late stages of the preparation of this paper.  B.~Skiff updated AUL 
with techniques to ensure that the coordinates and proper motions are modern and accurate.  Drs. 
Howard Bond and Jay Holberg provided suggestions regarding spectral types.  The appearance of this 
paper's Figures and Tables is due to the skills of James L.~Clem and Karen Richard, to whom AUL is 
very grateful.  This observational program has been supported by grants to AUL from the Air Force 
Office of Scientific Research (AFOSR) grant no.  82-0192, by STScI grant no.  CW-0004-85, and by 
NSF grants AST 9114457 and 0503871.}

\appendix

\section{Appendix I}
\label{sec:appendix1}

Knowledge of the sensitivity of a photomultiplier (any detector, for that matter) as a function of 
wavelength, as well as the transmission characteristics of the filters used in a photometric program is 
needed for the theoretical modeling of a photometric system.  Unfortunately, such information never was 
available or obtainable for the RCA 31034A-02 (KPNO no. H 18862) used to obtain the data described in 
this paper.  Such information for that brand photomultiplier may be found in \citet{Landolt1992} [ 
Please consult Table 11 and Figures 51 - 54, therein.].

\begin{figure}
\plotone{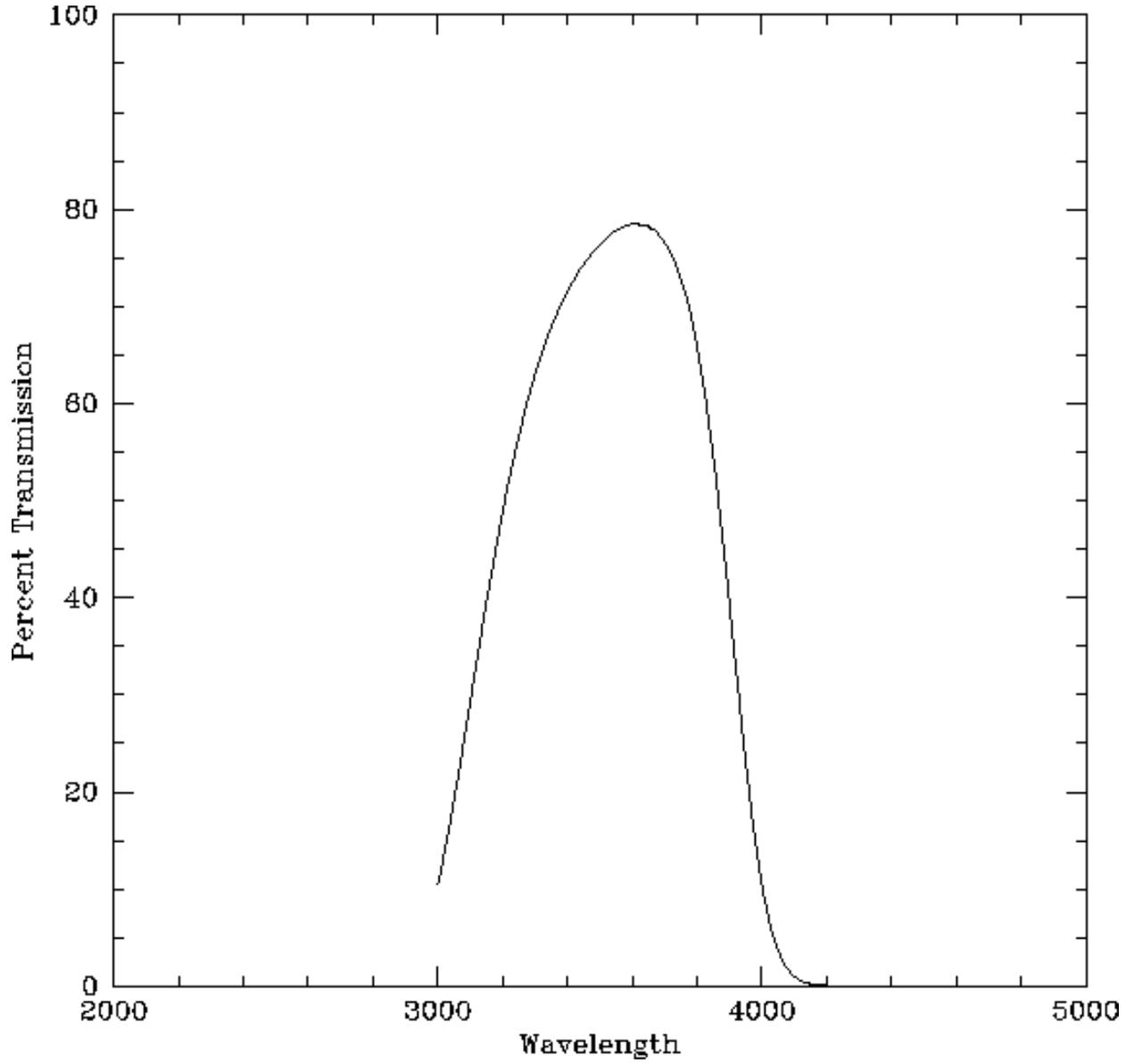}
\caption{The transmission characteristics of the $U$ filter:  1mm UG 2 + CuSO$_{4}$.}
\label{fig:figure44}
\end{figure}

\begin{figure}
\plotone{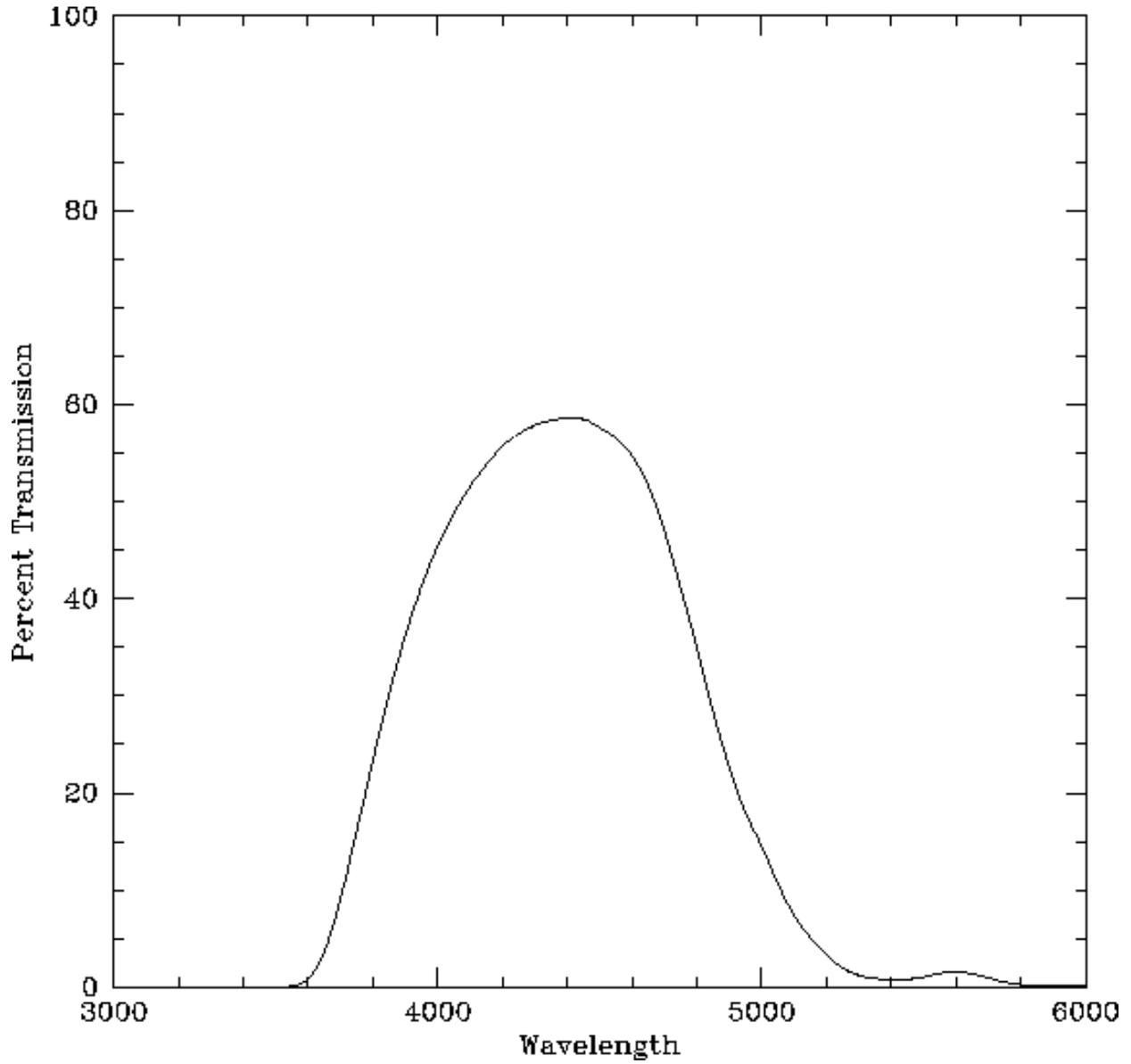}
\caption{The transmission characteristics of the $B$ filter:  2mm GG 385 + 1mm BG 12 + 1mm BG 18.}
\label{fig:figure45}
\end{figure}

\begin{figure}
\plotone{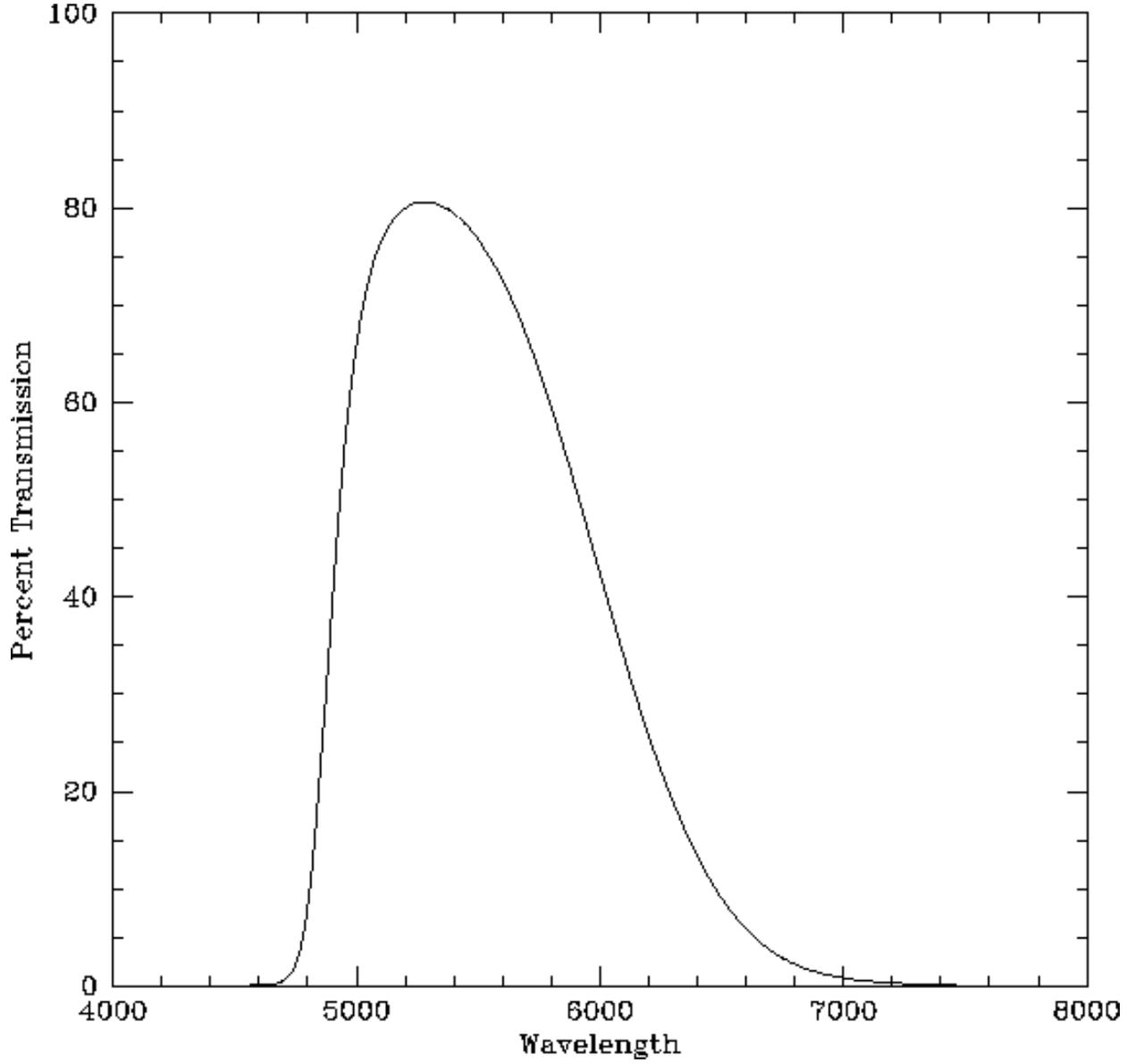}
\caption{The transmission characteristics of the $V$ filter:  2mm GG 495 + 1mm BG 18.}
\label{fig:figure46}
\end{figure}

\begin{figure}
\plotone{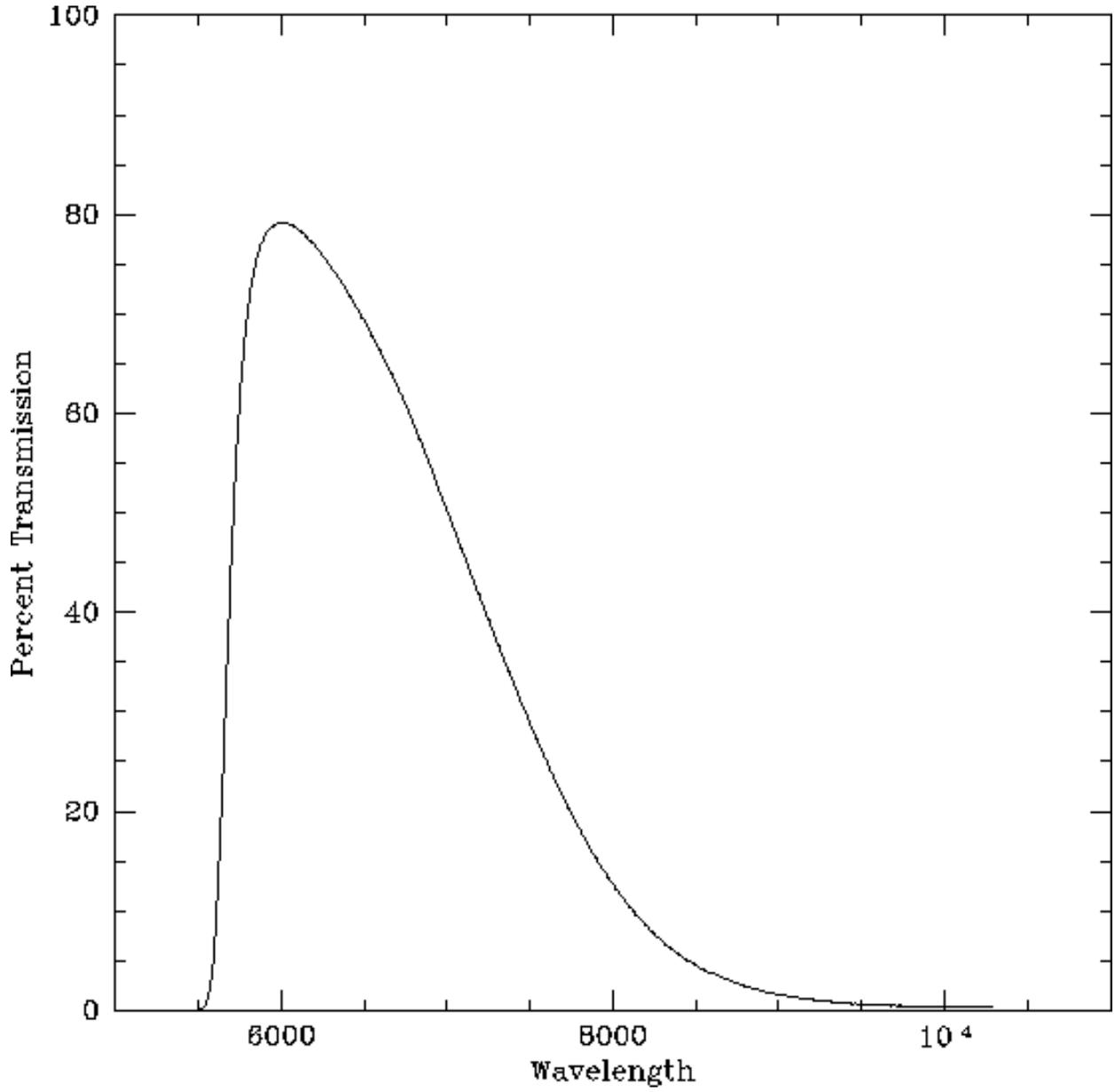}
\caption{The transmission characteristics of the $R$ filter:  2mm OG 570 + 2mm KG 3.}
\label{fig:figure47}
\end{figure}

\begin{figure}
\plotone{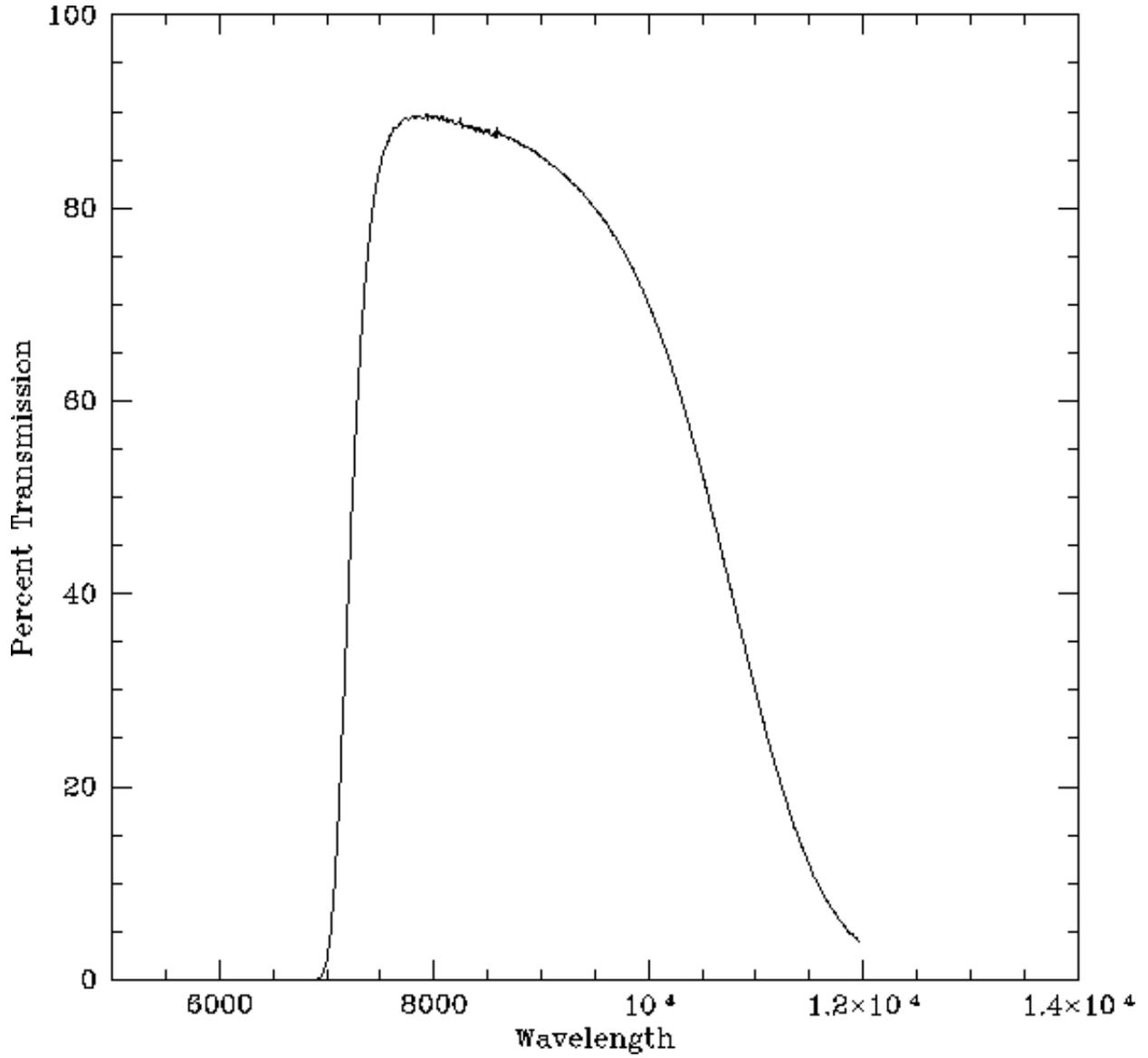}
\caption{The transmission characteristics of the $I$ filter:  3mm RGN 9.}
\label{fig:figure48}
\end{figure}

On the other hand, the transmission characteristics of the KPNO ``J" $UBVRI$ filter set were measured at 
KPNO by Mr. Ed Carder, using a Lambda 9 Spectrophotometer.  He used a slit that gave a resolution of 10 
Angstroms.  Tables \ref{tab:table5}-\ref{tab:table9} provide the measured transmission characteristics 
of the $UBVRI$ filters in the KPNO ``J" filter set.  Figures \ref{fig:figure44}-\ref{fig:figure48} 
illustrate the transmission characteristics for the KPNO ``J" filter set.

\section{Appendix II}
\label{sec:appendix2}

Observational programs sometimes demand the best available coordinate and motion information for 
standard stars.  Table \ref{tab:table10} provides the most recent  coordinates and proper motions for the 
program stars in Table \ref{tab:table4}.  All coordinates are for the epoch J2000.  The 2MASS-PSC 
positions come from The Two Micron All Sky Survey (2MASS) \citep{Skrutskie2006}.  The UCAC2 positions 
came from The Second US Naval Observatory CCD Astrograph Catalogue (UCAC2) \citep{Zacharias2004}.  
Representative spectral types are listed in Table \ref{tab:table10}.  The literature sources for these 
spectral types appear in the last column.


\begin{thebibliography}{}

\bibitem[Bartolini et al.(1982)]{Bartolini1982} Bartolini, C., Bonifazi, A., Pecci, F.~F., Oculi, L., 
Piccioni, A., Serra, R., \& Dantona, F.\ 1982, \apss, 83, 287 

\bibitem[Bessell(1979)]{Bessell1979} Bessell, M.~S.\ 1979, \pasp, 91, 589 

\bibitem[Bohlin et al.(1990)]{Bohlin1990} Bohlin, R.~C., Harris, A.~W., Holm, A.~V., \& Gry, C.\ 
1990, \apjs, 73, 413 

\bibitem[Carney \& Latham(1987)]{CarneyLatham1987} Carney, B.~W., \& Latham, D.~W.\ 1987, \aj, 93, 116 

\bibitem[Carilli et al.(1988)]{CarilliConner1988} Carilli, C., Conner, S., \& Green, D.~W.~E.\ 1988, 
\iaucirc, 4648, 2 

\bibitem[De Marco et al.(2004)]{DeMarco2004} De Marco, O., Bond, H.~E., Harmer, D., \& Fleming, 
A.~J.\ 2004, \apjl, 602, L93 

\bibitem[Eggen \& Greenstein(1965)]{EggenGreenstein1965} Eggen, O.~J., \& Greenstein, J.~L.\ 1965, \apj, 141, 83 

\bibitem[Filippenko \& Greenstein(1984)]{FilippenkoGreenstein1984} Filippenko, A., \& Greenstein, J.~L.\ 1984, 
\pasp, 96, 530 

\bibitem[Graham(1969)]{Graham1969} Graham, J.~A.\ 1969, in Low-Luminosity Stars, S.~S. Kumar, Editor, (Gordon \& Breach: New York), 139 

\bibitem[Hanson et al.(2004)]{Hanson2004} Hanson, R.~B., Klemola, A.~R., Jones, B.~F., \& Monet, 
D.~G.\ 2004, \aj, 128, 1430

\bibitem[Kazarovets et al.(1998)]{Kazarovets1998} Kazarovets, E.~V., Samus, N.~N., \& Durlevich, O.~V.\ 
1998, Informational Bulletin on Variable Stars, 4655, 1 

\bibitem[Kholopov et al.(1989)]{Kholopov1989} Kholopov, P.~N., Samus, N.~N., Kazarovets, B.~V., Frolov, 
M.~S., \& Kireeva, N.~N.\ 1989, Informational Bulletin on Variable Stars, 3323, 1 

\bibitem[Kilkenny(1977)]{Kilkenny1977} Kilkenny, D.\ 1977, \mnras, 181, 611 

\bibitem[Klemola et al.(1987)]{Klemola1987} Klemola, A.~R., Jones, B.~F., \& Hanson, R.~B.\ 1987, \aj, 94, 501 

\bibitem[Landolt(1967)]{Landolt1967} Landolt, A.~U.\ 1967, \aj, 72, 1012 

\bibitem[Landolt(1973)]{Landolt1973} Landolt, A.~U.\ 1973, \aj, 78, 959 

\bibitem[Landolt(1983)]{Landolt1983} Landolt, A.~U.\ 1983, \aj, 88, 439 

\bibitem[Landolt(1985)]{Landolt1985} Landolt, A.~U.\ 1985, \iaucirc, 4125, 2 

\bibitem[Landolt(1992)]{Landolt1992} Landolt, A.~U.\ 1992, \aj, 104, 340

\bibitem[Landolt(2007)]{Landolt2007} Landolt, A.~U.\ 2007, The Future of Photometric, Spectrophotometric, and 
Polarimetric Standardization, C. Sterken, Editor, ASP Conference Series, in press 

\bibitem[Lu et al.(1987)]{Lu1987} Lu, P.~K., Demarque, P., van Altena, W., McAlister, H., \& 
Hartkopf, W.\ 1987, \aj, 94, 1318 

\bibitem[Massey \& Gronwall(1990)]{MasseyGronwall1990} Massey, P., \& Gronwall, C.\ 1990, \apj, 358, 344

\bibitem[McCook \& Sion(2006)]{McCookSion2006} McCook, G.~P., \& Sion, E.~M.\ 2006,  
http://www.astronomy.villanova.edu/WDCatalog/index.html

\bibitem[Monet et al.(2003)]{Monet2003} Monet, D.~G., et al.\ 2003, \aj, 125, 984 

\bibitem[Napiwotzki(1993)]{Napiwotzki1993} Napiwotzki, R.\ 1993, Acta Astronomica, 43, 415 

\bibitem[Napiwotzki et al.(2001)]{Napiwotzki2001} Napiwotzki, R., Herrmann, M., Heber, U., \& Altmann, 
M.\ 2001, Post-AGB Objects as a Phase of Stellar Evolution, ed. R.~Szczerba \& S.~K.~Gorny (Dordrecht: 
Kluwer), 277 

\bibitem[Oke(1990)]{Oke1990} Oke, J.~B.\ 1990, \aj, 99, 1621 

\bibitem[Oke \& Gunn(1983)]{OkeGunn1983} Oke, J.~B., \& Gunn, J.~E.\ 1983, \apj, 266, 713 

\bibitem[Perryman et al.(1997)]{Perryman1997} Perryman, M.~A.~C., et al.\ 1997, \aap, 323, L49 

\bibitem[Roman(1955)]{Roman1955} Roman, N.~G.\ 1955, \apjs, 2, 195 

\bibitem[Schulte \& Crawford(1961)]{SchulteCrawford1961} Schulte, D., \& Crawford, D.~L., 1961, Kitt 
Peak National Observatory Contribution No. 10

\bibitem[Skrutskie et al.(2006)]{Skrutskie2006} Skrutskie, M.~F., et al.\ 2006, \aj, 131, 1163 

\bibitem[Thejll et al.(1995)]{Thejll1995} Thejll, P., Ulla, A., \& MacDonald, J.\ 1995, \aap, 303, 773 

\bibitem[Turnshek et al.(1990)]{Turnshek1990} Turnshek, D.~A., Bohlin, R.~C., Williamson, R.~L., II, 
Lupie, O.~L., Koornneef, J., \& Morgan, D.~H.\ 1990, \aj, 99, 1243 

\bibitem[Ulla \& Thejll(1998)]{UllaThejll1998} Ulla, A., \& Thejll, P.\ 1998, \aaps, 132, 1 

\bibitem[Urban et al.(2004)]{Urban2004} Urban, S.~E., Zacharias, N., \& Wycoff, G.~L.\ 2004, VizieR Online Data 
Catalog, 1294, 0 

\bibitem[Zacharias et al.(2004)]{Zacharias2004} Zacharias, N., Urban, S.~E., Zacharias, M.~I., Wycoff, G.~L., 
Hall, D.~M., Monet, D.~G., \& Rafferty, T.~J.\ 2004, \aj, 127, 3043 

\bibitem[Zuckerman \& Becklin(1988)]{ZuckermanBecklin1988} Zuckerman, B., \& Becklin, E.\ 1988, \iaucirc, 4652, 3

\end{thebibliography}

\newpage

\begin{deluxetable}{ll}
\tablecaption{KPNO $UBVRI$ Filter Set J}
\tablewidth{0pt}
\startdata
\tableline
\tableline
$V$  &  2mm GG 495 + 1mm BG 18 \\
$B$  &  1mm BG 12 + 2mm GG 385 + 1mm BG 18 \\
$U$  &  1mm UG 2 + CuSO$_{4}$ \\
$R$  &  2mm OG 570 + 2mm KG 3 \\
$I$  &  3mm RGN 9 \\
\enddata
\label{tab:table1}
\end{deluxetable}

\begin{deluxetable}{cccc}
\tablecaption{Extinction at Kitt Peak}
\tablewidth{0pt}
\tablehead{\colhead{Magnitude or}                 &
           \colhead{Coefficient}                  &
           \colhead{Average Coefficient}          &
           \colhead{Range in}                     \\ 
           \colhead{Color Index}                  &
           \colhead{Symbol}                       &
           \colhead{Value}                        &
           \colhead{Values}                       }
\startdata
$V$    &  $Q_y$    & $+0.162$ & $+0.081$ to $+0.256$ \\
       &           &          &                      \\
$B-V$  &  $k_1$    & $+0.119$ & $+0.046$ to $+0.223$ \\
       &  $k_2$    & $-0.020$ & $-0.042$ to $+0.000$ \\ 
$U-B$  &  $k_3$    & $+0.341$ & $+0.267$ to $+0.463$ \\
       &  $k_4$    & $-0.013$ & $-0.042$ to $+0.018$ \\
$V-R$  &  $k_5$    & $+0.043$ & $+0.021$ to $+0.082$ \\
       &  $k_6$    & $-0.003$ & $-0.038$ to $+0.017$ \\
$R-I$  &  $k_7$    & $+0.044$ & $-0.003$ to $+0.074$ \\
       &  $k_8$    & $+0.004$ & $-0.010$ to $+0.017$ \\
$V-I$  &  $k_9$    & $+0.087$ & $+0.047$ to $+0.127$ \\
       &  $k_{10}$ & $+0.001$ & $-0.028$ to $+0.014$ \\
\enddata
\label{tab:table2}
\end{deluxetable}

\begin{deluxetable}{ccccc}
\tablecaption{Extinction at Kitt Peak at Three Epochs}
\tablewidth{0pt}
\tablehead{\colhead{Magnitude or}               &
           \colhead{Coefficient}                &
           \colhead{ }                          &
           \colhead{ }                          &
           \colhead{ }                          \\
           \colhead{Color Index}                &
           \colhead{Symbol}                     &
           \colhead{1967}                       &
           \colhead{1969-1972}                  &
           \colhead{1985-1991}                  }
\startdata
$V$    &  $Q_y$   & $+0.171 \pm 0.016$  &  $+0.150$  &  $+0.162 \pm 0.040$ \\
       &          &                     &            &                     \\
$B-V$  &  $k_1$   & $+0.085 \pm 0.008$  &  $+0.080$  &  $+0.119 \pm 0.028$ \\
       &  $k_2$   & $-0.029 \pm 0.002$  &  $-0.030$  &  $-0.020 \pm 0.009$ \\
$U-B$  &  $k_3$   & $+0.326 \pm 0.025$  &  $+0.340$  &  $+0.341 \pm 0.037$ \\
       &  $k_4$   & $-0.020 \pm 0.005$  &  $-0.020$  &  $-0.013 \pm 0.014$ \\
\enddata
\label{tab:table3}
\end{deluxetable}

\begin{deluxetable}{ccccccccccccccccc}
\rotate
\tablecaption{$UBVRI$ Photometry of Spectrophotometric Standard Stars}
\tabletypesize{\tiny}
\tablewidth{0pt}
\tablehead{\colhead{ }                                                       &
           \multicolumn{2}{c}{$\alpha$~~~~(2000)~~~~$\delta$}                &
           \multicolumn{8}{c}{ }                                             &
           \multicolumn{6}{c}{Mean Error of the Mean}                        \\
           \colhead{Star}                    &
           \colhead{\hms}                    &
           \colhead{\dms}                    &
           \colhead{$V$}                     &
           \colhead{$B-V$}                   &
           \colhead{$U-B$}                   &
           \colhead{$V-R$}                   &
           \colhead{$R-I$}                   &
           \colhead{$V-I$}                   &
           \colhead{$n$}                     &
           \colhead{$m$}                     &
           \colhead{$V$}                     &
           \colhead{$B-V$}                   & 
           \colhead{$U-B$}                   &
           \colhead{$V-R$}                   &
           \colhead{$R-I$}                   &
           \colhead{$V-I$}                   }
\startdata
G158 100               & 00 33 54 & $-$12 07 57 & 14.891 & $+$0.681 & $-$0.061 & $+$0.424 & $+$0.418 & $+$0.840 & 33 & 12 & 0.0021 & 0.0047 & 0.0059 & 0.0030 & 0.0063 & 0.0068 \\
BPM 16274              & 00 50 03 & $-$52 08 17 & 14.206 & $-$0.049 & $-$0.803 & $-$0.104 & $-$0.119 & $-$0.222 & 37 & 17 & 0.0016 & 0.0025 & 0.0043 & 0.0028 & 0.0074 & 0.0081 \\
HZ 4                   & 03 55 22 & $+$09 47 19 & 14.506 & $+$0.086 & $-$0.675 & $-$0.074 & $-$0.060 & $-$0.136 & 51 & 21 & 0.0028 & 0.0015 & 0.0028 & 0.0020 & 0.0042 & 0.0046 \\
LB 227                 & 04 09 29 & $+$17 07 54 & 15.323 & $+$0.055 & $-$0.718 & $-$0.085 & $-$0.108 & $-$0.192 & 47 & 21 & 0.0035 & 0.0032 & 0.0034 & 0.0042 & 0.0131 & 0.0143 \\
HZ 2                   & 04 12 44 & $+$11 51 50 & 13.877 & $-$0.090 & $-$0.884 & $-$0.107 & $-$0.111 & $-$0.217 & 34 & 15 & 0.0014 & 0.0015 & 0.0029 & 0.0017 & 0.0024 & 0.0024 \\
                       &          &             &        &          &          &          &          &          &    &    &        &        &        &        &        &        \\
G191 B2B               & 05 05 31 & $+$52 49 54 & 11.781 & $-$0.326 & $-$1.205 & $-$0.149 & $-$0.181 & $-$0.327 & 48 & 23 & 0.0023 & 0.0014 & 0.0026 & 0.0016 & 0.0017 & 0.0025 \\
G193 74                & 07 53 27 & $+$52 29 36 & 15.674 & $+$0.256 & $-$0.563 & $+$0.163 & $+$0.161 & $+$0.324 & 41 & 20 & 0.0055 & 0.0037 & 0.0039 & 0.0055 & 0.0064 & 0.0094 \\
\enddata
\label{tab:table4}
\end{deluxetable}

\newpage
\begin{deluxetable}{cccccccccc}
\tablecaption{Transmission characteristics of the $U$ filter in KPNO's $UBVRI$ filter set J}
\tabletypesize{\footnotesize}
\tablewidth{0pt}
\tablehead{\colhead{$\lambda$ }                  &
           \colhead{\% T}                        &
           \colhead{$\lambda$ }                  &
           \colhead{\% T}                        &
           \colhead{$\lambda$ }                  &
           \colhead{\% T}                        &
           \colhead{$\lambda$ }                  &
           \colhead{\% T}                        &
           \colhead{$\lambda$ }                  &
           \colhead{\% T}                        }
\startdata
3000 & 10.495 & 3260 & 58.085 & 3520 & 77.045 & 3780 & 69.440 & 4040 & 4.785 \\
3010 & 12.090 & 3270 & 59.380 & 3530 & 77.320 & 3790 & 67.845 & 4050 & 3.785 \\
3020 & 13.805 & 3280 & 60.665 & 3540 & 77.640 & 3800 & 66.085 & 4060 & 2.980 \\
3030 & 15.530 & 3290 & 61.890 & 3550 & 77.805 & 3810 & 64.185 & 4070 & 2.315 \\
3040 & 17.360 & 3300 & 63.010 & 3560 & 78.005 & 3820 & 62.125 & 4080 & 1.790 \\
3050 & 19.300 & 3310 & 64.110 & 3570 & 78.185 & 3830 & 59.920 & 4090 & 1.365 \\
3060 & 21.225 & 3320 & 65.060 & 3580 & 78.305 & 3840 & 57.460 & 4100 & 1.030 \\
3070 & 23.250 & 3330 & 66.060 & 3590 & 78.435 & 3850 & 54.860 & 4110 & 0.770 \\
\enddata
\label{tab:table5}
\end{deluxetable}

\newpage
\begin{deluxetable}{ccccccccccc}
\tablecaption{Transmission characteristics of the $B$ filter in KPNO's $UBVRI$ filter set J}
\tabletypesize{\footnotesize}
\tablewidth{0pt}
\tablehead{\colhead{$\lambda$ }                  &
           \colhead{\% T}                        &
           \colhead{$\lambda$ }                  &
           \colhead{\% T}                        &
           \colhead{$\lambda$ }                  &
           \colhead{\% T}                        &
           \colhead{$\lambda$ }                  &
           \colhead{\% T}                        &
           \colhead{$\lambda$ }                  &
           \colhead{\% T}                        }
\startdata
3520 &  0.000 & 4100 & 51.560 & 4680 & 49.205 & 5260 & 1.830 & 5840 & 0.100 \\
3530 &  0.010 & 4110 & 52.030 & 4690 & 48.175 & 5270 & 1.655 & 5850 & 0.085 \\
3540 &  0.030 & 4120 & 52.505 & 4700 & 47.140 & 5280 & 1.495 & 5860 & 0.070 \\
3550 &  0.065 & 4130 & 52.930 & 4710 & 46.030 & 5290 & 1.360 & 5870 & 0.055 \\
3560 &  0.125 & 4140 & 53.380 & 4720 & 44.925 & 5300 & 1.245 & 5880 & 0.050 \\
3570 &  0.220 & 4150 & 53.760 & 4730 & 43.765 & 5310 & 1.140 & 5890 & 0.040 \\
3580 &  0.360 & 4160 & 54.150 & 4740 & 42.575 & 5320 & 1.055 & 5900 & 0.035 \\
3590 &  0.565 & 4170 & 54.580 & 4750 & 41.390 & 5330 & 0.980 & 5910 & 0.030 \\
\enddata
\label{tab:table6}
\end{deluxetable}

\newpage
\begin{deluxetable}{ccccccccccc}
\tablecaption{Transmission characteristics of the $V$ filter in KPNO's $UBVRI$ filter set J}
\tabletypesize{\footnotesize}
\tablewidth{0pt}
\tablehead{\colhead{$\lambda$ }                  &
           \colhead{\% T}                        &
           \colhead{$\lambda$ }                  &
           \colhead{\% T}                        &
           \colhead{$\lambda$ }                  &
           \colhead{\% T}                        &
           \colhead{$\lambda$ }                  &
           \colhead{\% T}                        &
           \colhead{$\lambda$ }                  &
           \colhead{\% T}                        }
\startdata
4500 &  0.025 & 5100 & 76.505 & 5700 & 66.865 & 6300 & 19.130 & 6900 & 1.310 \\
4510 &  0.030 & 5110 & 77.025 & 5710 & 66.260 & 6310 & 18.525 & 6910 & 1.230 \\
4520 &  0.035 & 5120 & 77.525 & 5720 & 65.565 & 6320 & 17.920 & 6920 & 1.165 \\
4530 &  0.040 & 5130 & 77.950 & 5730 & 64.870 & 6330 & 17.320 & 6930 & 1.100 \\
4540 &  0.045 & 5140 & 78.375 & 5740 & 64.170 & 6340 & 16.735 & 6940 & 1.045 \\
4550 &  0.050 & 5150 & 78.755 & 5750 & 63.430 & 6350 & 16.145 & 6950 & 0.975 \\
4560 &  0.060 & 5160 & 78.990 & 5760 & 62.750 & 6360 & 15.590 & 6960 & 0.930 \\
4570 &  0.065 & 5170 & 79.315 & 5770 & 61.980 & 6370 & 15.045 & 6970 & 0.880 \\
\enddata
\label{tab:table7}
\end{deluxetable}

\newpage
\begin{deluxetable}{ccccccccccc}
\tablecaption{Transmission characteristics of the $R$ filter in KPNO's $UBVRI$ filter set J}
\tabletypesize{\footnotesize}
\tablewidth{0pt}
\tablehead{\colhead{$\lambda$ }                  &
           \colhead{\% T}                        &
           \colhead{$\lambda$ }                  &
           \colhead{\% T}                        &
           \colhead{$\lambda$ }                  &
           \colhead{\% T}                        &
           \colhead{$\lambda$ }                  &
           \colhead{\% T}                        &
           \colhead{$\lambda$ }                  &
           \colhead{\% T}                        }
\startdata
5470 &  0.000 & 6440 & 71.125 & 7410 & 32.765 & 8380 & 5.770 & 9350 & 0.640 \\
5480 &  0.005 & 6450 & 70.895 & 7420 & 32.310 & 8390 & 5.690 & 9360 & 0.610 \\
5490 &  0.015 & 6460 & 70.550 & 7430 & 31.880 & 8400 & 5.525 & 9370 & 0.590 \\
5500 &  0.030 & 6470 & 70.310 & 7440 & 31.500 & 8410 & 5.455 & 9380 & 0.580 \\
5510 &  0.065 & 6480 & 69.935 & 7450 & 31.035 & 8420 & 5.300 & 9390 & 0.600 \\
5520 &  0.130 & 6490 & 69.650 & 7460 & 30.690 & 8430 & 5.195 & 9400 & 0.545 \\
5530 &  0.250 & 6500 & 69.465 & 7470 & 30.270 & 8440 & 5.085 & 9410 & 0.530 \\
5540 &  0.450 & 6510 & 69.115 & 7480 & 29.920 & 8450 & 4.940 & 9420 & 0.545 \\
\enddata
\label{tab:table8}
\end{deluxetable}

\newpage
\begin{deluxetable}{ccccccccccc}
\tablecaption{Transmission characteristics of the $I$ filter in KPNO's $UBVRI$ filter set J}
\tabletypesize{\footnotesize}
\tablewidth{0pt}
\tablehead{\colhead{$\lambda$ }                  &
           \colhead{\% T}                        &
           \colhead{$\lambda$ }                  &
           \colhead{\% T}                        &
           \colhead{$\lambda$ }                  &
           \colhead{\% T}                        &
           \colhead{$\lambda$ }                  &
           \colhead{\% T}                        &
           \colhead{$\lambda$ }                  &
           \colhead{\% T}                        }
\startdata
6860 & 0.000 & 7890 & 89.435 & 8920 & 86.015 & 9950 & 71.760 & 10980 & 31.660 \\
6870 & 0.000 & 7900 & 89.440 & 8930 & 85.950 & 9960 & 71.535 & 10990 & 31.250 \\
6880 & 0.010 & 7910 & 89.335 & 8940 & 85.910 & 9970 & 71.315 & 11000 & 30.810 \\
6890 & 0.020 & 7920 & 89.545 & 8950 & 85.885 & 9980 & 71.005 & 11010 & 30.370 \\
6900 & 0.035 & 7930 & 89.695 & 8960 & 85.780 & 9990 & 70.755 & 11020 & 29.885 \\
6910 & 0.060 & 7940 & 89.655 & 8970 & 85.710 & 10000 & 70.460 & 11030 & 29.525 \\
6920 & 0.105 & 7950 & 89.215 & 8980 & 85.610 & 10010 & 70.210 & 11040 & 29.030 \\
6930 & 0.165 & 7960 & 89.625 & 8990 & 85.535 & 10020 & 69.940 & 11050 & 28.610 \\
\enddata
\label{tab:table9}
\end{deluxetable}

\newpage
\begin{deluxetable}{lrccccccc}
\tablecaption{Accurate Coordinates and Proper Motions for Stars in Table \ref{tab:table4}.}
\tabletypesize{\tiny}
\tablewidth{0pt}
\tablehead{\colhead{Star Name}                                &
           \colhead{2MASS-PSC}                                &
           \multicolumn{2}{c}{$\alpha$~~~~(2000)~~~~$\delta$} &
           \colhead{$\mu_{\alpha}$}                           &
           \colhead{$\mu_{\delta}$}                           &
           \colhead{$\mu$ Ref.}                               &
           \colhead{Sp. Type}                                 &
           \colhead{Sp. Type Ref.}                            \\
           \colhead{ }                                        &
           \colhead{UCAC2}                                    &
           \colhead{\hms}                                     &
           \colhead{\dms}                                     &
           \colhead{(mas/yr)}                                 &
           \colhead{(mas/yr)}                                 &
           \colhead{}                                         &
           \colhead{}                                         &
           \colhead{}                                         }
\tablenotetext{~}{1.  \citep{Klemola1987} [Lick NPM1 Catalogue]}
\tablenotetext{~}{2.  \citep{Monet2003} [USNO-B1.0]}
\tablenotetext{~}{3.  \citep{Hanson2004} [Lick NPM2 Catalogue]}
\tablenotetext{~}{4.  \citep{Urban2004} [UCAC2-BSS]}
\tablenotetext{~}{5.  \citep{Zacharias2004} [UCAC2]}
\tablenotetext{~}{6.  \citep{McCookSion2006}}
\tablenotetext{~}{7.  \citep{Bohlin1990}}
\tablenotetext{~}{8.  \citep{Oke1990}}
\tablenotetext{~}{9.  \citep{Roman1955}}
\startdata
G158 100             &    00335445-1207588 & 00 33 54.46  & -12 07 58.8  &   155.6  &  -192.9  &      1  &  sdG  &  8  \\
                     &                     &              &              &   154    &  -192    &      2  &       &     \\
BPM 16274            &    00500367-5208155 & 00 50 03.68  & -52 08 15.6  &   132    &    44    &      2  &  DA2  &  6  \\
                     &                     &              &              &          &          &         &       &     \\
HZ 4                 &    03552198+0947180 & 03 55 21.99  & +09 47 18.0  &   168    &     4    &      2  &  DA4  &  6  \\
                     &                     &              &              &          &          &         &       &     \\
LB 227               &    04092888+1707542 & 04 09 28.89  & +17 07 54.3  &   104    &   -24    &      2  &  DA3  &  6  \\
                     &                     &              &              &          &          &         &       &     \\
\enddata
\label{tab:table10}
\end{deluxetable}

\clearpage

\clearpage

\clearpage

\clearpage

\clearpage

\clearpage

\clearpage

\clearpage

\clearpage

\clearpage

\clearpage
 
\clearpage
 
\clearpage
 
\clearpage
 
\clearpage
 
\clearpage
 
\clearpage
 
\clearpage
 
\clearpage
 
\clearpage
 
\clearpage
 
\clearpage
 
\clearpage
 
\clearpage
 
\clearpage
 
\clearpage
 
\clearpage
 
\clearpage
 
\clearpage
 
\clearpage
 
\clearpage
 
\clearpage
 
\clearpage
 
\clearpage
 
\clearpage
 
\clearpage
 
\clearpage
 
\clearpage
 
\clearpage
 
\clearpage
 
\clearpage

\clearpage

\clearpage

\clearpage

\clearpage

\clearpage

\end{document}